\documentclass[aps,showpacs,nofootinbib,preprintnumbers,amsmath,amssymb,
 floatfix]{revtex4-1}
\usepackage{caption}
\usepackage{amsmath, amssymb, slashed, color, graphicx}
\usepackage{epsfig}
\usepackage{hyperref,enumerate}
\usepackage{pstricks}
\usepackage{slashed}
\usepackage{amsfonts}
\usepackage{url}
\usepackage{color}
\usepackage[T1]{fontenc}
\usepackage{lmodern}
\usepackage{subfig}
\hypersetup{colorlinks,citecolor= nicegreen,linkcolor= nicered}
\definecolor{nicered}{rgb}{0.7,0.1,0.1}
\definecolor{nicegreen}{rgb}{0.1,0.5,0.1}

\usepackage[normalem]{ulem}

\def\({\left(}
\def\){\right)}
\def\[{\left[}
\def\]{\right]}

\begin{document}

\begin{flushright}  \mbox{\normalsize \rm CUMQ/HEP 190, HIP-2016-20/TH,  TIFR/TH/16-21}
       \end{flushright}
\vskip -5cm      
\title{Probing Higgs-radion mixing in warped models through complementary searches at the LHC
and the ILC}
\vskip 2cm

\author{
~Mariana Frank,$^1$
              \footnote{\tt{Electronic address: mariana.frank@concordia.ca}}%
~Katri Huitu,$^2$
              \footnote{\tt{Electronic address: katri.huitu@helsinki.fi}}%
~Ushoshi Maitra,$^3$
              \footnote{\tt{Electronic address: ushoshi@theory.tifr.res.in}}              
~Monalisa Patra$^4$
              \footnote{\tt{Electronic address: mpatra@irb.hr}}}

\affiliation{$^1$Department of Physics, Concordia University,
7141 Sherbrooke St. West, Montreal, Quebec, Canada H4B 1R6 \\
$^2$Department of Physics and Helsinki Institute of Physics, P.O. Box 64 (Gustaf H\"allstr\"ominkatu 2),
FIN-00014 University of Helsinki, Helsinki, Finland \\
$^3$ Department of Theoretical Physics, Tata Institute of Fundamental Research, Mumbai 400 005, India \\ 
$^4$Ru\dj jer Bo\v{s}kovi\'c Institute, Division of Theoretical Physics, Bijeni\v{c}ka 54, HR-10000 Zagreb, Croatia
             }
\thispagestyle{myheadings}


\vskip 5cm

\begin{abstract}
We consider the  Higgs-radion mixing in the context of warped space extra dimensional models with custodial symmetry 
and investigate the prospects of detecting the mixed radion.  Custodial symmetries allow the Kaluza-Klein excitations 
to be lighter, and protect $Zb {\bar b}$ to be in agreement with experimental constraints. We 
perform a complementary study of discovery reaches of the Higgs-radion mixed state
at the 13 and 14 TeV LHC and at the 500 and 1000 GeV ILC. We  carry out a comprehensive analysis
of the most significant production and decay modes of the mixed radion in the 80 GeV $-$ 1 TeV 
mass range, and indicate the parameter space that can be probed at the LHC and the  
ILC. There exists a region of the parameter space which can be probed, at the LHC, through the diphoton channel
even for a relatively low luminosity of 50 fb$^{-1}$. The reach of the 4-lepton final state,
in probing the parameter space is also studied in the context of 14 TeV LHC, for a luminosity of 1000 fb$^{-1}$.
At the ILC, with an integrated luminosity of 500 fb$^{-1}$, we analyze the $Z$-radion associated production and
the $WW$ fusion production, followed by the radion decay into $b{\bar b}$ and $W^+W^-$. The $WW$ fusion production is 
favored over the $Z$-radion associated channel in probing regions of the parameter space beyond the LHC reach.  The complementary 
study at the LHC and the ILC is useful both for the discovery of the radion and the understanding of its mixing sector. 

\end{abstract}

\pacs{12.60.-i, 11.10.Kk, 14.80.Ec, 14.80.Rt} 
\maketitle

\section{Introduction}

The Standard Model (SM) of particle physics has been successful in describing the fundamental 
particles of our world and is currently in good agreement with almost all the experimental results. However,
under closer scrutiny the SM reveals that there are well motivated reasons for expecting  
new physics (NP). The most discussed topic in the area of theoretical shortcomings of
the SM is the so-called hierarchy
problem. In a nut-shell, the problem is that the SM fails to explain why the Higgs mass is light 
(electroweak scale), while calculations within the SM framework allow it to grow indefinitely to Planck scales. 
Many theoretical frameworks 
were proposed  to take into account and resolve the hierarchy problem. Supersymmetry and 
extra dimensional models are the most studied scenarios in the literature. In this analysis, 
we will work in the context of Randall--Sundrum (RS) warped extra dimensional model.
 
The solution to the gauge hierarchy problem in the context of extra dimensions, earlier
suggested by Arkani-Hamed, Dimopoulos and Dvali (ADD)~\cite{ArkaniHamed:1998rs,ArkaniHamed:1998nn} 
allowed only gravity to access
the extra dimension(s). 
 The ADD scenario explained the weakness of the gravity 
compared to other forces by the fact that gravity becomes diluted in the volume 
of large extra dimensions. However, this scenario transferred the gauge hierarchy
problem onto the problem of the discrepancy between the large size of extra 
dimensions, R $\approx$ 1 mm (fixed from experimental constraints) and the natural 
value of R $\approx$ 10$^{-33}$ cm.
An ambitious proposal to the hierarchy problem
was then put forward by Randall and 
Sundrum. In the original RS model~\cite{Randall:1999ee}, there are two 3-branes embedded in 
the 5-dimensional(5-D) anti-de Sitter (AdS) space, with all the SM particles localized on the visible 3-brane, 
and only the graviton propagating in the bulk. This model offers a simple and natural
solution to the hierarchy problem. The separation between the 
two 3-branes leads directly to the existence of an additional 
scalar called the radion, corresponding to the quantum fluctuations of the  distance between the two 
3-branes. 
The radion is massless in the limit of the backreaction going to zero, but acquires
a mass with a suitable stabilizing mechanism \cite{Goldberger:1999uk}. The radion can be much 
lighter than the massive gravitons. 
For different phenomenological studies, the mass of the radion $m_\varphi$ is usually considered in the 
range of $\mathcal{O}(10~{\rm GeV}) \leq m_\varphi \leq  \mathcal{O}(\rm{TeV})$~\cite{Kribs:2001ic}.
The radion couples with the matter via the trace of the energy momentum tensor.
Thus, the structure of the coupling of the radion with the SM
fields is  similar to that of the Higgs boson. However, for massless gauge bosons, there is an enhancement coming
from the trace anomaly term.  General covariance allows a possibility of mixing between the radion and the Higgs boson.
The dedicated analysis of the Higgs sector at the LHC will help in constraining the 
Higgs-radion mixing hypotheses. The phenomenology of the Higgs-radion mixed 
sector has been studied thoroughly in the literature in the context of the electroweak 
precision measurements and using the Higgs results from the
LHC~\cite{Giudice:2000av,Chaichian:2001rq,Huitu:2011zh,Dominici:2002jv,Rizzo:2002bt,Gunion:2003px,Battaglia:2003gb,Toharia:2008tm,
deSandes:2011zs,Grzadkowski:2012ng,Kubota:2012in,Cho:2013mva,Desai:2013pga,Kubota:2014mma}.

In an effort to improve the predictions of the model, and to render it suitable for phenomenology,  the RS model was modified 
allowing the gauge bosons to propagate in the bulk, so as to facilitate the gauge coupling 
unification~\cite{Agashe:2002pr}. Allowing the gauge bosons in the bulk in turn introduces
towers of gauge Kaluza-Klein(KK) states on the TeV brane. The SM fermions and the Higgs, 
localized on the TeV brane, couple maximally to these KK gauge states, leading 
to large corrections to the Peskin-Takeuchi ($S$ and $T$) parameters, respectively. This in turn 
leads to strong constraints on the 5-D theory from the electroweak
precision observables. The bounds on the KK masses are of the order of 
30 TeV~\cite{Davoudiasl:1999tf,Chang:1999nh,Huber:2000fh,Csaki:2002gy,Delaunay:2010dw} 
in the models with only the SM gauge bosons and the gravitons in the bulk. 
Putting the SM fermions in the 
bulk along with the massive SM gauge bosons solves the flavor violation problems of the theory \cite{Gherghetta:2000qt}. 
The Higgs boson is localized on the IR brane in this modified model, to account for the gauge hierarchy 
problem. This approach naturally explains the hierarchy problem 
in the Yukawa sector, with the fermions being localized at different points 
in the bulk. Localizing  lighter fermions nearer to the Planck brane, with 
the top and the Higgs localized on the IR brane, leads to weakening of constraints 
on the KK gauge states from the oblique $S$ parameter. However the 
strong constraints from the $T$ parameter still persist~\cite{Delgado:2007ne}, 
with the electroweak precision results imposing a bound of around 10 TeV  
on the mass of lightest spin one resonances.  This strong constraint from $T$ 
parameter  in these 5D warped bulk scenarios can be mitigated by 
various methods, discussed in~\cite{Casagrande:2008hr}. One of 
the possible cures for the increased $T$ parameter is to extend the symmetry of the model to the custodial $SU(2)_R$ 
symmetry, first discussed in~\cite{Agashe:2003zs}. The bulk gauge 
symmetry in this case is  $SU(3)_c \times SU(2)_L \times SU(2)_R \times U(1)_X$. 
The tree-level $S$ and $T$ parameters in this scenario are well behaved,
with the $T$ parameter likely to vanish and the constraints mainly 
coming from the $S$ parameter. This in turn leads into a lower 
bound of about 6 TeV on the first KK mode of the gauge bosons. 
The second possible solution which limits the masses of the first 
KK gauge bosons to the order of 5 TeV, are the ones with large 
brane localized kinetic terms for the gauge 
fields~\cite{Davoudiasl:2002ua,Carena:2003fx}. The third possibility
is in the form of the models with deformed metric~\cite{Falkowski:2008fz,Cabrer:2011fb,Frank:2013un,Frank:2014aca}. 
In this framework  the SM particles, including the Higgs, are in the bulk, the space  
departs from AdS$_5$ near the IR brane, while being similar to RS near the UV brane. In these scenarios KK gauge boson modes  as low as 1-2 TeV can be consistent with the electroweak 
precision tests, for a suitable choice of model parameters. Recent works discussing 
the bounds on the lightest KK gauge boson, in the context of the above scenarios, with
up to date electroweak fits are in~\cite{Iyer:2015ywa}.

Here we choose to work in the framework of custodial RS model and concentrate on the phenomenological aspect 
of the Higgs-radion mixing, where all the SM particles except the Higgs bosons are in the bulk. The 
fermions are allowed to be localized anywhere in the bulk, leading to the natural generation of the Yukawa coupling 
hierarchies. The fermion masses depend on the bulk mass parameters $c_L$ (for the doublets) and $c_R$  
(for right-handed fields) characterizing the profiles of the zero mode  fermions. The values of $c_L$ and 
$c_R$ are fixed depending on the wavefunction of the fermion on the Planck brane or the TeV brane.
They are chosen to match the fermion mass hierarchy fixed through the physical quark masses and mixing at the weak scale.
The custodial RS model, compared to the original RS model, predicts the existence
of 5 additional gauge bosons, 3 of which are neutral and 2 of which are charged.
The coupling of the radion with the SM gauge bosons (massive and massless) in custodial
RS model is similar to the non-custodial case. The top quark being heavy (localized on 
the TeV brane like the new KK modes) couples strongly to the new heavy gauge bosons. The bottom 
quark ($b_L$) being in the same electroweak doublet as $t_L$ is also affected, 
and in turn modifies the $Z b_L \bar{b}_L$ coupling, which must be in accordance
with the SM prediction at 0.25\% level. This calls for adjusting the profiles for 
$(t,b)_L$ and $t_R$, so as to protect the $Zb\bar{b}$ coupling. There are many analyses 
which have looked into different fermion representations, with the custodial symmetry protecting
$Zb\bar{b}$~\cite{Agashe:2006at,Agashe:2007ki,Agashe:2008jb}. These scenarios consider the first
two generations of the quarks and the leptons as doublets
under the $SU(2)_L$. The third generation left-handed quarks are in the doublet representation of both
$SU(2)_L$ and $SU(2)_R$, which leads to non-SM fermions with no zero modes. These fermions are very heavy and
therefore not relevant for our analysis. 
In order to accommodate the large top and bottom mass difference, $t_R$ can be either in a singlet or in
a triplet representation of $SU(2)_R$ (for details see~\cite{Agashe:2006at,Agashe:2008jb}).
Overall the top and the bottom quark can have various possible quantum numbers and profiles in the 
extra dimension, so as to satisfy the constraints from flavor violation and precision tests.  
Different possibilities include the localization of $t_R$ very close to the TeV brane with $(t,b)_L$ having a profile
close to flat, or the contrary case with $(t,b)_L$ very close to the TeV brane and
$t_R$ close to flat. The intermediate case with the profile of $(t,b)_L$ and $t_R$ being near, but not close to the 
TeV brane is also possible. It was shown in Ref.~\cite{Carena:2007ua,Agashe:2007ki} that 
the electroweak fits favor a close to flat profile for $(t,b)_L$, with $t_R$ peaked near the TeV brane. Therefore 
we consider this case, with $c^{(t,b)}_L$ = 0.4 - 0.3 and $c^t_R$ = 0. All the other fermions including $b_R$, are 
assumed to be localized on or very near the Planck brane, with $c^i_L~>~1/2,~c^i_R~<~-1/2$. 
The specific choice of representations of the third generation quarks will not affect our results for 
$gg,\gamma\gamma, W^+W^-$ final states, while the values for $c^{(t,b)}_L$ and $c^{(b)}_R$ would influence 
the analysis in the $b\bar{b}$ final state, but not by more than $\mathcal{O}(1)$ factor.   

Within the present scenario, with the latest constraints coming from the LHC 8 TeV results, 
it is worthwhile to study the current status of the Higgs-radion mixing. 
As an artifact of the trace anomaly, the production of the radion at the LHC proceeds 
dominantly through the gluon fusion. The radion
can also be produced through vector boson fusion as well as in association with a 
single massive gauge boson or $t{\bar t}$ pair.
Once produced, it can decay to a pair of gluon, photon or $b\bar{b}$ final state
with sizable cross section. The decay to a pair of massive gauge bosons as well as a pair of Higgs bosons, 
if kinematically allowed, can also be observed in the LHC. One can observe a light radion at the LHC in the 
diphoton channel~\cite{Davoudiasl:2012xd,Bhattacharya:2014wha}. In the presence of the Higgs-radion mixing,
the coupling of both the scalars to the SM particle is altered.
There exists a particular value of mixing, where one of the scalars couples maximally to the gluon and the photon, and 
almost decouples from the massive SM particles. 
As will be discussed later, the parameter space involving the scenario of the Higgs mixed with the radion 
mostly is restricted by the current LHC bounds coming from the heavy scalar searches in the $WW$ and the $ZZ$ channel.
Therefore only this particular value of mixing, which can only be probed through the diphoton final state, is still allowed.
We first perform a comprehensive analysis of this region of the parameter space (where the radion only couples to 
the gluon and the photon), for the case of both 13 and 14 TeV LHC, with the radion mass varying from 80 GeV $-$ 1 TeV.
The $WW$ and $ZZ$ decay modes dominate over most of the mixing parameter space, once kinematically
allowed. The hadronic decay channels of these gauge bosons are difficult to observe at the LHC. We therefore
repeat our analysis with the radion being produced through gluon fusion, and decaying to $ZZ$, with the $Z$
decaying leptonically. We do not consider the $W$ decay channel, because the leptonic decay of $W$ is accompanied
by missing energy, making it difficult for the scalar mass reconstruction. 
The mixed radion  can also be produced at the LHC via vector boson fusion, but the production rate are suppressed by the vacuum expectation value
(VEV) of the radion.

The proposed International Linear Collider(ILC)~\cite{Behnke:2013xla,Baer:2013cma} will be the next generation
$e^+e^-$ collider, designed to operate at the center-of-mass (c.m.) energies of 250, 500 and 1000 GeV,
with integrated luminosity of 250 fb$^{-1}$, 500 fb$^{-1}$, and 1000 fb$^{-1}$, respectively.
Compared to the LHC, the leptonic linear collider has moderate hadronic backgrounds
and a tunable but restricted centre of mass energy. The ILC is mainly intended for precision measurements
of the masses and couplings of the SM particles. Therefore if there is any evidence of new physics 
at the LHC, a thorough precision study in the ILC will be necessary to pinpoint the validity of the
new scenario.  The ILC will also has the additional advantage of initial beam polarization, 
both longitudinal and transverse (60\% for $e^+$ and 90\% for $e^-$). 
At the ILC, the direct search for the radion $\varphi$ can be  via $e^+e^-\rightarrow Z \varphi$ and $e^+e^-\rightarrow \nu \bar{\nu} \varphi$,
with the radion decaying to either $b\bar{b}$ or $WW/ZZ$. The analysis through these decay channels will help the ILC to probe those regions of the parameter space which will be difficult 
to explore at the LHC. We have therefore also performed a through analysis, for radion of
mass in the range 100 GeV $-$ 1 TeV, in case of ILC, considering the above production channels.
The $b\bar{b}$ and the hadronic decay channel of the $W$ are considered. 
The main purpose of this work is to show the synergy of the LHC and the ILC in exploring 
the Higgs-radion mixed scenario, in the context of warped  model with custodial symmetry.

This paper is organized as follows. In Sec.~\ref{sec:RS}, we review the 
Randall-Sundrum model and the emergence of the radion. The coupling of the Higgs and the radion
to the SM particles, prior to mixing, is reviewed in Sec.~\ref{sec:non-mixing} and the 
mixing case is discussed in Sec.~\ref{sec:mixing}.   The detailed study in the LHC for the favored parameter space is explored in Sec.~\ref{sec:lhc_analysis}, 
after accounting for the constraints on the Higgs-radion parameter space, from the latest LHC results  
in Sec.~\ref{subsec:constraints}.
The complementary study in Sec.~\ref{sec:ilc_analysis} is devoted to our systematic analysis in the ILC.
Finally we summarize our findings and conclude in Sec.~\ref{sec:conclusion}.

\section{The Randall Sundrum model}\label{sec:RS}

The original version of the RS model  has one extra dimension,  
compactified on a circle. This compactification has a topology of 
$S^1/\mathcal{Z}_2$ orbifold, where $S_1$ is a sphere in one dimension 
and $\mathcal{Z}_2$ is the multiplicative group $\{-1,1\}$. By construction,
the final picture is of two 3-branes, separated by a distance and enclosing 
a 5D bulk. The branes are located at the orbifold fixed points, 
with $\phi$ = 0, $\pi$. The two branes are required to have opposite tensions, 
which cancel the 5D bulk cosmological constant to yield a
vanishing 4D cosmological constant. These 3-branes support the 
3+1 dimensional theory and are called the visible (TeV) and the hidden 
(Planck) branes. The fundamental action describing the above part, excluding 
the 3-branes, is
\begin{equation}
\mathcal{S}=\int d^4x\int_{-\pi}^{\pi}d\phi \sqrt{-G} (2 M_5^3 {\cal R}[G] -\Lambda),
\end{equation}
where $G$ is the determinant of the five-dimensional metric, $M_5$ is the 
fundamental 5D mass scale, ${\cal R}$ is the Ricci scalar and $\Lambda$ is the bulk cosmological constant. The 5-D metric, 
respecting the four-dimensional Poincare invariance in the $x^\mu$ direction, 
takes the form
\begin{equation}
ds^2=e^{-2 k r_c |\phi|} \eta_{\mu\nu}dx^\mu dx^\nu + r_c^2 d\phi^2, \qquad -\pi \le \phi \le \pi,
\end{equation}
with $r_c$  the compactification radius and $k$  the bulk curvature.
As mentioned before, in the original RS model, the SM particles are
localized on the brane. Therefore the four dimensional effective action is obtained 
by integrating out the extra dimension. All the fields have initially masses near 
the 4D Planck scale, and the fundamental mass parameter is exponentially 
suppressed, as $M_{\rm{TeV}}=e^{-k r_c \pi} M_{\rm{Planck}}$, with the physical mass 
being warped down to the weak scale. Since $M_{\rm{TeV}} \approx 10^{-16} M_{\rm{Planck}}$,
the size of the extra dimension is given by $k r_c \pi \approx 35$. The compactification radius, $r_{c}$ is 
arbitrary and is treated as a free parameter in the theory. 
It can be considered as a fluctuation in the extra dimension, which results to the existence of a massless scalar field called the radion.
Goldberger and Wise~\cite{Goldberger:1999un,Goldberger:1999wh} proposed 
an interesting solution to fix the size of extra dimension by introducing a massive scalar field in the bulk with 
an associated potential, along with interaction terms on the two 3-branes. Taking into account 
the backreaction of the geometry due to the scalar field, an effective potential 
 is generated, which stabilizes the compactified radius. This effective potential generates mass and 
VEV of the radion.
 The 5-D metric is then subsequently 
expanded taking into account the scalar perturbations $F(x,\phi)$ due to the effect
of the radion field. 
\begin{equation}
ds^2=e^{-2(kr_c|\phi | +F(x,\phi))}\eta_{\mu\nu} dx^\mu dx^\nu-(1+2F(x,\phi))^2 r_c^2 d\phi^2,
\end{equation}
where $F(x,\phi) = \varphi(x)R(\phi)$~\cite{Csaki:2000zn}. Here $R(\phi)$ is determined by requiring that 
the metric solves Einstein's equations, whereas $\varphi(x)$ is the canonically 
normalized 4D scalar field obtained after integrating out the extra dimension. When 
the backreaction of the metric background due to the scalar field is small, 
the wavefunction is then
\begin{equation}
F(x,\phi)=\frac{\varphi(x)}{\Lambda_\phi}e^{2kr_c(\phi-\pi)},
\end{equation}
where $\Lambda_\phi=\sqrt{6}M_{\rm{Planck}}e^{-kr_c\pi}$  is the VEV of the radion. The mass of the 
radion depends on the mass of the bulk scalar and can be 
smaller than 1 TeV. As the other KK fluctuations lie in the scale of at least 2-3 TeV, the radion is 
the lightest scalar in this scenario that can be detected directly at the LHC.

\section{Radion and Higgs boson couplings with the SM fermions and gauge bosons}
\label{sec:non-mixing}
The model  considered here is the  5D electroweak group incorporated
 in a custodial $SU(2)$ symmetry. The gauge group in the bulk is 
 $SU(2)_L \times SU(2)_R  \times U(1)_{B-L}$, with the breaking of 
 $SU(2)_L \times SU(2)_R \rightarrow SU(2)_D$ on the TeV brane, and 
 $SU(2)_R  \times U(1)_{B-L} \rightarrow U(1)_Y$ on the Planck brane. 
 In this picture, the matter and the gauge fields 
propagate in the bulk. The electroweak symmetry  breaking 
on the TeV brane is achieved via a localized Higgs as discussed in~\cite{Agashe:2003zs}. 
The zero modes of the bulk fields are identified with the SM particles.
The fermions are assumed to be localized
near the Planck brane, for $c^i_L~>~1/2,~c^i_R~<~-1/2$, and near the IR brane
for $c^i_L~<~1/2,~c^i_R~>~-1/2$, where $i$ is the flavor index. In order to generate the Yukawa couplings hierarchy,
lighter fermions are assumed to be localized near Planck brane and $t_{R},t_{L},b_{L}$ are close to IR brane.
Assuming that the Yukawa 
coupling matrix is diagonal, the electron Yukawa coupling leads to 
$c^e~\simeq~0.64$ and the top Yukawa coupling constrains $c^t~\simeq~-0.5$.
Therefore for the remaining fermions the corresponding Yukawa couplings are
obtained for $c^t~\simeq~c^i~\simeq~c^e$~\cite{Huber:2000ie,Gherghetta:2010cj}. 
In our analysis, the $c^i$ for all the fermions except the bottom and
the top quark are fixed
such that the coupling of the radion to the fermions is similar to the 
SM Yukawa coupling ($c_L-c_R = 1$), that is, with the couplings being 
proportional to the mass of the fermions. 
The only difference arises from the fact that the couplings with the radion are suppressed by $\Lambda_\phi$ (the radion VEV)
instead of $v$ (the Higgs VEV). For the bottom quark, 
the profile parameters as discussed before are $c^b_L$ = 0.3 and $c^b_R < -0.5$. 
Similarly for the top quark we have $c^t_L$ = 0.3 and $c^t_R$ = 0. The interaction with fermions is then, for the Higgs and the radion
\begin{eqnarray}
 \mathcal{L}_h^{ff} &=& \frac{h}{v}\left(m_f \bar{\psi} \psi \right), \\
\mathcal{L}_\varphi^{ff} &=& \frac{\varphi}{\Lambda_\phi}\left(m_f (c^f_L - c^f_R) \bar{\psi} \psi \right),
\end{eqnarray}
where $(c^f_L - c^f_R)~\sim~1$ for all the fermions except the $b$ and $t$ quark. 

The radion coupling to the gauge bosons differs from the coupling of the gauge bosons
to the Higgs. We first discuss the couplings to massive gauge bosons, where in the case of the SM Higgs the coupling
is proportional to the mass. In case of the radion, due to the propagation of the 
gauge bosons in the bulk, there is an additional tree level coupling of the radion 
to the bulk kinetic term of the massive gauge bosons. This additional tree level coupling
is also present for the massless gauge bosons propagating in the bulk. The couplings for the Higgs boson and radion with the gauge bosons are, respectively,
\begin{eqnarray}
\mathcal{L}_h^{WW,ZZ} &=& \frac{h}{v}\left(2 M_W^2 W_\mu^+ W^{\mu -} + M_Z^2 Z_\mu Z^\mu \right), \\ 
 \mathcal{L}_\varphi^{WW,ZZ} &=& \frac{\varphi}{\Lambda_\phi} 
\left[2 M_W^2 \left(1-\frac{3k r_c \pi M_W^2}{\Lambda_\phi^2}\right)W_\mu^+ W^{\mu-} 
+ M_Z^2 \left(1-\frac{3k r_c \pi M_Z^2}{\Lambda_\phi^2} \right) Z_\mu Z^\mu \right].  
\end{eqnarray}
In the case of massless
gauge bosons, along with the tree level coupling, the effects of the localized trace anomalies 
on the TeV brane are also included, and these are proportional to the $\beta$-function coefficient
of the light fields localized on the TeV brane~\cite{Csaki:2000zn,Csaki:2007ns}. In addition to these couplings,
the SM fermions enter in triangle diagrams involving  decays into $\gamma\gamma$ and $gg$ and the 
$W$ boson in  $\gamma\gamma$ diagrams.  These triangle diagrams are similar to the massless gauge
boson coupling to the Higgs and calculated similarly. There can be additional
contributions from the KK fermions, and the KK $W$ boson in the triangle loop. We 
have neglected the contributions of the KK modes in our 
 analysis\footnote{The contributions of the KK modes for $W$ and $f$ depend on the 
 cutoff scale $\Lambda_\phi$ and additionally $KKf$ depend on the fermion bulk mass
 parameter. The contributions $KKW$ and $KKf$ are suppressed for large cut-off scales, 
 with $KKW$ contribution amounting to 0.06 the SM one,  and $KKt$ being  0.012 the SM one,  for $\Lambda_\phi$ = 4 TeV~\cite{Kubota:2012in}.
 The  $KKW$ contributions to $gg$ amount to $<5 \%$~\cite{Carena:2012fk}. These contributions are suppressed with respect to the trace anomaly part and hence, we have ignored them.}, as they are
very massive and therefore give negligible contributions. The  loop-induced couplings for Higgs and radion are, respectively,
\begin{eqnarray}
 \mathcal{L}_h^{gg,\gamma\gamma} &=& \frac{h}{4v}\left(\frac{\alpha_s}{2\pi}b^h_{QCD}G^a_{\mu\nu}
 G^{a \mu\nu} + \frac{\alpha}{2 \pi}b^h_{EM}F_{\mu\nu}F^{\mu\nu}\right), \\
 \mathcal{L}_\varphi^{gg,\gamma\gamma} &=& \frac{\varphi}{4\Lambda_\phi} 
 \left[\left(\frac{1}{k r_c \pi}+
 \frac{\alpha_s}{2\pi}b^R_{QCD}\right)G^a_{\mu\nu} G^{a \mu\nu} 
 +\left(\frac{1}{k r_c \pi} + \frac{\alpha}{2 \pi}b^R_{EM}\right)F_{\mu\nu}F^{\mu\nu} \right],
\end{eqnarray}
where $\alpha_s$ and $\alpha$ are the strong and the electromagnetic coupling constants and we define
\begin{eqnarray}
b^h_{QCD} &=&   \tau_f(1+(1-\tau_f)f(\tau_f)), \\
b^h_{EM} &=&  \frac{8}{3}\left[\tau_f(1+(1-\tau_f)f(\tau_f))\right]
-\left[2+3 \tau_W + 3 \tau_W (2-\tau_W)f(\tau_W)\right],  \\
b^R_{QCD} &=&  7+ \tau_f(1+(1-\tau_f)f(\tau_f)), \\
b^R_{EM} &=&-\frac{11}{3}+ \frac{8}{3}\left[\tau_f(1+(1-\tau_f)f(\tau_f))\right]
-\left[2+3 \tau_W + 3 \tau_W (2-\tau_W)f(\tau_W)\right],  \\
f(\tau) &=&  \left(\sin^{-1}\frac{1}{\sqrt{\tau}}\right)^2~ ({\rm{for}}~ \tau > 1), \\
f(\tau) &=& -\frac{1}{4}\left(\log \frac{\eta_+}{\eta_-}-i \pi\right)^2~ ({\rm{for}}~ \tau < 1), \\
\eta_\pm &=& 1\pm \sqrt{1- \tau},~ \tau_i = \left(\frac{2 m_i}{m_S}\right)^2. \label{hrfin}
\end{eqnarray}
In Eq.~\ref{hrfin} $m_i$ is the mass of the particle in the loop and $m_S$ is either mass of the
radion or the Higgs, depending on the coupling. The $\tau_f=m_f^2/m_S^2$ in the above equations denotes fermion mass ratios squared,
 whereas $\tau_W=m_W^2/m_S^2$ is for the $W$ gauge boson. Through its interaction with
the stress-energy momentum tensor, the radion couples with the Higgs boson as
\begin{eqnarray}\label{eq:rhh_nomix}
\mathcal{L}_\varphi^{hh} = \frac{\varphi}{\Lambda_\phi} (-\partial_\mu h \partial^\mu h + 2 m_h^2 h^2).
\end{eqnarray}

\section{The Higgs-radion mixing}
\label{sec:mixing}

\begin{figure}[htb]
\begin{center}
\includegraphics[width=7.5cm, height=6cm]{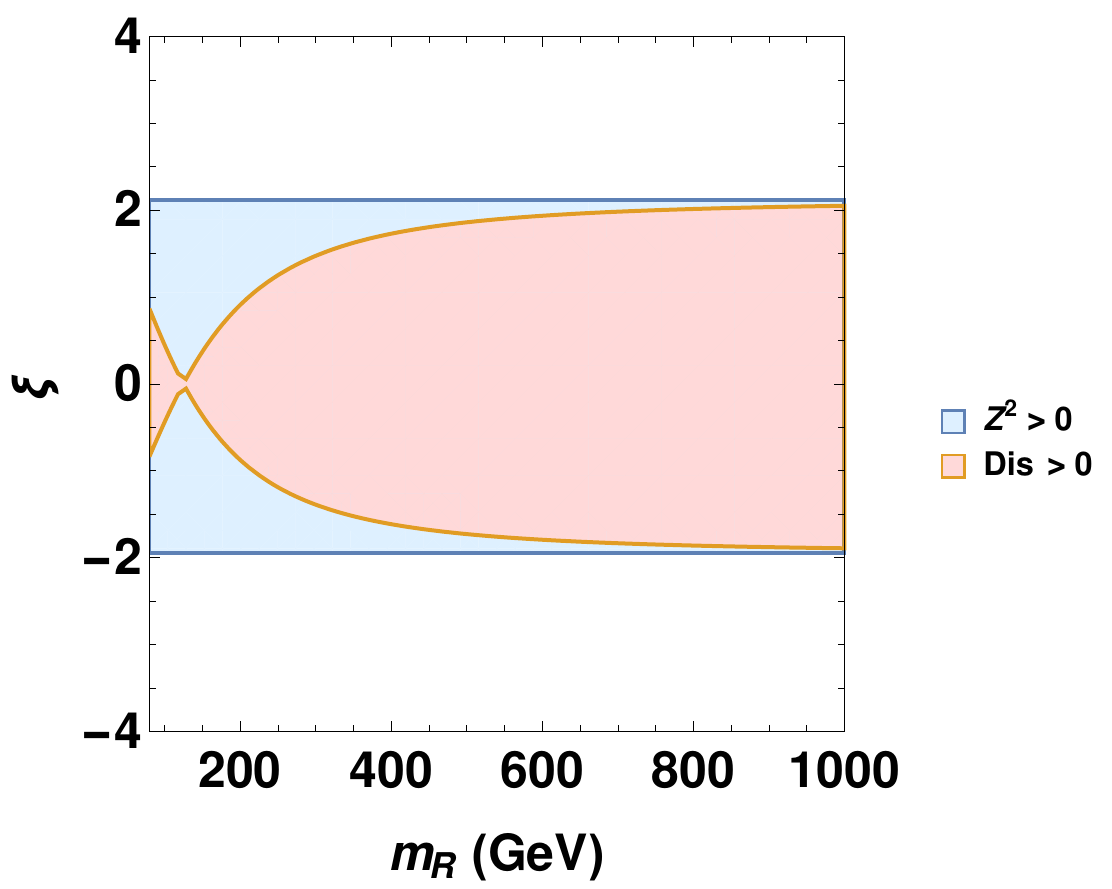}
\includegraphics[width=7.5cm, height=6cm]{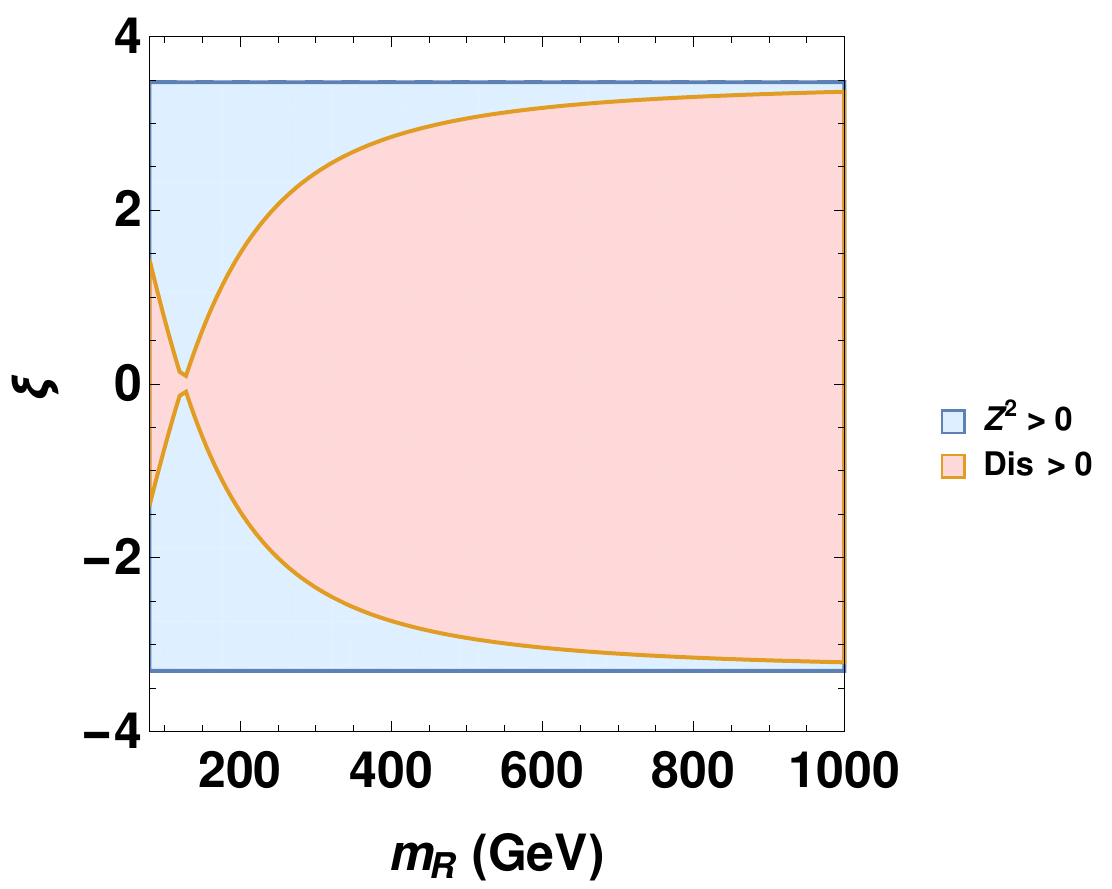}
\end{center}
\caption{ The allowed parameter region 
for $\xi$ satisfying Eq.~\ref{eq:cond1} (in blue for $Z^2 >0$, in pink for Dis $>0$), as a function 
of the radion mass $m_R$. We consider
$\Lambda_\phi$ = 3 TeV (left panel) and 5 TeV (right panel), and take $m_H$ = 125 GeV.}
\label{fig:xi_range}
\end{figure}

The operator giving rise to mixing between the radion and the Higgs boson  follows from 
the principle of general covariance~\cite{Dominici:2002jv}:
\begin{equation}
\label{eqn:action}
 S_{\xi} = \xi \int d^{4}x \sqrt{g_{vis}} {\cal R}(g_{vis}) \hat{H}^{\dagger} \hat{H}.
\end{equation}
Here $\xi$ is the mixing parameter, $g_{vis}^{\mu \nu}$ is
the metric induced on the visible brane and ${\cal R}(g_{vis})$ is a four dimensional Ricci scalar
of the induced metric. After expanding the radion field about its VEV and keeping only the terms 
containing bilinear fields, we get
\begin{equation}
\label{eqn:mix_lag}
 {\cal{L}}_{mix} = -\frac{1}{2} \left(1 + 6 \gamma^{2} \xi \right) \varphi \Box \varphi 
 - \frac{1}{2} \varphi m_{\varphi}^{2} \varphi - \frac{h}{2} \left(\Box + m_{h}^{2} \right) h 
 - 6 \xi \gamma h \Box \varphi,
\end{equation}
where $m_\varphi$ and $m_h$ are the radion and the Higgs mass.
In order to diagonalize the kinetic part of $\mathcal{L}_{mix}$, 
we consider the following transformation,
\begin{eqnarray}
\label{eqn:diag}
 h & = & (\cos{\theta} + \frac{6 \xi \gamma}{Z} \sin{\theta}) h_{m}
 +  (\sin{\theta} - \frac{6 \xi \gamma}{Z} \cos{\theta}) r_{m} 
 = a_\theta h_m + b_\theta r_m, \\ \nonumber
 \varphi & = & \frac{-\sin{\theta}}{Z} h_{m} + \frac{\cos{\theta}}{Z} r_{m} 
 = c_\theta h_m + d_\theta r_m,
\end{eqnarray}
where $ Z^{2} ~=~ 1 + 6 \xi \gamma^{2} - 36 \xi^{2} \gamma^{2}$ and the mixing angle is
\begin{equation}
\label{eq:tan2theta}
 \tan{2 \theta} = \frac{12 \xi \gamma Z m_{h}^{2} }
 { [m_{h}^{2}(Z^2 - 36 \xi^{2} \gamma^{2})- m_{\varphi}^{2}]}, 
\end{equation}
The corresponding mass eigenvalues of the physical fields ($h_{m}~(r_{m})$) are
then given by 
\begin{equation}\label{eq:massev}
m_{\pm}^{2} = \frac{1}{2Z^{2}} \left[ m_{\varphi}^{2}+ \beta m_{h}^{2}
 \pm \sqrt{( m_{\varphi}^{2}+ \beta m_{h}^{2})^2-4 Z^2 m_{h}^{2} m_{\varphi}^{2}}\right], 
\end{equation}
where $\beta = 1+ 6 \gamma^2 \xi$ and we denote the larger of $[m_{h_m},m_{r_m}]$ with $m_+$. 
Eq.~\ref{eq:massev} can be inverted to express $(m_h,m_\varphi)$ in terms of $m_\pm$.
\begin{equation}\label{eq:massev2}
[\beta m_{h}^{2}, m_\varphi^2] = \frac{Z^{2}}{2} \left[ m_{+}^{2}+ m_{-}^{2}
 \pm \sqrt{(m_{+}^{2}+ m_{-}^{2})^2-\frac{4 \beta m_+^2 m_-^2}{Z^2}}\, \right]. 
\end{equation}
Thus, the Higgs-radion mixed system can be described by four independent parameters, i.e the mixing parameter $\xi$, 
the radion VEV $\Lambda_{\phi}$, and the physical masses of the two mixed scalars $(m_{h_m})$ and $(m_{r_m})$.
In order to remove the quadratic ambiguity in the equation for physical masses, we adopt the convention
where $H$ is the Higgs-like and $R$ is the radion-like scalar
\footnote{Higgs-like, because it behaves like the SM Higgs boson in the limit $\xi~\rightarrow~0$. }.
From now on we will refer to the scalars $h_m$ and $r_m$ as $H$ and $R$ and similarly their masses 
will be denoted by $m_H$ and $m_R$, respectively. 
Moreover for given values of physical states masses $m_H$ and $m_R$, there is an additional constraint on the 
mixing parameter $\xi$, obtained by demanding that the quantity inside the square root of
Eq.~\ref{eq:massev2} be positive. Therefore the parameter $\xi$ must satisfy
\begin{equation}
\label{eq:cond1}
 (m_{H}^{2}+ m_{R}^{2})^2-\frac{4 \beta m_{H}^2 m_{R}^2}{Z^2} > 0~{\rm{and}}~ Z^2 > 0. 
\end{equation}
We assume that the Higgs-like
mixed scalar has mass $m_H$ = 125 GeV, and the radion-like mixed scalar $R$ should satisfy the
current experimental limits (to be discussed later).  In Fig.~\ref{fig:xi_range} we plot the allowed range 
of $\xi$ as a function of $m_R$, for $\Lambda_\phi$ = 3 and 5 TeV satisfying Eq.~\ref{eq:cond1}, 
with $m_H$ fixed to 125 GeV. The expression Dis in Fig.~\ref{fig:xi_range} refers to $(m_{H}^{2}+ m_{R}^{2})^2-
(4 \beta m_H^2 m_R^2)/Z^2$ in Eq.~\ref{eq:cond1}. The area enclosed by the 
blue region has $Z^2~>~0$ and the red region shows the allowed parameter region
with Dis $>~0$. The theoretically allowed parameter space increases with 
 $\Lambda_{\phi}$ as can be seen in ~Fig.~\ref{fig:xi_range}.

\subsection{Decays of the mixed scalars ($H$, $R$) to SM particles}
\label{subsec:decays}

\begin{figure}[htb]
\begin{center}
\includegraphics[width=7.5cm, height=6cm]{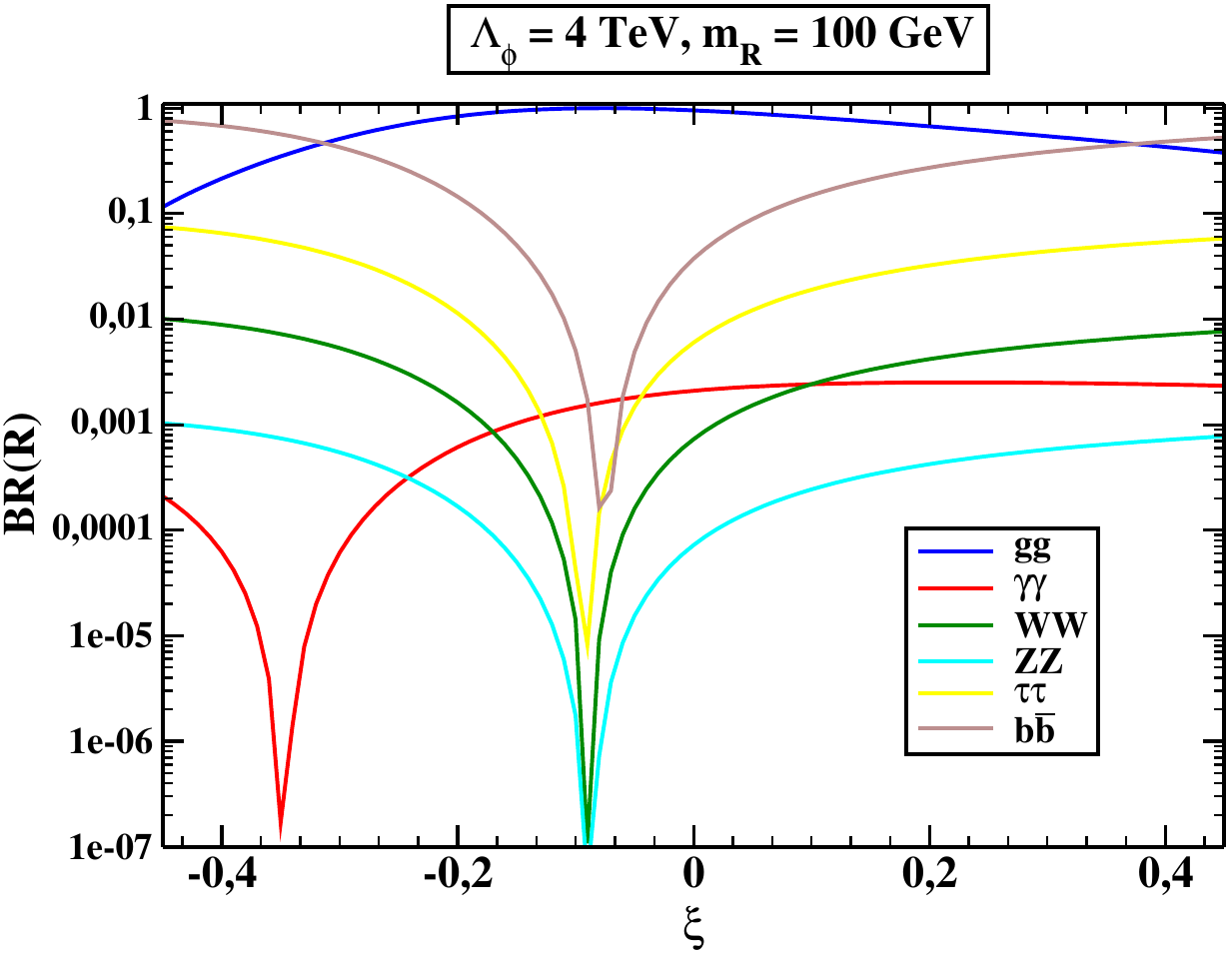}
\includegraphics[width=7.5cm, height=6cm]{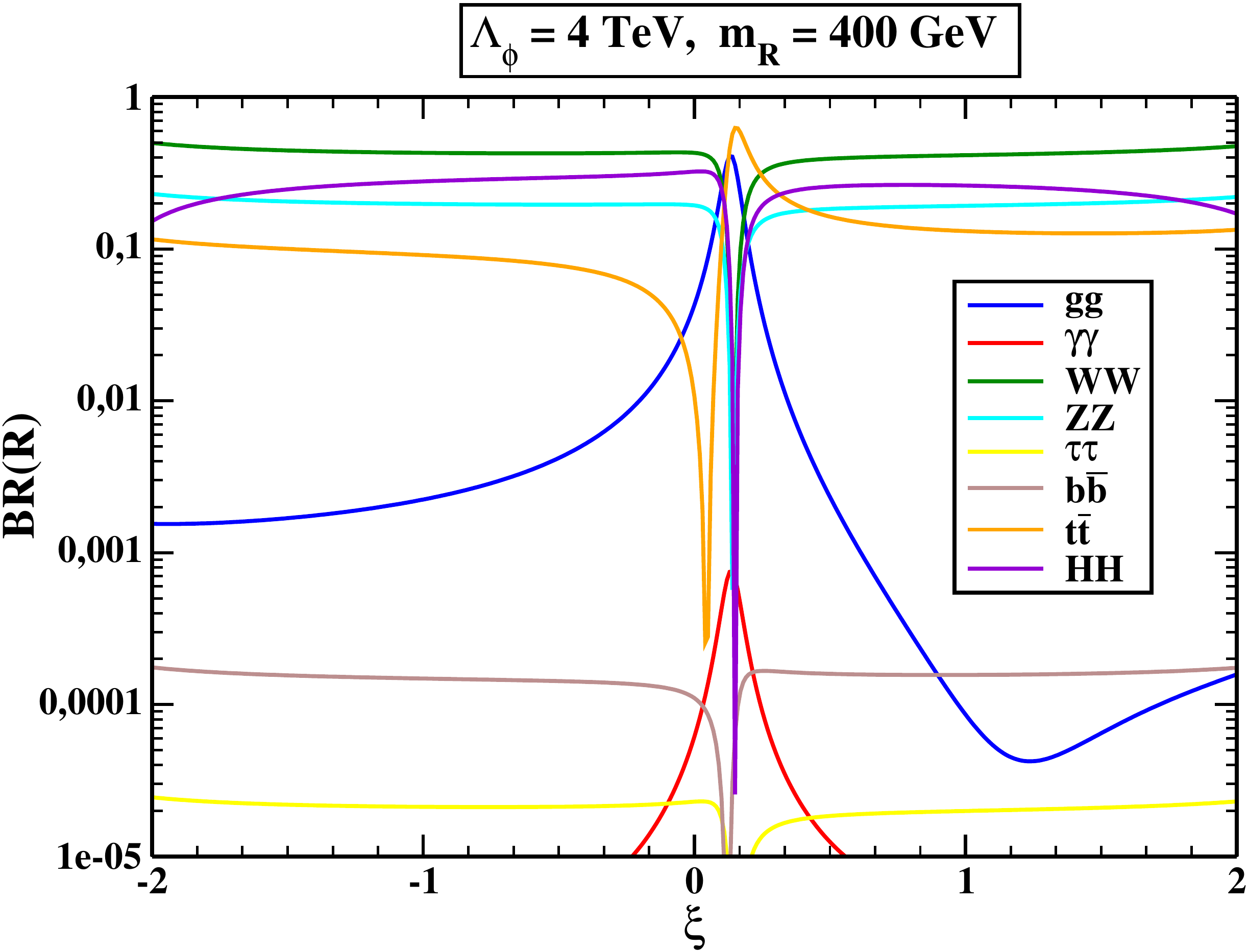}
\end{center}
\caption{The branching ratio of the mixed radion state to  kinematically allowed final states as a function of mixing
parameter ($\xi$), for $\Lambda_\phi$ = 4 TeV, and for $m_R$ = 100 GeV (left panel) 
and 400 GeV (right panel).}
\label{fig:br_mix}
\end{figure}

Below we list the interactions of the mixed scalars  with the SM particles. The expressions 
can be easily obtained  using Eq.~\ref{eqn:diag}, for the interactions listed in the 
previous section, Sec.~\ref{sec:non-mixing}.
From this point onwards, we will refer to the mixed Higgs $H$ as the Higgs and mixed radion $R$ as the radion.
The effective Lagrangians for the decay of the Higgs and radion into a pair of massive
gauge boson and into fermion pairs are given by, respectively
\begin{eqnarray}
\label{HWWZZ}
\mathcal{L}_H^{WW,ZZ} &=& \frac{H}{v}\left\{\left[a_\theta+ \gamma c_\theta
\left(1-\frac{3k r_c \pi M_W^2}{\Lambda_\phi^2}\right)\right]2 M_W^2 W_\mu^+ W^{\mu -} \right. \nonumber \\
&+& \left. \left[a_\theta+ \gamma c_\theta \left(1-\frac{3k r_c \pi M_Z^2}{\Lambda_\phi^2} \right)
\right] M_Z^2 Z_\mu Z^\mu \right \}, \\ 
\label{RWWZZ}
 \mathcal{L}_R^{WW,ZZ} &=& \frac{R}{v}\left \{\left[b_\theta+ \gamma d_\theta
\left(1-\frac{3k r_c \pi M_W^2}{\Lambda_\phi^2}\right)\right]2 M_W^2 W_\mu^+ W^{\mu -} \right. \nonumber \\
&+& \left. \left[b_\theta+ \gamma d_\theta \left(1-\frac{3k r_c \pi M_Z^2}{\Lambda_\phi^2} \right)
\right] M_Z^2 Z_\mu Z^\mu \right\}, \\
\label{Hff}
\mathcal{L}_H^{ff} &=& \frac{H}{v} \left[a_\theta+ \gamma c_\theta \frac{1}{2}
(c_{L}-c_{R}) \right]
\left(m_f \bar{\psi} \psi \right), \\
\label{Rff}
\mathcal{L}_R^{ff} &=& \frac{R}{v} \left[b_\theta+ \gamma d_\theta \frac{1}{2}
(c_{L} - c_{R}) \right]
\left(m_f \bar{\psi} \psi \right).
\end{eqnarray}
Similarly the interaction terms of the radion and the Higgs with the massless gauge bosons 
in the mixing scenario are given by,
\begin{eqnarray}
 \mathcal{L}_H^{gg,\gamma\gamma} &=& \frac{H}{4v}\left\{\left[a_\theta
 \frac{\alpha_s}{2\pi}b^h_{QCD} + \gamma c_\theta\left(\frac{1}{k r_c \pi}+
 \frac{\alpha_s}{2\pi}b^R_{QCD}\right)\right]G^a_{\mu\nu} G^{a \mu\nu} \right. \nonumber \\
 &+& \left.\left[a_\theta
 \frac{\alpha}{2\pi}b^h_{EM} + \gamma c_\theta\left(\frac{1}{k r_c \pi}+
 \frac{\alpha}{2\pi}b^R_{EM}\right)\right]F_{\mu\nu}F^{\mu\nu}\right \}, \\ 
 \mathcal{L}_R^{gg,\gamma\gamma} &=&  \frac{R}{4v}\left \{\left[b_\theta
 \frac{\alpha_s}{2\pi}b^h_{QCD} + \gamma d_\theta\left(\frac{1}{k r_c \pi}+
 \frac{\alpha_s}{2\pi}b^R_{QCD}\right)\right]G^a_{\mu\nu} G^{a \mu\nu} \right. \nonumber \\
 &+& \left.\left[b_\theta
 \frac{\alpha}{2\pi}b^h_{EM} + \gamma d_\theta\left(\frac{1}{k r_c \pi}+
 \frac{\alpha}{2\pi}b^R_{EM}\right)\right]F_{\mu\nu}F^{\mu\nu}\right \}, 
\end{eqnarray}
where $a_\theta,~b_\theta,~c_\theta,~d_\theta$ are defined in Eq.~\ref{eqn:diag},  
and $\gamma=v/\Lambda_\phi$. 
Finally, when $m_{R} > 2 \,m_{H}$,  the heavier scalar 
can decay into a pair of lighter scalars. The coupling between the two scalars
in the mixed scenario comes from three basic sources, 
($a$) the interaction of the radion with the trace of the SM Higgs field, also 
present in the unmixed case Eq.~\ref{eq:rhh_nomix},
($b$) the trilinear term in the Higgs potential and ($c$) a contribution from 
the Higgs-radion mixing term in Eq.~\ref{eqn:action}. The interaction Lagrangian of 
the radion with two Higgs bosons is therefore given by
\begin{eqnarray}
\mathcal{L}_R^{HH} &=& \frac{R}{\Lambda_\phi} \Big \{\left[ - a_\theta^2 d_\theta \left ( 
\partial_\mu H \partial^\mu H + 2 m_H^2 H H  \right) \right] 
+  4 a_\theta b_\theta c_\theta m_H^2 H H \Big\}  \nonumber \\
&-& 3\frac{m_{H}^{2}}{2 v} \left(a_\theta^{2} b_\theta R H^{2}\right)
-\frac{3 \xi}{\Lambda_{\phi}}\left[
\left(a_\theta^2 d_\theta H^2 \Box{R} + 2 a_\theta b_\theta c_\theta H \Box{H}R\right) \right ] \nonumber \\
&-& 6 \xi \frac{v}{\Lambda_\phi}\left[
\left(a_\theta c_\theta d_\theta H^2 \Box{R} + (a_\theta d_\theta
+ b_\theta c_\theta)c_\theta H \Box{H}R\right)\right].  
\end{eqnarray}
We plot the branching ratios of the radion to different allowed final states in Fig.~\ref{fig:br_mix}, as a function
of the mixing parameter, $\xi$, for two radion masses, 100 and 400 GeV.
We can see from Fig.~\ref{fig:br_mix} that 
for a certain value of $\xi$ ($\xi <0$ for $m_R=100$ GeV, $\xi >0$ for $m_R=400$ GeV), the decays of the radion $R$ to the SM  leptons, quarks, and massive gauge bosons are  suppressed. At this point the expression appearing in Eq. \ref{RWWZZ} and \ref{Rff}, $b_{\theta} + \gamma d_{\theta}~\sim~0$, whereas
the radion coupling to $\gamma\gamma$ and $gg$ have extra  contributions coming from trace anomaly 
and bulk kinetic term, and hence these are the dominant decay modes. Due to  propagation in the bulk, the $WW$ and $ZZ$ radion couplings also have additional small terms, but their effect is very small and can be ignored.
\newline
Hence, we divide our analysis into two regions, 
\begin{itemize}
 \item \textbf{Region-1 [the LHC region]:} In this region the radion is coupled maximally 
to the massless gauge bosons. This region of parameter space is ideally suited for study at the LHC, 
where we can focus on the production of the mixed scalar via gluon fusion production and its 
subsequent decay to diphoton.
\item \textbf{Region-2 [the ILC region]:} Although this region of parameter space can be 
studied in the LHC with the radion being produced through the vector boson fusion followed by decay to massive vector bosons, 
the production  cross sections for the associate channels are suppressed by the VEV of the radion, $\Lambda_{\phi}$.
We considered the gluon fusion production of the heavy scalar and its decay to vector bosons. This region of 
parameter space can also be probed by the associated production of the heavy scalar and the top or massive vector 
bosons, and its decay to $b\bar{b}$.
But the $b\bar{b}$ final states are  mostly accompanied by large hadronic background in the LHC 
and this is difficult to probe. Thus, this scenario is better suited for study at the ILC. 
We concentrate on the production of the radion by its associated production with $Z$ and the vector boson fusion, followed by 
and its decay to $b\bar{b}$ and massive gauge bosons, depending on the mass of the radion. 
\end{itemize}
We will discuss these regions in detail in two sections, Sec. \ref{sec:lhc_analysis} and \ref{sec:ilc_analysis}.

\section{Prospects for the searches at the 13 and 14 TeV LHC}
\label{sec:lhc_analysis} 

In this section, we first discuss the present limits derived on the mass of the radion and 
its mixing parameter $\xi$, then incorporate these into an investigation of the parameter space which 
can be probed at the LHC at $\sqrt{s}=$ 13 and 14 TeV.

\subsection{Constraints on the $\xi$ and $m_R$ values from the LHC data}
\label{subsec:constraints}

\begin{figure}[htb]
\begin{center}
\includegraphics[width=8 cm, height= 6cm]{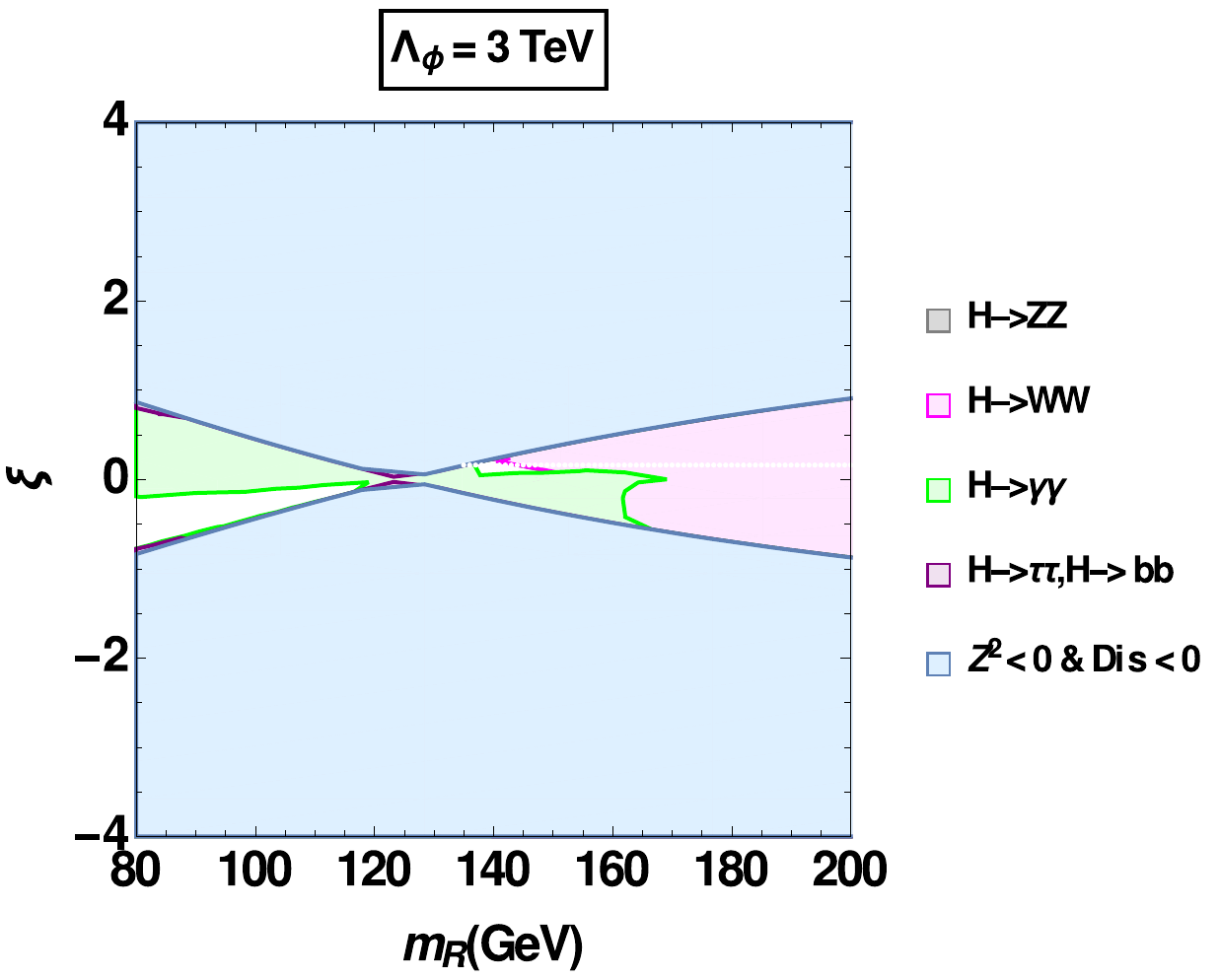}
\vspace{0.5cm}
\includegraphics[width=8 cm, height= 6cm]{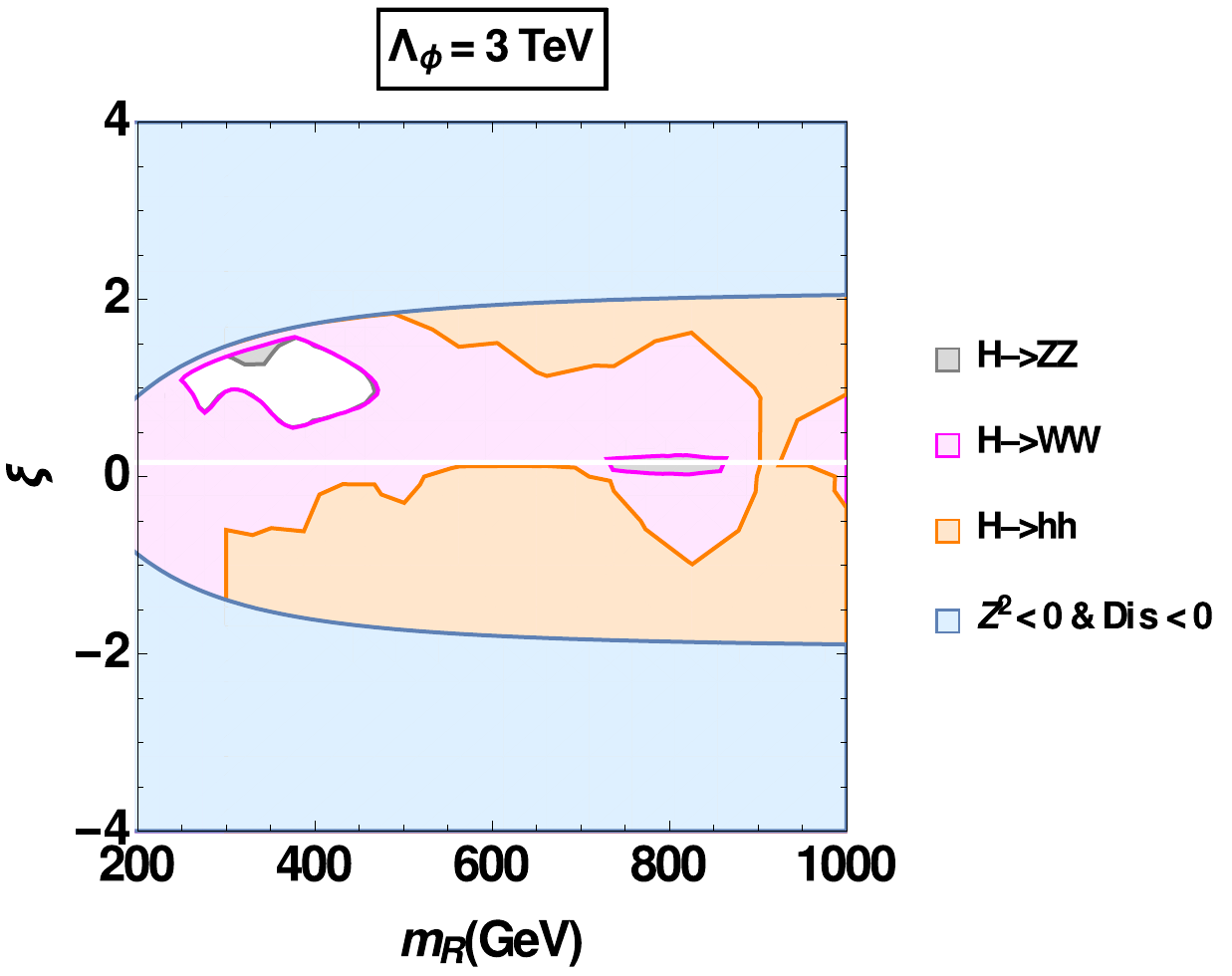}
\vspace{0.5cm}
\includegraphics[width=8 cm, height= 6cm]{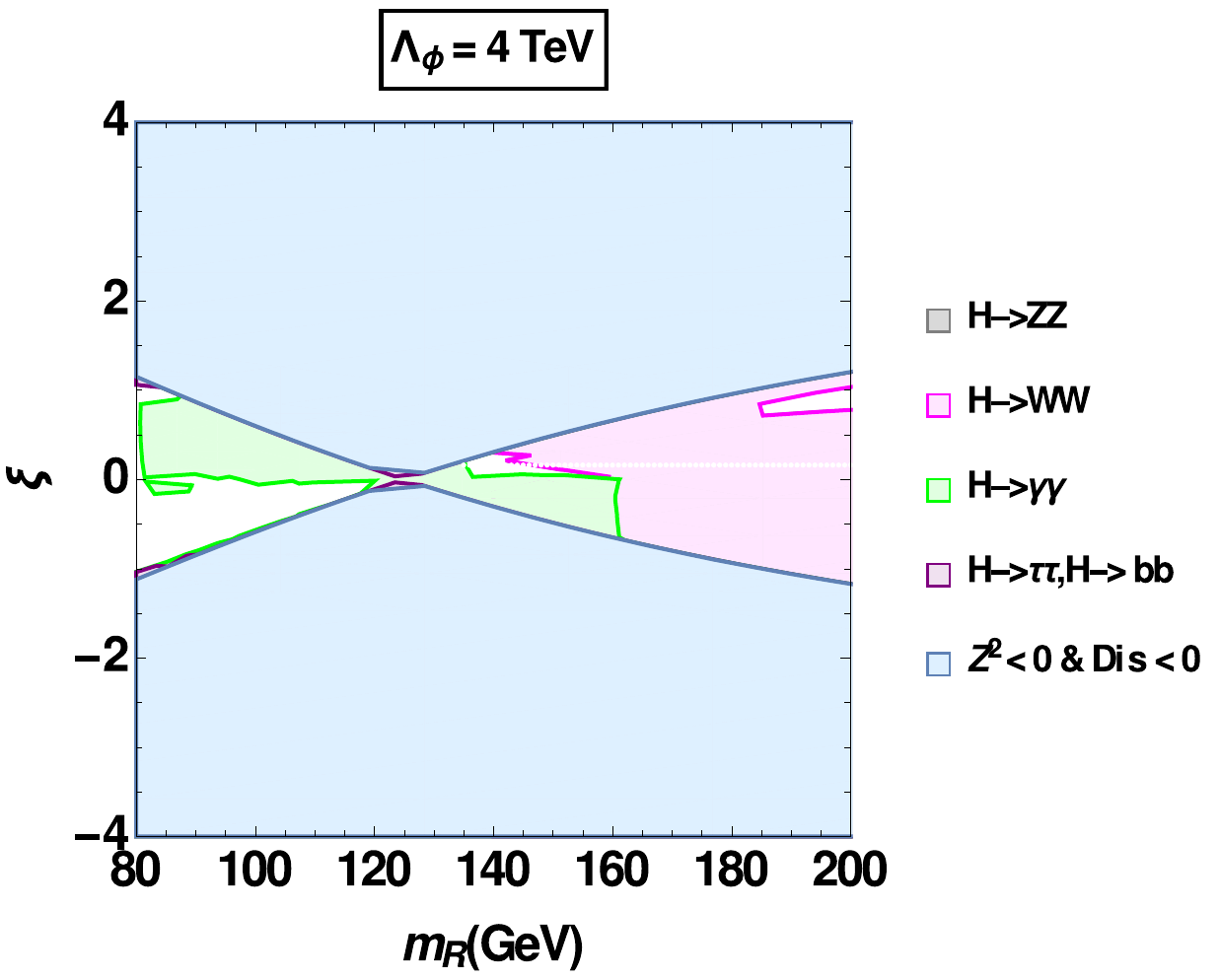}
\vspace{0.5cm}
\includegraphics[width=8 cm, height= 6cm]{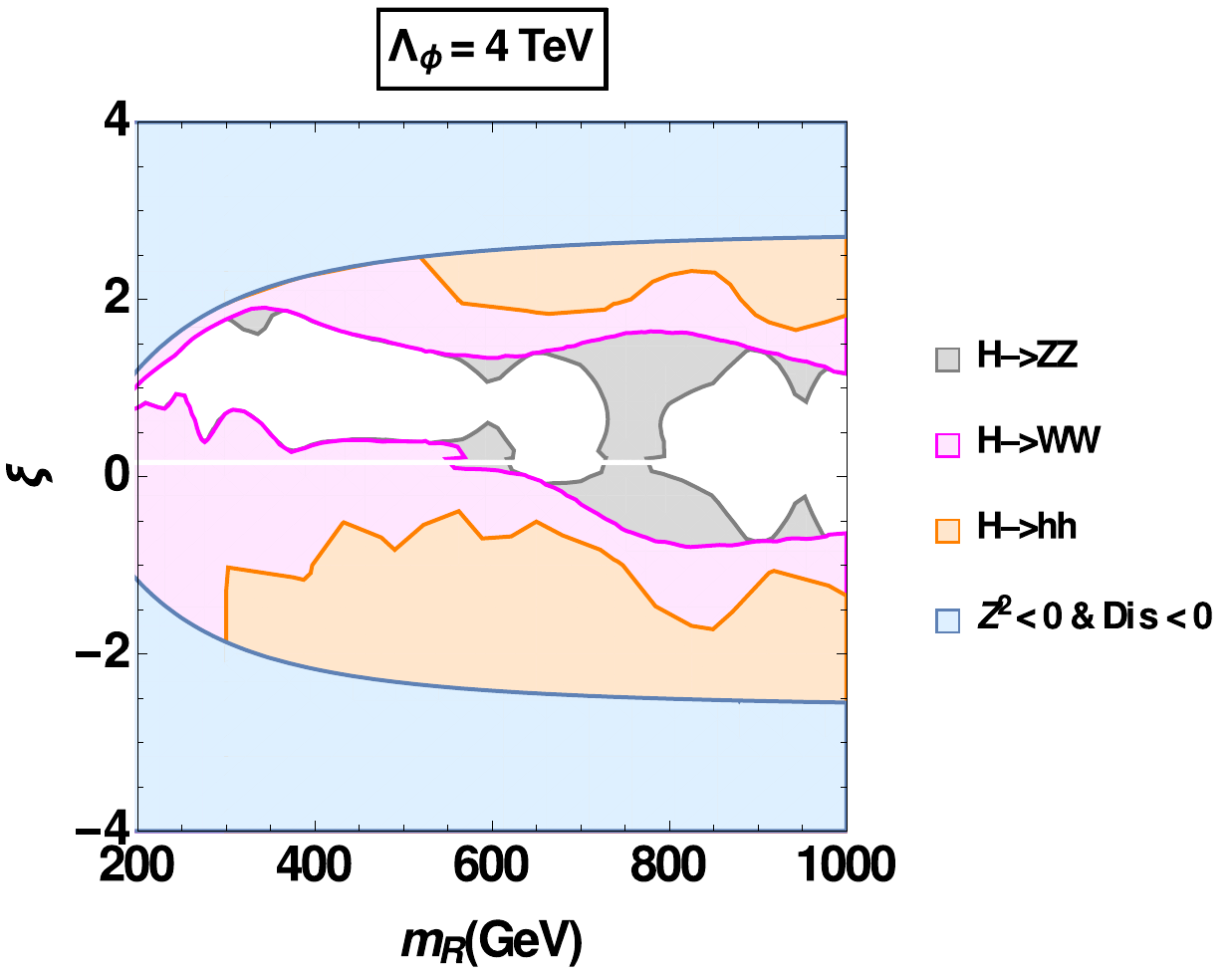}
\vspace{0.5cm}
\includegraphics[width=8 cm, height= 6cm]{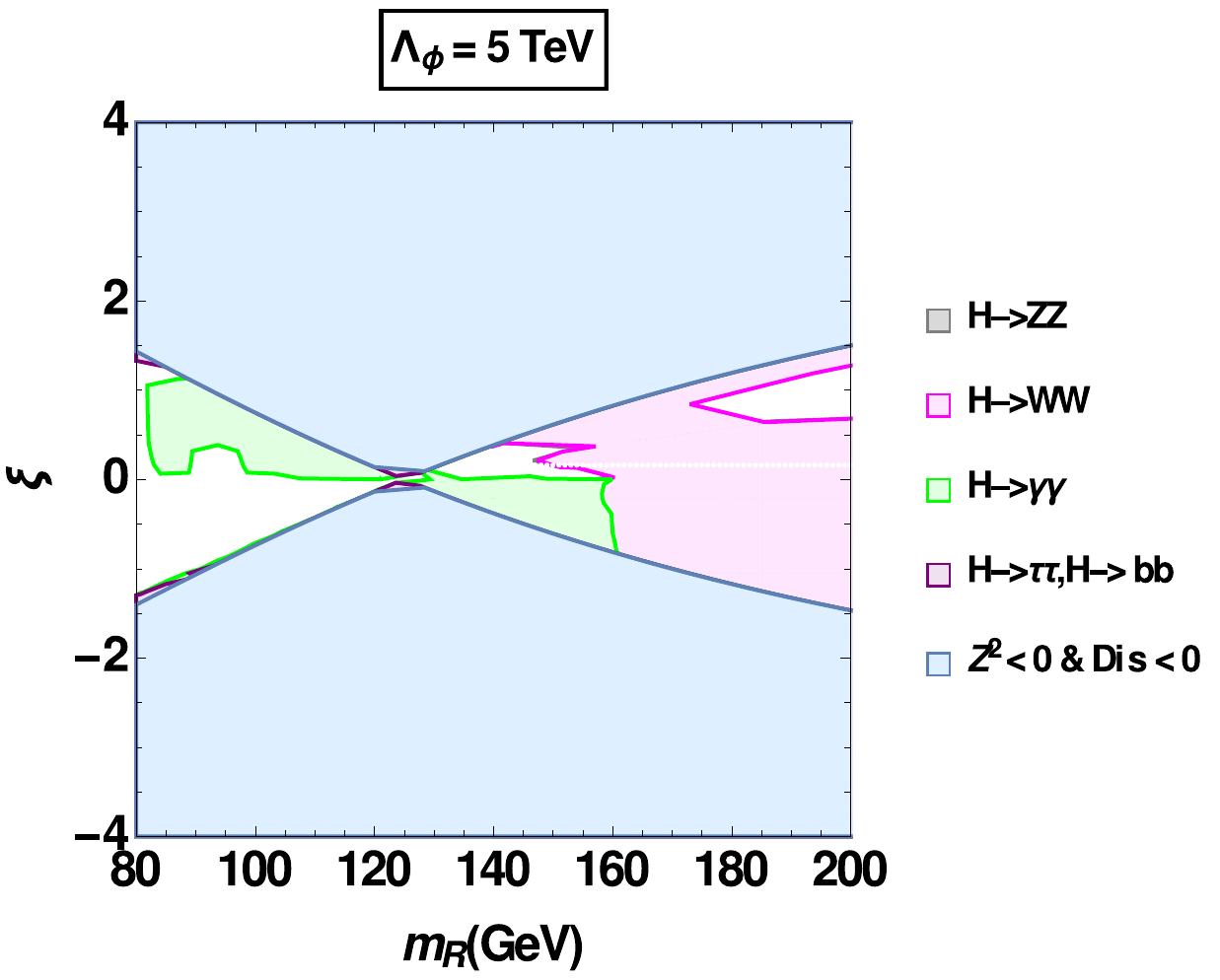}
\vspace{0.5cm}
\includegraphics[width=8 cm, height= 6cm]{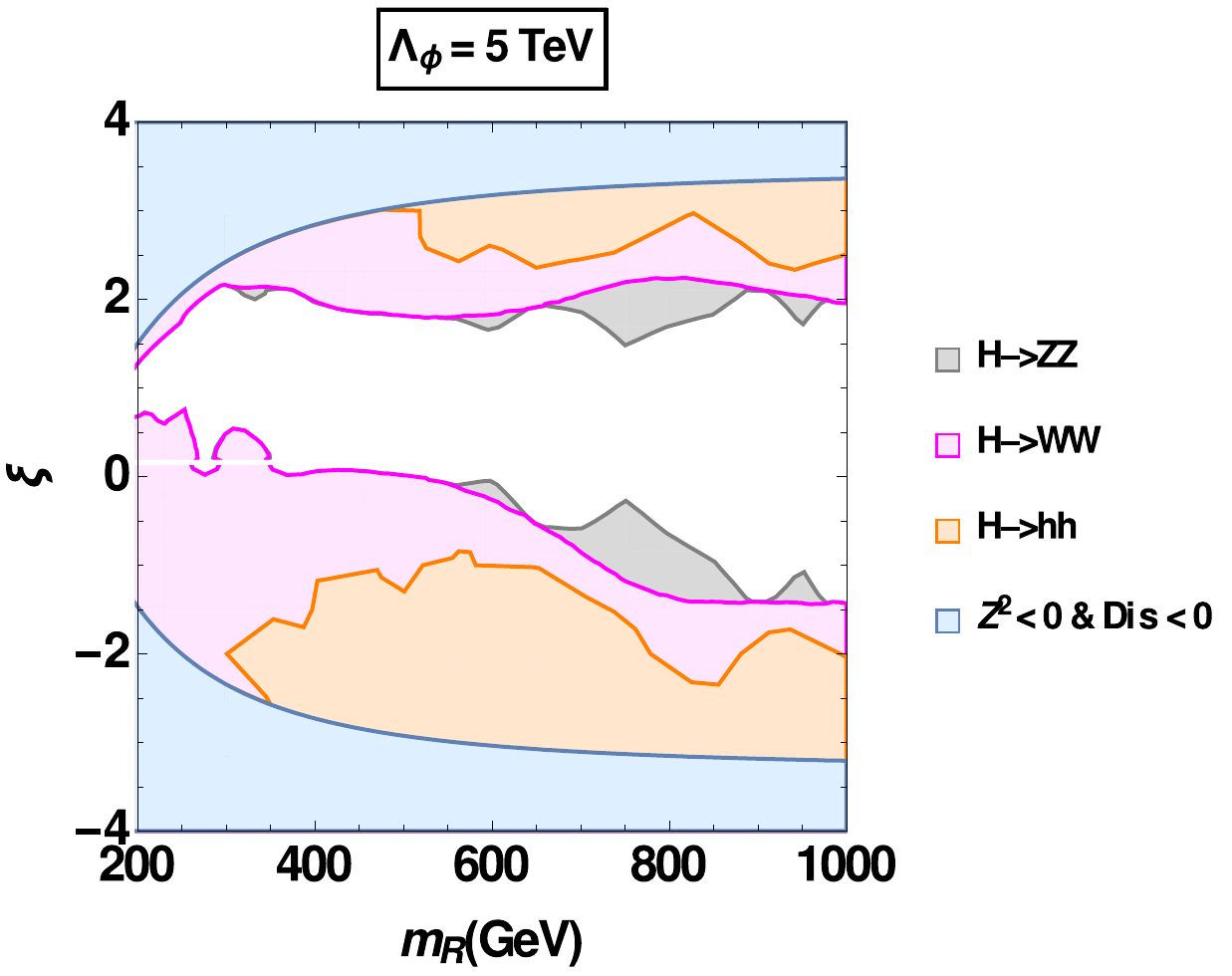}
\caption{The exclusion plot in the $\xi-m_R$ plane for $\Lambda_\phi$ = 3, 4 and 5 TeV from additional 
scalar ($H$) searches in the 8 TeV LHC for different decay channels for (left panel) light states, and (right panel), heavy states. Here $H$ states  represent
the additional scalar. We show: the theoretical restrictions (in blue); the additional restrictions from decays 
into $W^+W^-$ (in pink), $ZZ$ (in grey) $\tau^+ \tau^-$ and $b {\bar b}$ (in purple) $hh$ (in orange) 
and $\gamma \gamma$ (in green).
The region in white is allowed as surviving all restrictions.}
\label{fig:exclusion_all_mR}
\end{center}
\end{figure}

During run 1, the LHC experiment has looked for additional scalar particles, other 
than the Higgs boson, decaying through narrow resonances into different final states. We assumed the scenario 
where the mixed-Higgs scalar ($H$) mimics the scalar at 125 GeV discovered in LHC-run 1. The non-observation 
of the other scalar in LHC-run 1 puts constraints on the values of $\Lambda_{\phi}$ and $\xi$.
We show our results for $m_R$ in the mass range of 80 GeV to 1 TeV  in Fig.~\ref{fig:exclusion_all_mR}.
The experimental data used to constrain the $\xi-m_R$ parameter space
comes from the diphoton~\cite{Aad:2014ioa,CMS:2015ocq,Khachatryan:2015qba},  
$WW$~\cite{Khachatryan:2015cwa,Aad:2015agg},  $ZZ$~\cite{Aad:2015kna}, $b\bar{b}$~\cite{Aad:2014xzb}
and $\tau^+\tau^-$~\cite{CMS:2015mca} searches at the LHC. When the mass of the radion is greater than 250 GeV,
the heavy scalar can decay into a pair of SM Higgs bosons.
We have included the bounds coming from the LHC searches for Higgs boson pair production, in the 4$b$ final 
state~\cite{Aad:2015xja,Khachatryan:2015yea}. We scanned the $\xi-m_R$ parameter space for 
$\Lambda_\phi$ =  3, 4 and 5 TeV\footnote{We started with $\Lambda_{\phi}~>~2.5$ TeV, which is compatible with the minimum bound on the mass of the spin-1 KK resonance.}, taking into account all of the above experimental 
results, and  also imposed the theoretical bound, stated in Eq.~\ref{eq:cond1}.
The LHC search channels, which constrain the $\xi-m_R$ parameter space, are more restrictive than the 
theoretical bound, as shown in  Fig.~\ref{fig:exclusion_all_mR}, where  
the allowed parameter space is plotted in the $\xi-m_R$ plane, for different $\Lambda_\phi$, for both heavy and light states.
The parameter space is mostly constrained by the heavy Higgs data from 
$WW$ (region in pink) and $ZZ$ (region in pink+grey) final state. 
When $m_{R} < 125$ GeV, the parameter space is mostly excluded by diphoton searches (region in green) and 
from Higgs signal strength measurement, while for $m_R>250$ GeV, the decay into two Higgs 
bosons (4$b$ final states) restricts the parameter space further (region in orange). 
Finally, the region in blue represents the theoretical restrictions on the parameter space.

\begin{figure}[htb]
\begin{center}
\includegraphics[width=7 cm, height= 6cm]{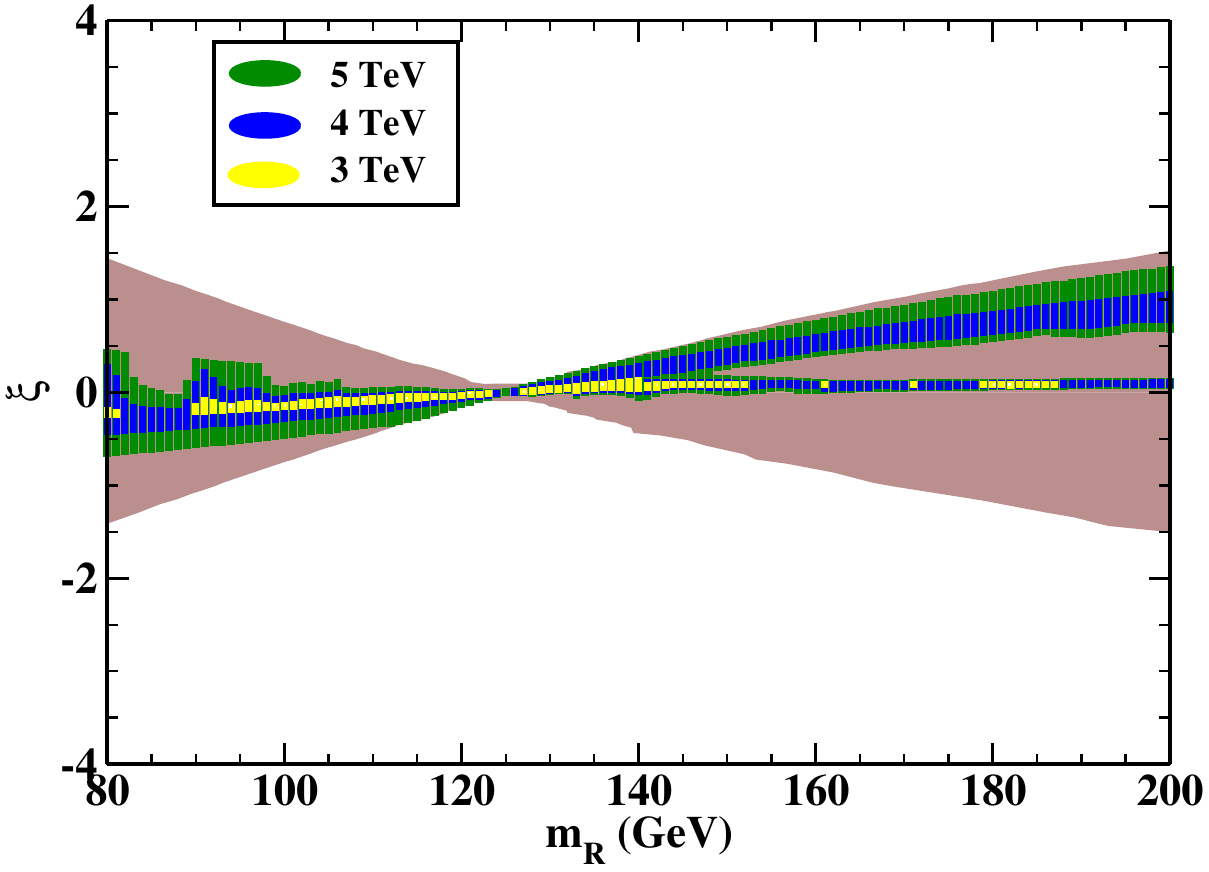}
\vspace{0.5cm}
\includegraphics[width=7 cm, height= 6cm]{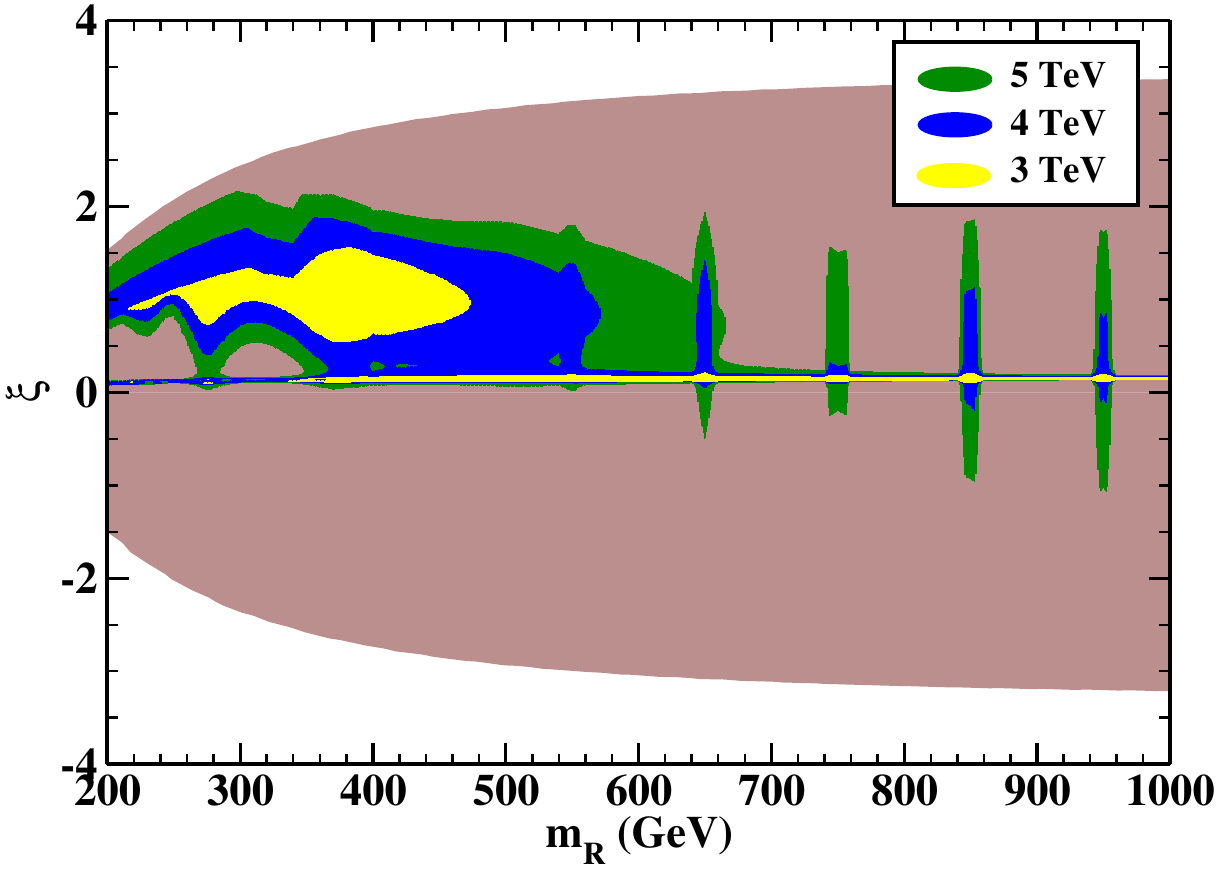}
\end{center}
\caption{The allowed parameter space in the $\xi-m_R$ plane for $\Lambda_\phi$ = 3, 4 and 5 TeV for (left panel), light states  and (right panel), heavy states. 
The region in brown is the theoretically allowed one, 
while yellow is the allowed region for $\Lambda_\phi$ = 3 TeV, yellow and blue for $\Lambda_\phi$ = 4 TeV, 
and green in addition to the previous colors, for $\Lambda_\phi$ = 5 TeV.}
\label{fig:allowed_all_mR}
\end{figure}

The white region in Fig.~\ref{fig:exclusion_all_mR} is the one currently allowed by the LHC experiments 
and the theory. We highlight this region in Fig.~\ref{fig:allowed_all_mR}  for $\Lambda_\phi$ = 3, 4 and 5 TeV.  
In these plots, apart from the LHC heavy Higgs data mentioned above,
we have also included the constraints from the Higgs signal strength 8 TeV 
LHC data~\cite{Atlas-Cms} as well as the constraints from the LHC 13 TeV data, showing an 
excess of 3.9$\sigma$~(ATLAS~\cite{Atlas_gg})
and $2.6\sigma$~(CMS~\cite{CMS:2015dxe}) in the diphoton final state at $m_{\gamma\gamma}$ = 750 GeV. 
We find that the negative $\xi$ region is ruled out by the LHC data for 
$m_R >$125 GeV. The region in brown is the theoretically allowed one, whereas the regions surviving 
after taking into account the additional constraints 
for each scale are shown in green for $\Lambda_\phi$ = 5 TeV, in blue for $\Lambda_\phi$ = 4 TeV, 
and in yellow for $\Lambda_\phi$ = 3 TeV. The values for different $\Lambda_\phi$ are superimposed
i.e. for example, for $\Lambda_\phi$ = 5 TeV the 
parameter space allowed covers the regions in green, blue and yellow,  for $\Lambda_\phi$ = 4 TeV the allowed
parameter space  covers the regions in blue and yellow whereas for $\Lambda_\phi$ = 3 TeV, the region
in yellow is the allowed one. We also find that there is a narrow region 
about $\xi~\sim~0.1~(-0.1)$ for $m_{R}~>~200$ GeV ($m_{R}~<~100$ GeV), which is 
allowed for almost all mass ranges. This region corresponds to $b_{\theta} + \gamma d_{\theta} \sim 0$ 
and can only be probed through the diphoton channel at the LHC. 

\subsection{Signal analysis for search at the LHC}
\label{subsec:LHC_search}

We performed our analysis for the LHC, with the mass of the radion varied from 80 GeV to 1 TeV.
The radion can be produced via gluon fusion, vector boson fusion, and associated production, with the
top or the vector bosons produced along with the radion. The production mechanism of the radion is similar 
to that of the SM Higgs bosons. As discussed before, compared to the Higgs, in case of radion
the $\gamma\gamma$ and $gg$ decay channels are enhanced from the trace anomaly. One of the unique 
features of the Higgs-radion mixing is that there exists a particular value of mixing still allowed by LHC-run 1, 
where the radion can decay only to a pair of photons or to a pair of gluons, referred as {\bf region-1} in our analysis\footnote{Several analyses were performed, which use this particular mixing value to explain the 750 GeV diphoton excess.
Here  $\Lambda_{\phi}>2.5$ TeV and therefore, for this case the cross section times decay rates are much lower than 
the observed value of 5-15 fb.}. As the diphoton final states are very clean to reconstruct at the LHC, 
we first perform our analysis through this channel. The $\xi$ parameter space has regions, where the 
coupling to the massive gauge bosons is maximal, leading to the $WW$ and $ZZ$ decay mode.
The $WW$ and the $ZZ$ will either decay leptonically or hadronically. 
The leptonic final states of $W$ are accompanied by missing energy and hence, mass reconstruction is not possible.
Thus, this channel can not be used as the channel for discovery. On the other hand, 
the leptonic final state of $ZZ$ can be used for mass reconstruction. However, the cross section of the production 
of radion via vector boson fusion and its decay to four leptons mediated by $Z$ is suppressed by the vev of the radion, 
$\Lambda_{\phi}$. We therefore 
consider the production of the radion through gluon fusion and its decay to four leptons mediated by $Z$. 

\subsubsection{Analysis in the $gg \rightarrow R \rightarrow \gamma\gamma$ decay channel}
\label{subsubsec:gammagamma}

\begin{figure}[htb]
$\begin{array}{ccc}
\hspace{-1.2cm}
 \includegraphics[width=6 cm, height= 6cm]{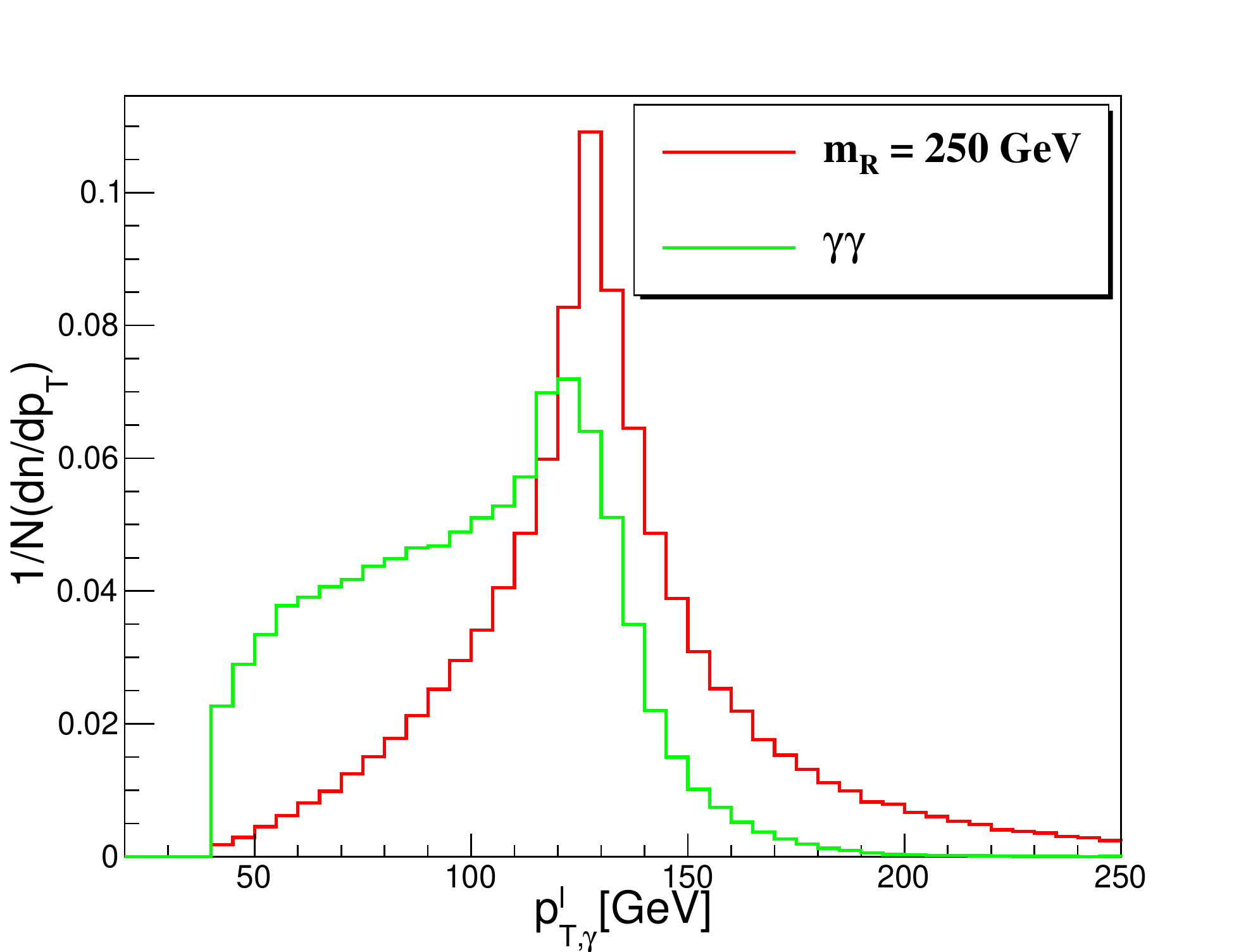} &
  \includegraphics[width=6 cm, height= 6cm]{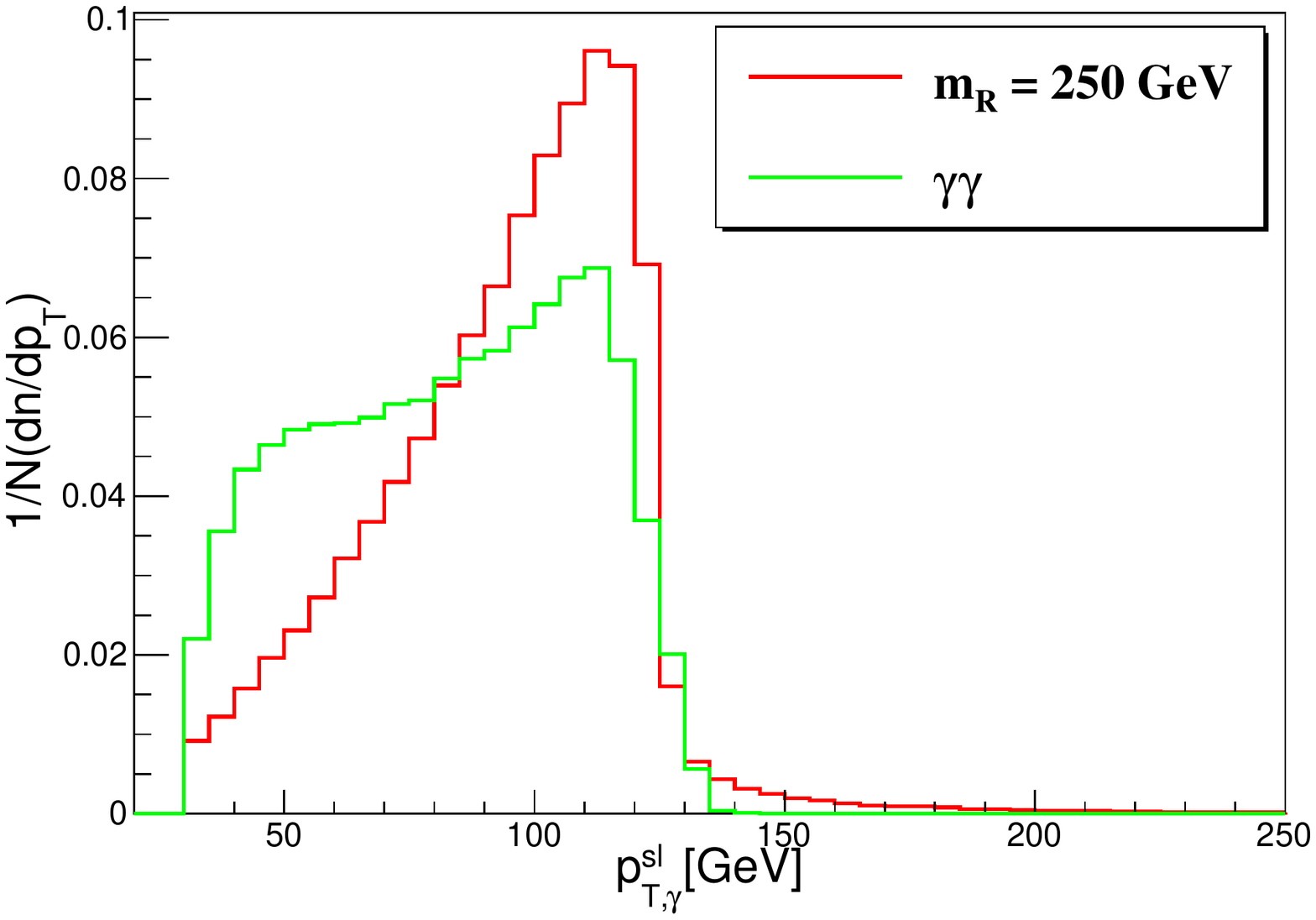} &
    \includegraphics[width=6 cm, height= 6cm]{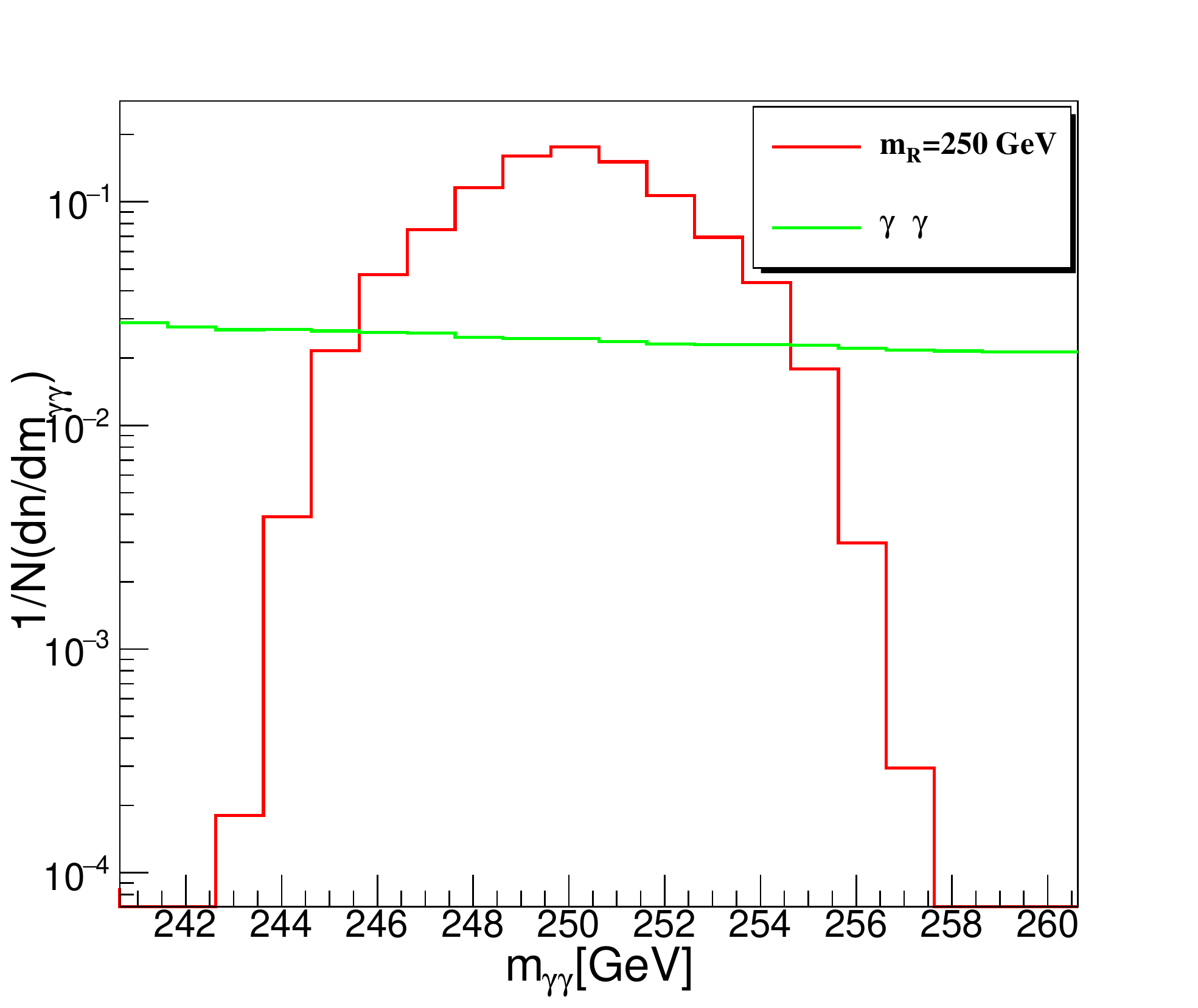}
    \end{array}$
  \caption{The $p_{T}$ distribution of the leading $p_T^l$ (left panel) and 
  subleading photon $p_T^{sl}$ (middle panel), and the invariant mass distribution of the (leading + subleading) 
  photons (right panel), for a radion with mass of 250 GeV.  The red line represents the signal, the green the ($\gamma \gamma$) background.  }
  \label{fig:pT_and_inv_mass}
\end{figure}

\begin{figure}[htb]
$\begin{array}{ccc}
\hspace{-1.2cm}
 \includegraphics[width=6 cm, height= 6cm]{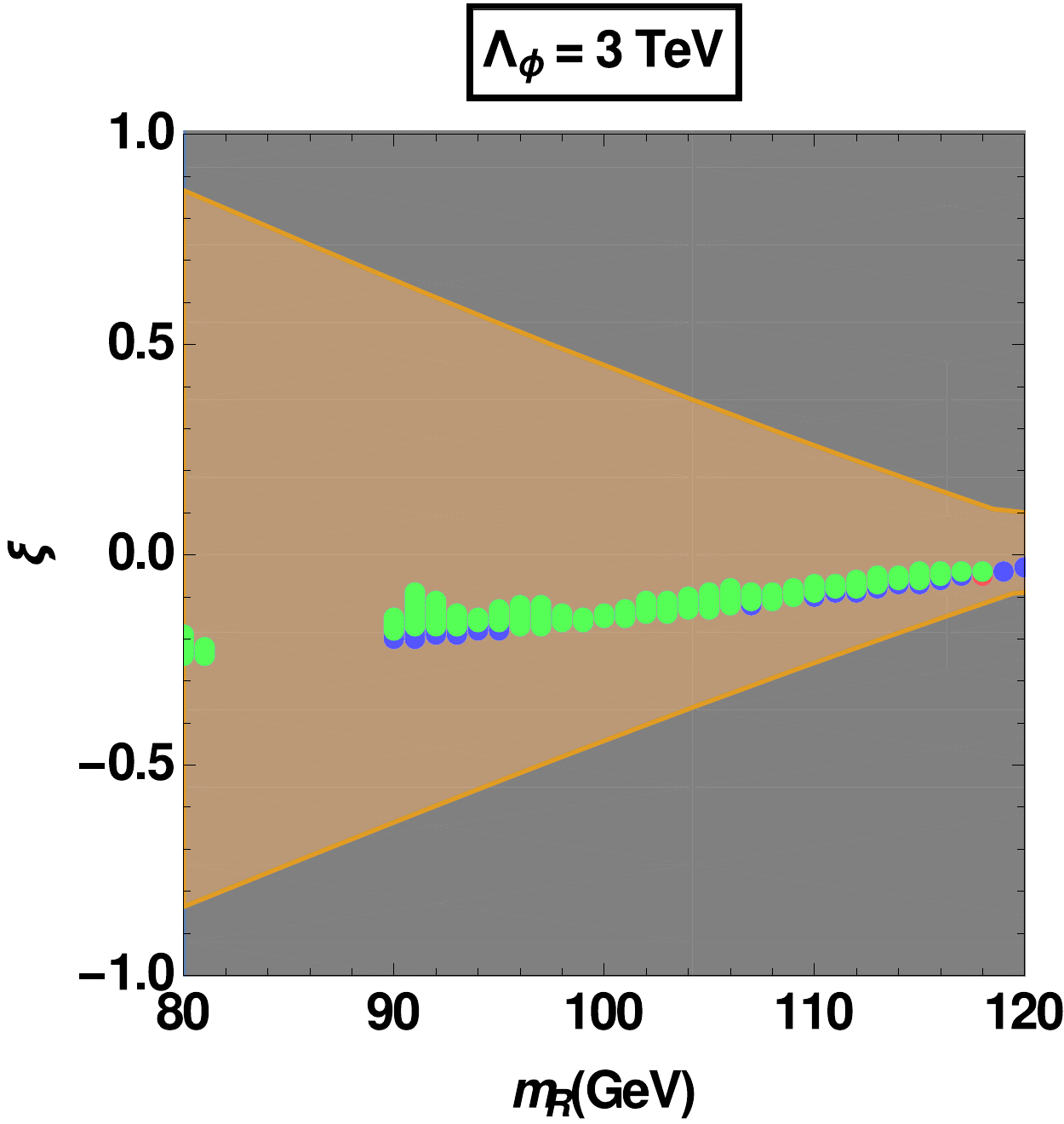}
&  \includegraphics[width=6 cm, height= 6cm]{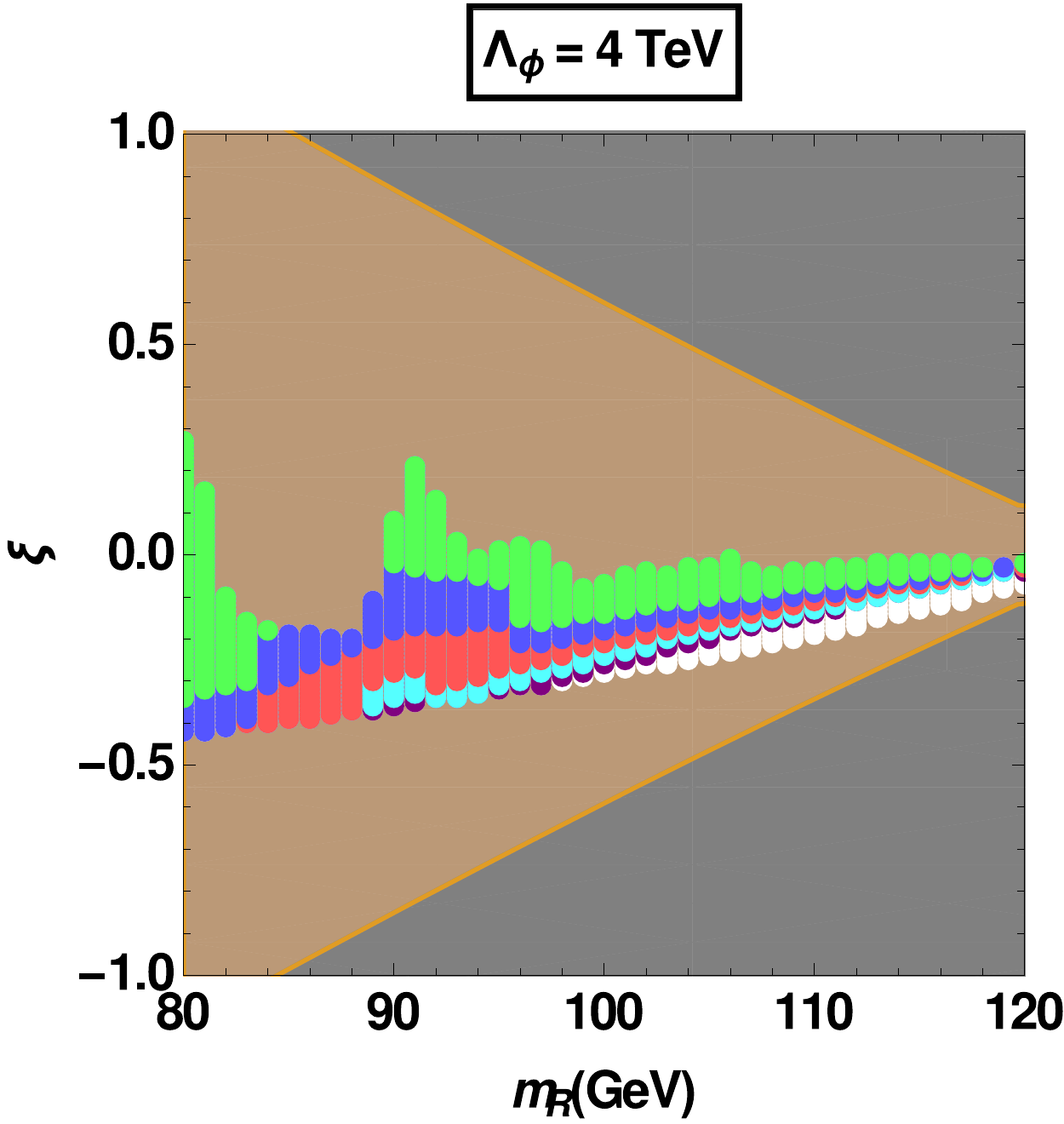}
& \includegraphics[width=6 cm, height= 6cm]{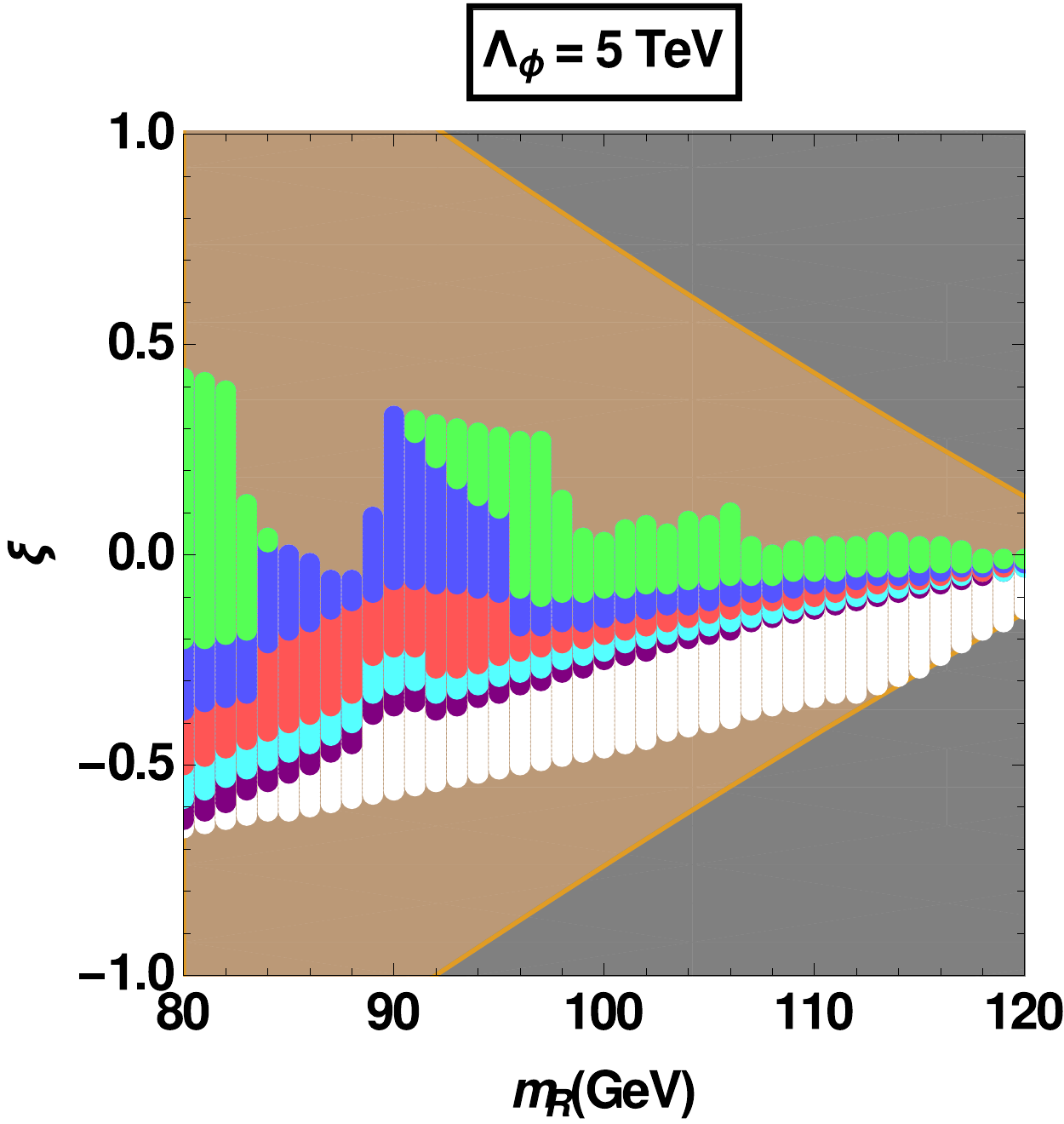}
\end{array}$
 \caption{Projected {\bf region-1} of the parameter space  for, $\Lambda_{\phi} = 3,\, 4,\,5$ TeV,  
 for a light mixed state. The brown region is the theoretically allowed region, while 
 the dark grey region is theoretically disallowed.  Green, blue, red, cyan and violet 
 colored regions represent regions which can be probed with 50, 150 fb$^{-1}$,  both for 13 TeV LHC, and 300, 1000, 3000 fb$^{-1}$ integrated 
 luminosity for 14 TeV LHC, respectively. The interior white region (nonexistent for $\Lambda_{\phi} =3$ TeV, 
 small for $\Lambda_{\phi} =4$ TeV, but larger for $\Lambda_{\phi} = 5$ TeV) represents the parameter region which
 cannot be probed.}
 \label{fig:lowmass}
\end{figure}

\begin{figure}
 \includegraphics[width=7 cm, height= 6cm]{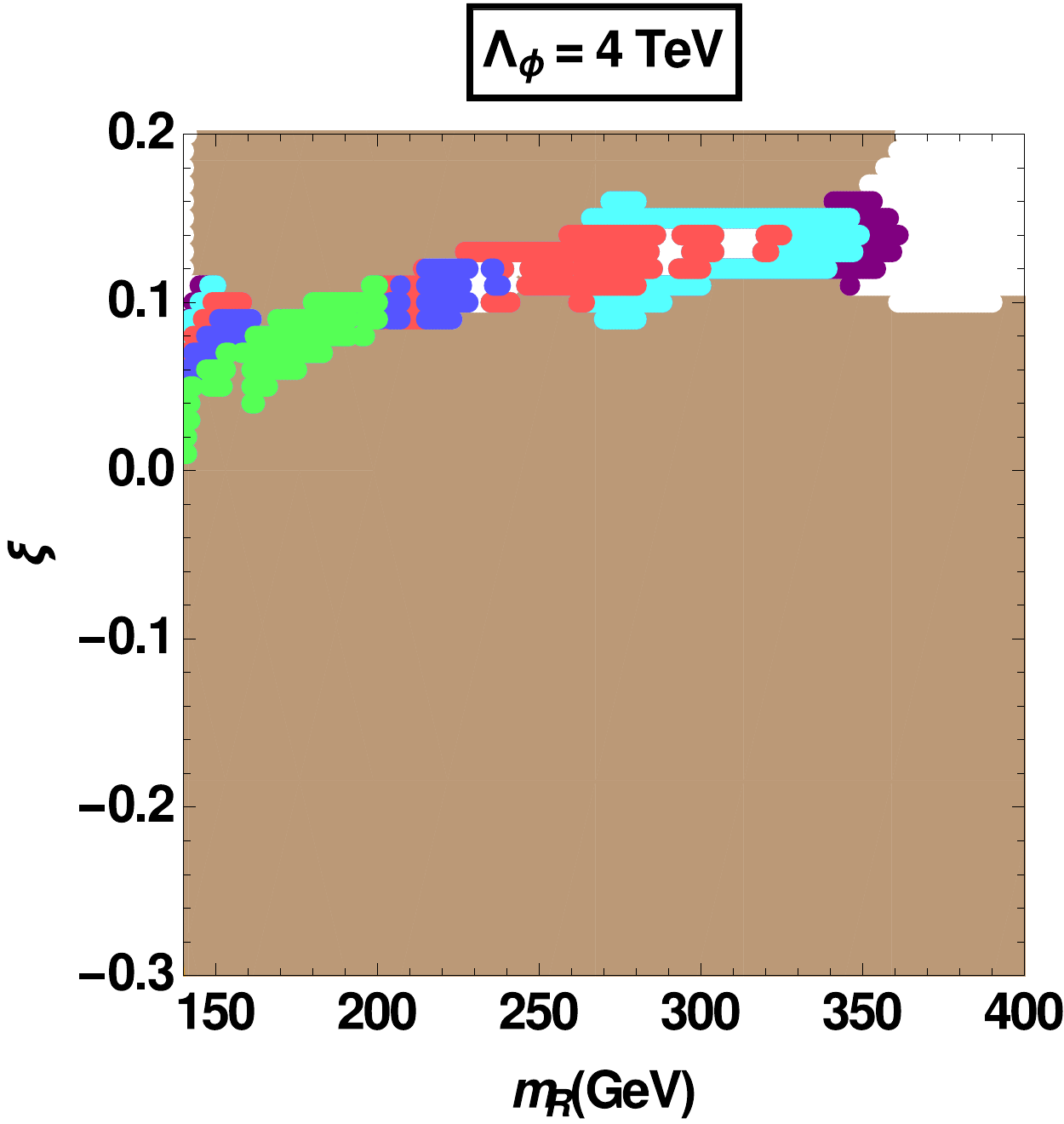}
  \includegraphics[width=7 cm, height= 6cm]{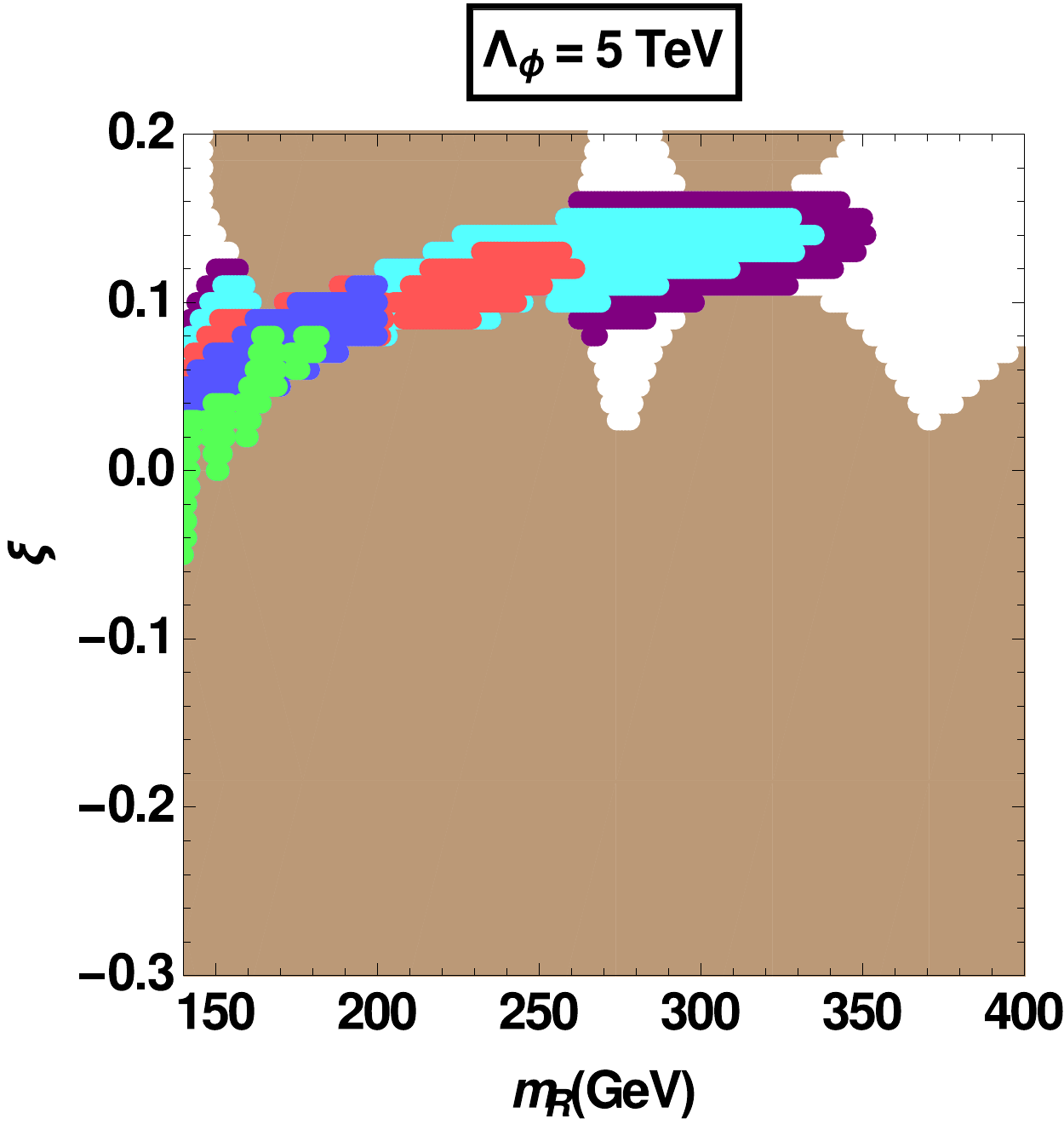}
  \caption{Projection of {\bf region-1} parameter space at 13 TeV and 14 TeV for $\Lambda_{\phi} = 4$ TeV (left panel) 
  and  for $\Lambda_{\phi} =5$ TeV (right panel), for a heavier mixed state. 
  The brown region is the theoretically allowed region. The color coding is as in Fig.~\ref{fig:lowmass}.}
  \label{fig:highmass}
\end{figure}

We first focus on the diphoton channel at the LHC, since this is the cleanest signal. We consider two
isolated photons as our signal. The final state with two isolated photons can be mimicked by several SM processes:
\begin{itemize}
\item Irreducible diphoton background: This background is produced from two prompt photons coming from tree 
level $q\bar{q}$ annihilation as well as gluon-gluon  box diagram with quarks propagating in the loop.
The production rate from the box diagram is comparable to that from the tree level process,
due to a high gluon flux at small $x$, where $x$ is energy fraction carried by each partons.
These photons are mostly isolated.
\item Reducible background: There are three backgrounds in this category:
\begin{enumerate}
\item $j\gamma$ background: The $\pi^{0}$, $\eta$ and $\rho$ inside a jet can decay into two collimated photons 
which can be identified as a single photon candidate in electromagnetic calorimeter (ECAL).
However, the photons are mostly non-isolated and hence can  be suppressed by using photon isolation criteria.
\item $j j$ background: Similar  to $j\gamma$, each of the jets can produce two collimated photons and hence can mimic
two photon states. However, this background can also be completely removed by photon isolation.
\item Drell-Yan background: The electrons can fake a photon when the tracks of the electrons are not properly 
reconstructed by inner tracking chamber. We considered the Drell-Yan background with a track 
mis-measurement efficiency of about $5\%$. This background is comparable to the irreducible background near the $Z$ mass.
\end{enumerate}
\end{itemize}
The signal and the background events with showering and hadronization are generated at the leading order in 
{\tt PYTHIA 8} \cite{Sjostrand:2014zea}. We have used CTEQ6l1 \cite{Dulat:2015mca} as our 
parton density function (PDF). The renormalization and factorization scales for both the signal and the background 
are kept at their default values. In order to enhance the statistics for the signal over the background events,
we divided our analysis into different phase space regions, depending upon the mass of the radion.
We classified different regions of $\hat{m}$ (the invariant mass
of the outgoing partons) depending on the mass of the radion\footnote{We imposed $|\hat{m}$ - $m_{R}|~ <$  15 GeV.}.
In order to have robust signal-background estimations, we implemented an algorithm which approximates the 
clustering procedure in an electromagnetic calorimeter. The electromagnetic shower from an unconverted photon 
is contained within a $5 \times 5$ crystal matrix around the seed (the actual hit spot). For a converted photon, 
the region is even wider. We considered a cluster of such hits within a cone of $\Delta R$ = 0.09.
The momentum of the photon candidate is defined as the vector sum of all the photons and the electrons 
within the cone of $\Delta$ R of 0.09 around the seed, which is the 
direct photon or electron.  To account for the detector resolution, we  smeared photons, electrons and
jets with Gaussian functions. 
We selected photon candidates with $p_{T}^{\rm {leading\,(subleading)}}~>~30\,(25)$ GeV and
lying within $|\eta|~<2.5$, while considering signals for $m_{R} < 200$ GeV. For $m_{R}~>~$ 200 GeV,
we selected photon candidates with $p_{T}^{\rm {leading\,(subleading)}}~>~40\,(30)$ GeV.
Jets are reconstructed with $|\eta|<4.7$ and $p_{T}~>~25\,(20)$ GeV.
Photons with $1.44~<~|\eta|~<1.55$ are not considered. The triggered photons are 
required to have minimal hadronic activity. This has been ensured by demanding that the 
total scalar sum of the transverse energy within the cone of $\Delta R~=~0.3$ should be less than 5 GeV. 
We further demand that the two photon candidates should be separated by at least $\Delta R~=0.4$. 
Surviving events with two such 'isolated' photons qualify for our further analysis\footnote{When 
mass of the radion is less than 200 GeV, we found that absolute isolation works better, however, 
beyond 200 GeV, both the isolation criteria  have almost the same efficiency.}. 
We found that $jj$ background is completely removed by demanding two such isolated photons. 
The $p_{T}$ distribution of the photons coming from the background peaks in the lower side than the signal.
Also, as $p_{T}$ of the photon increases, the misidentification rate for $\gamma-j$ decreases. Thus, the background
can be separated from the signal by applying a relatively harder $p_{T}$ cut.
After applying the $p_{T}$-cut on the leading ($l$) and sub-leading ($sl$) photon, we finally select 
only those events that lie within 10 GeV invariant mass bin about the radion mass. 
The kinematic cuts considered for this analysis are 
$$p_{T}^{l}~>~0.5 \,m_{R} + 5.0~~{\rm and}~~p_{T}^{sl}~>~p_{T}^{l} - 5.0,~|m_{\gamma \gamma}-m_{R}| < 3.0
~~~{\rm  when}~m_{R} \leq 200~{\rm GeV}$$ and
$$p_{T}^{l}~>~0.5 \,m_{R} - 10.0~~{\rm and}~~p_{T}^{sl}~>~p_{T}^{l} - 5.0,~|m_{\gamma \gamma}-m_{R}| < 5.0
~~~{\rm when}~m_{R} > 200~{\rm GeV}.$$ 
The hard $p_{T}$ cut on the leading as well as the sub-leading photon helps to get rid 
of $\gamma-j$ background, in the low mass region. Although we loose some signal events from this cut,
 the significance increases appreciably. The gluon fusion cross-section decreases 
with the increase in mass of the radion, hence, we applied a relatively softer $p_{T}$ cut on the photons.
As the production rate for diphoton background falls with $p_{T}$ of the photon, we still achieve higher significance,
as seen in Fig.~\ref{fig:pT_and_inv_mass}, where we plot 
the normalized $p_{T}$ distribution of the leading and subleading 
photon, as well as the invariant mass for the (leading and subleading) photon pair, 
for a radion mass of 250 GeV.
The prospects for restricting the mixed Higgs radion state parameter space at the LHC operating at 13 and 14 TeV 
are shown in  Fig.~\ref{fig:lowmass}. We consider integrated luminosities of 50 fb$^{-1}$ and 150 fb$^{-1}$  
for the 13 TeV LHC and  300, 1000 and 3000 fb$^{-1}$ for the 14 TeV LHC. The green and the green$+$blue region indicate the 
values of the mixing parameter $\xi$ that can be observed with more than 3$\sigma$ significance level
at 50  and 150 fb$^{-1}$ in the diphoton channel at 13 TeV center of mass energy. 
The area enclosed by the red, cyan and violet are the regions  can be probed by the 14 TeV LHC,
with an integrated luminosity of 300, 1000 and 3000 fb$^{-1}$ respectively.
We find that most of the $\xi$ region can be completely probed at 13 TeV using diphoton channel. 
With the increase of the center of mass energy from 13 TeV to 14 TeV,
we found that higher mass of {\bf region-1} can be probed effectively. In Fig.~\ref{fig:highmass},
we show the regions of the $\xi$ parameter space, that can be probed for a heavy radion. The color
coding is similar to the Fig.~\ref{fig:lowmass}. 
We find that the  diphoton channel can probe most of the parameter space of {\bf region-1}.
The production rate of radion, decaying into two photons beyond 400 GeV  is too low and can not be 
probed even at 3000 fb$^{-1}$. We  next discuss the prospect of probing {\bf region-2} in $ZZ$ decay channel of the radion.

\subsubsection{Analysis in the $gg \rightarrow R \rightarrow ZZ, Z\rightarrow l\bar{l}$ decay channel}
\label{subsubsec:ggZZ}

\begin{figure}[htb]
$\begin{array}{cc}
\hspace{-1.2cm}
 \includegraphics[width=6 cm, height= 6cm]{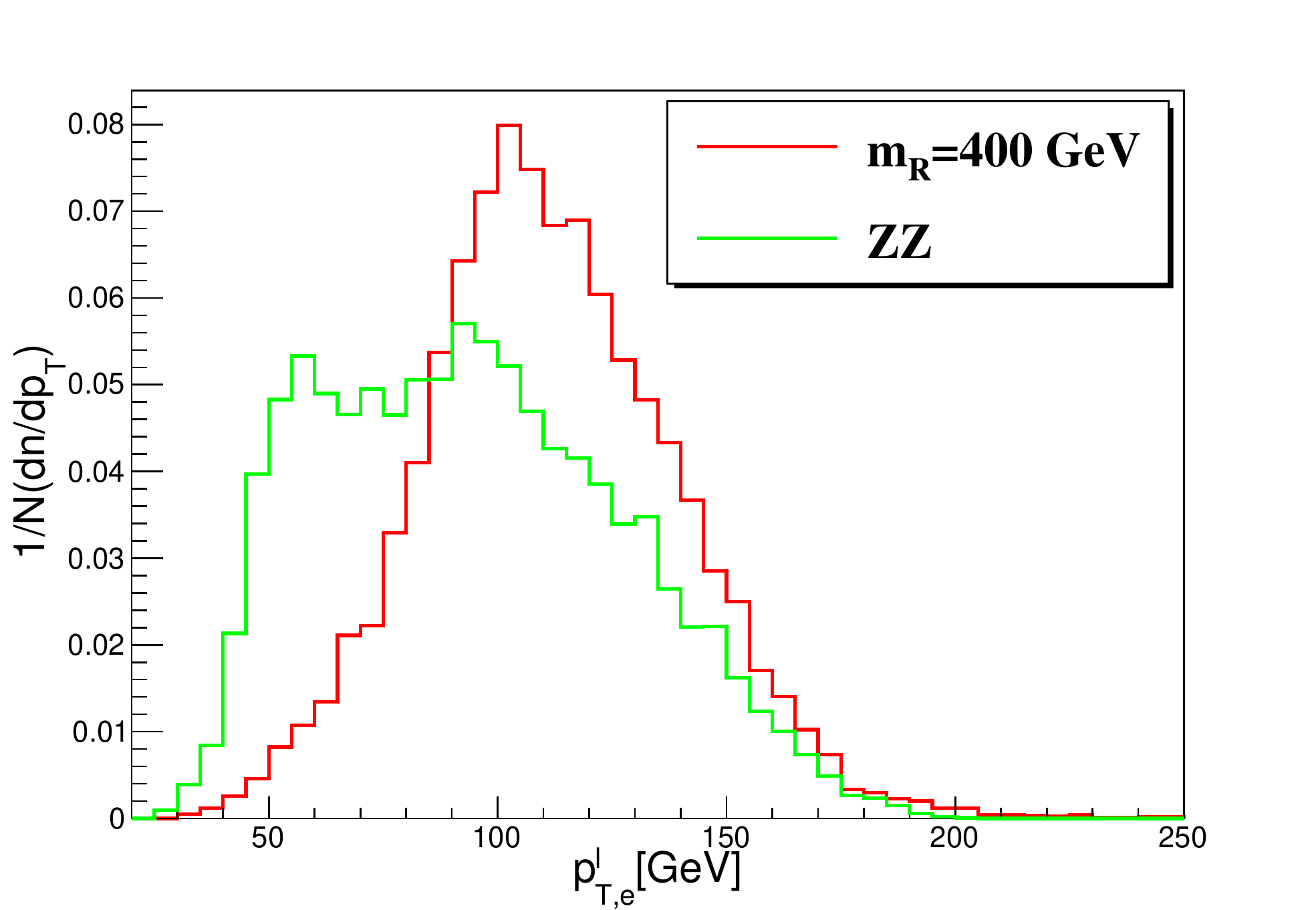} &
  \includegraphics[width=6 cm, height= 6cm]{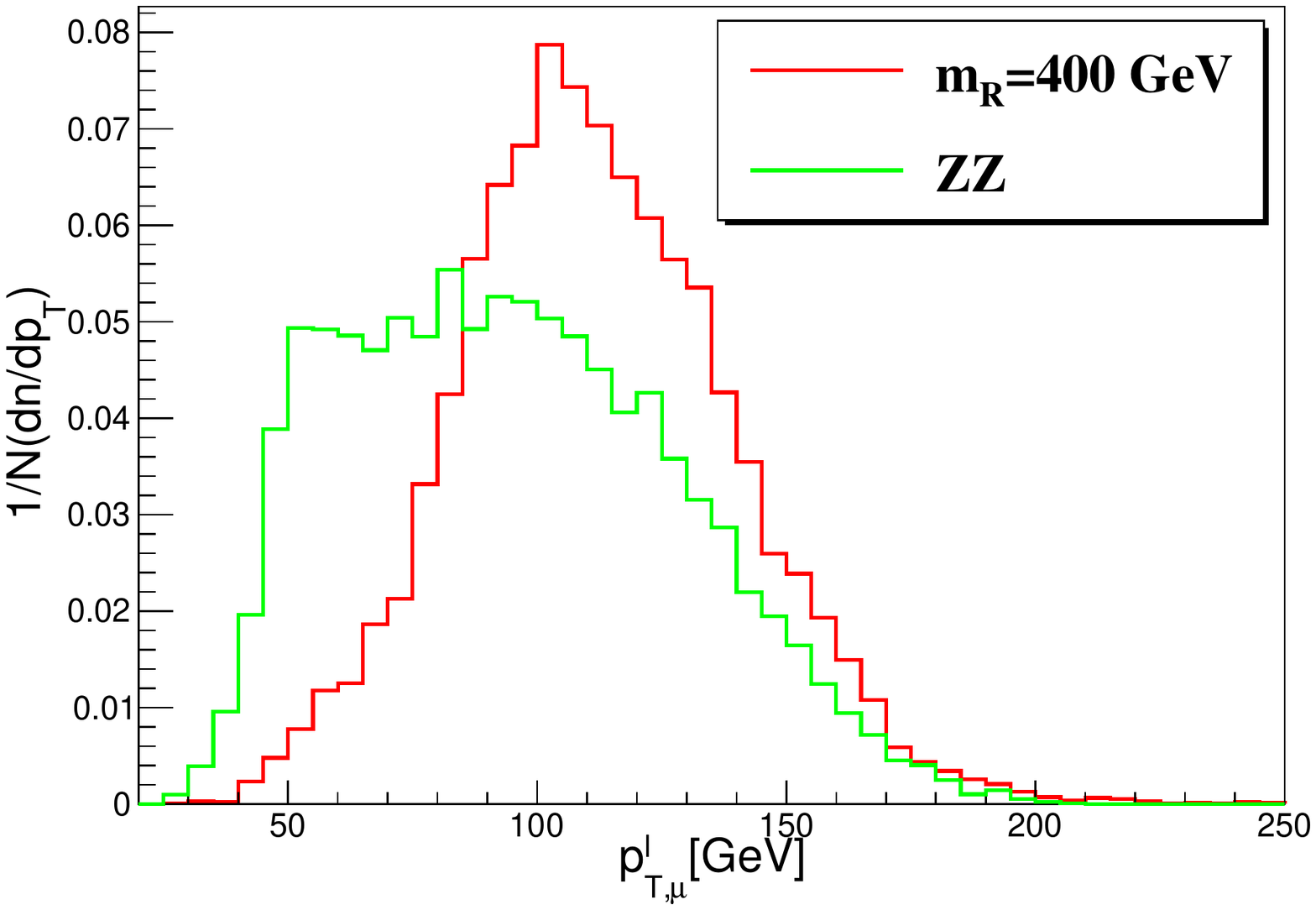}\\
  \hspace{-1.2cm} \includegraphics[width=6 cm, height= 6cm]{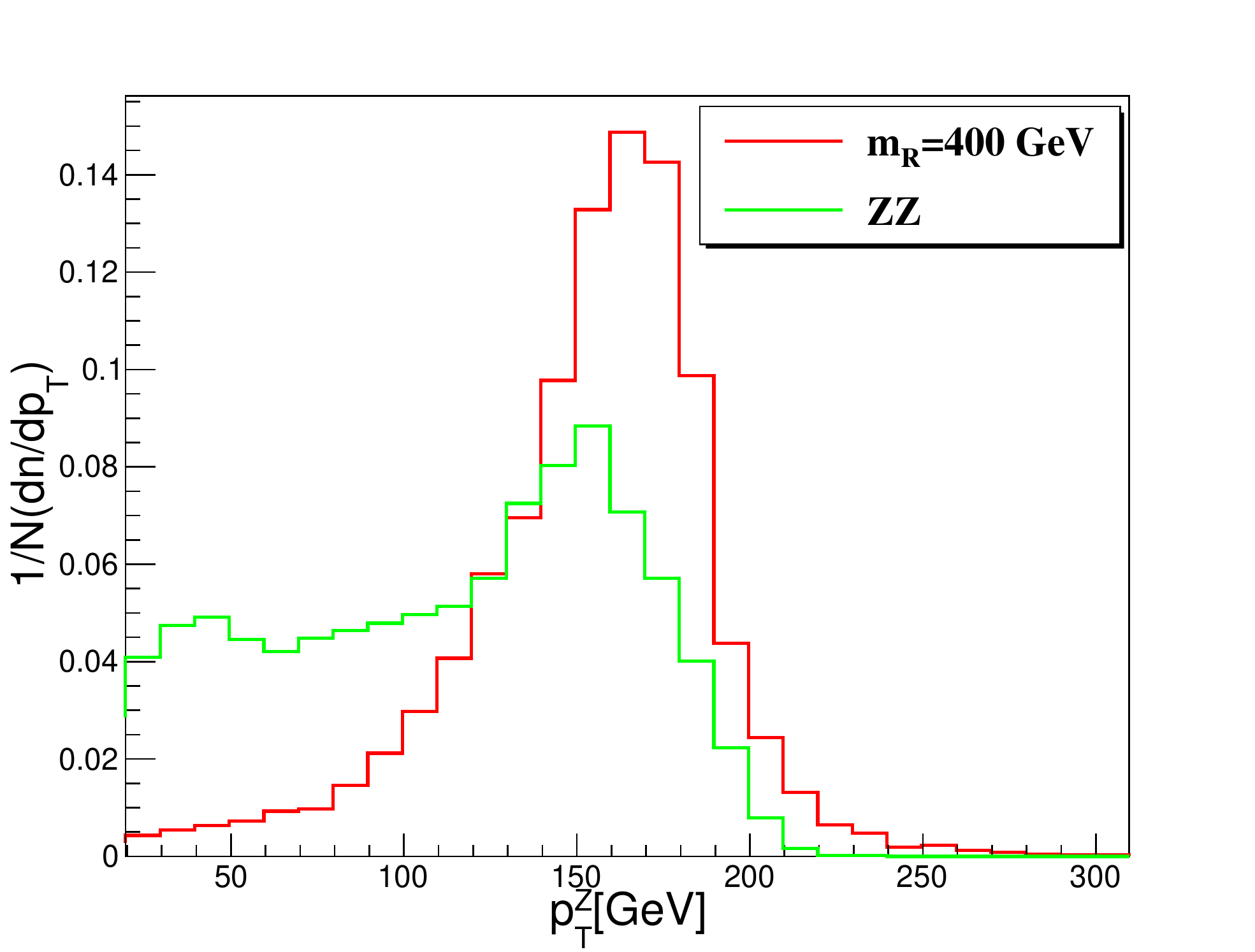} &
    \includegraphics[width=6 cm, height= 6cm]{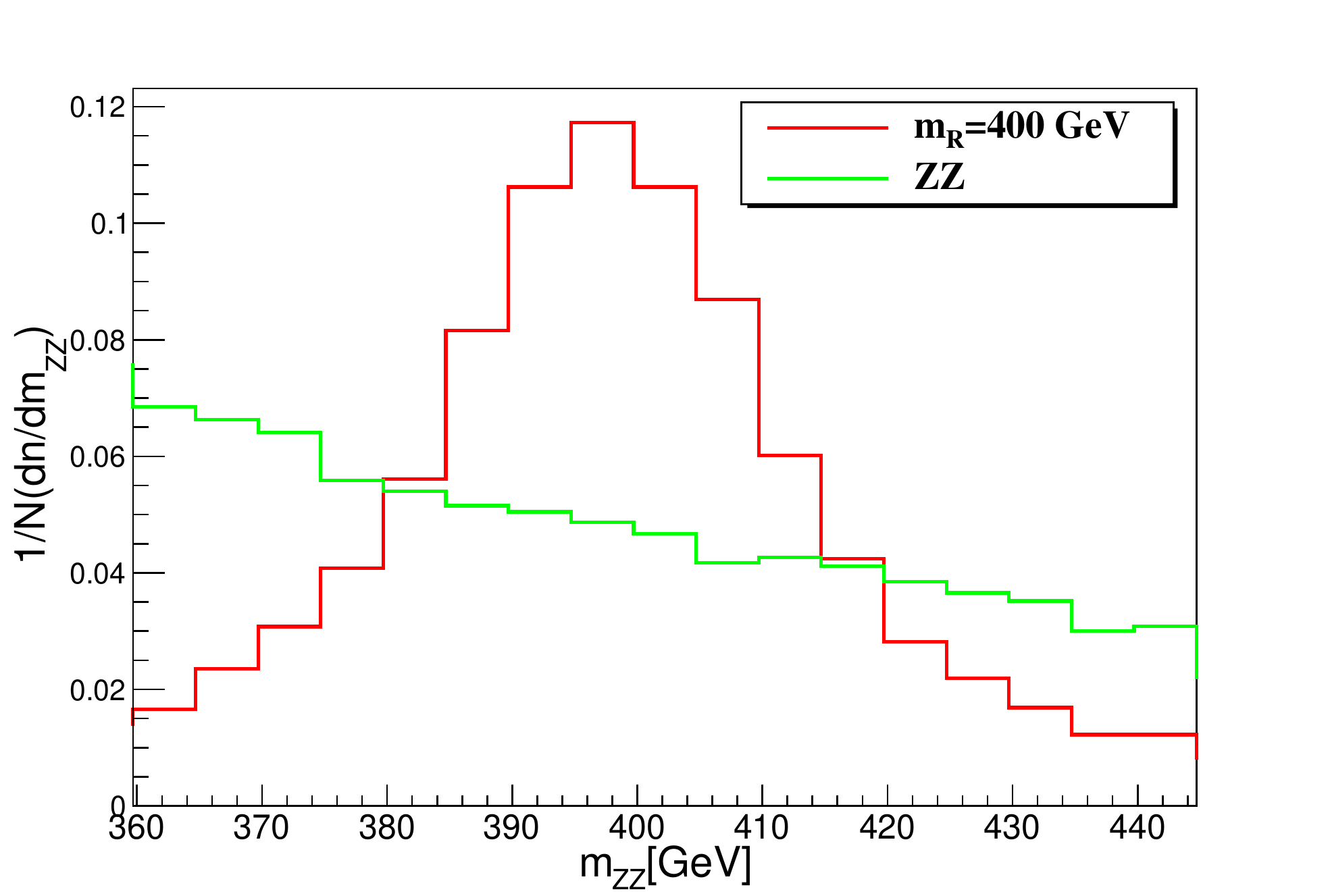}
    \end{array}$
  \caption{The $p_{T}$ distribution of the leading electron (top-left panel),
  leading muon (top-right panel), of the reconstructed Z (bottom-left panel), and the invariant mass distribution of the  
  Z  boson (bottom-left panel), for a radion of mass of 400 GeV.  The red line represents the signal, the green the background.  }
  \label{fig:pT_and_inv_mass_ZZ}
\end{figure}

When mass of the radion is greater than 200 GeV, one can consider its decay to $ZZ$ and subsequently into four leptons.
This channel also offers a clean reconstruction of the scalar mass. We considered gluon 
fusion of the scalar and its decay to $ZZ$. Although the production rate is not enough to probe the scenario at 13 TeV LHC, one can probe 
{\bf region-2}, where the coupling of the
heavy scalar to massive gauge bosons is large at LHC 14 TeV with higher integrated luminosity. To suppress combinatorial leptonic
background, we consider a pair of
electrons and a pair of muons as our signal, following the analysis given in ~\cite{CMS:2016noo}. 
The main irreducible background comes from the SM $ZZ$ production.
As before, we have generated the signal and the background events with showering and hadronization at the leading order in 
{\tt PYTHIA 8}. We have used CTEQ6l1 as our 
parton density function (PDF). The renormalization and factorization scales for both the signal and the background 
are kept at their default values.
We selected events with no associated jets. The jet-veto cut helps us to get rid of the background coming from $t\bar{t}Z$.
We consider two electrons having $p_{T}~>~17~\rm{GeV}$ and $\eta~<~2.5$, and two muons 
with $p_{T}~>~17~\rm{GeV}$ and $\eta~<~2.1$. The leptons are considered 'isolated' if the scalar sum of the
$p_{T}$ deposited within the cone of $\Delta~R$=0.3 about the lepton is less than 10$\%$ of the $p_{T}$ of the lepton.
We considered  such pairs of isolated electron and muon with $p_{T}~>~25~\rm{GeV}$. We ensure that the pairs of same flavor leptons
reconstruct $Z$-mass. Events with transverse momentum of the same flavor dilepton system greater than 55 GeV are selected. 
This selection cut helps us to control the irreducible background as shown in the 
Fig.~\ref{fig:pT_and_inv_mass_ZZ}(bottom-left). These four leptons are further considered in our analysis.

\begin{figure}
 \includegraphics[width=7 cm, height= 7cm]{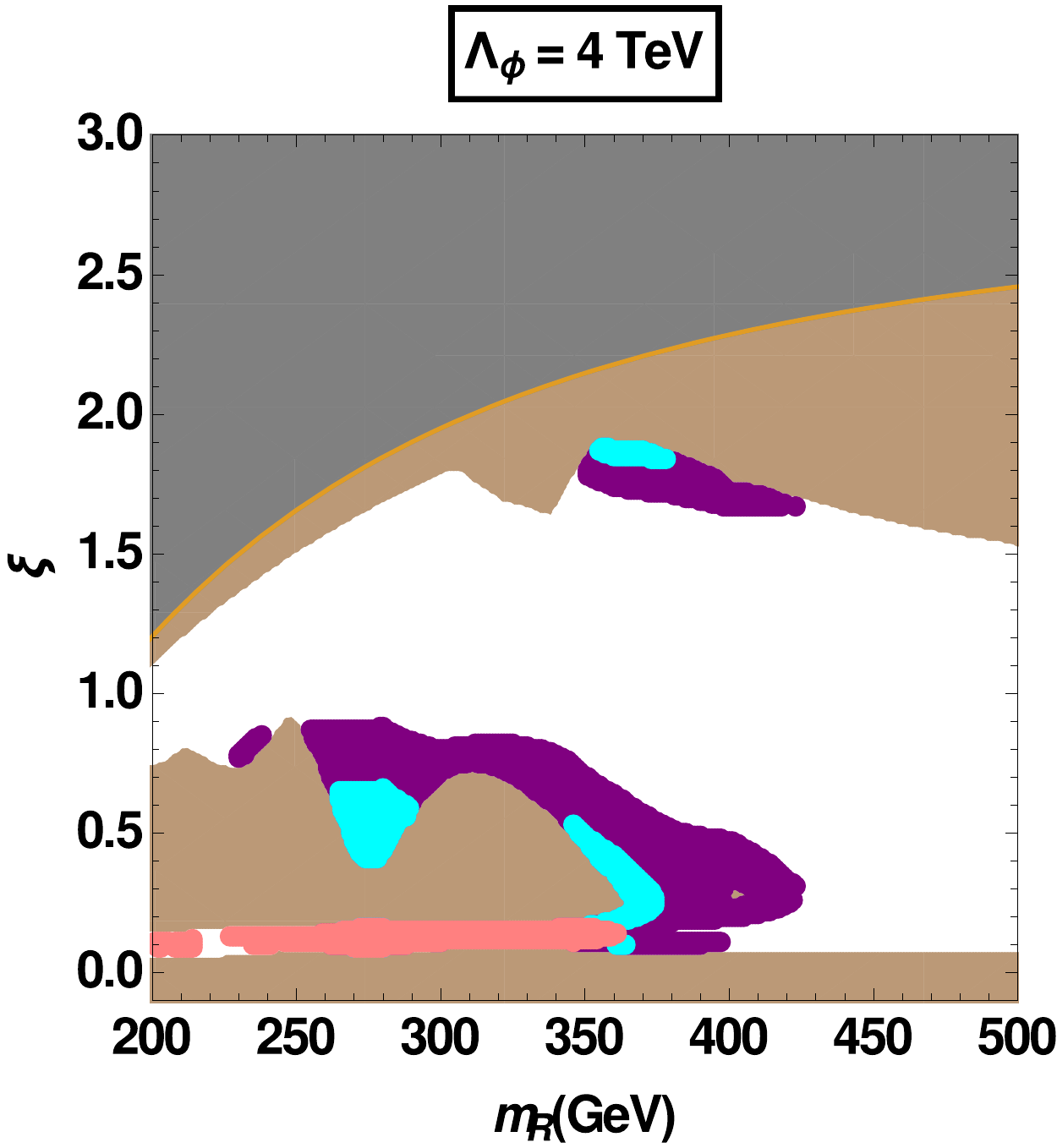}
  \includegraphics[width=7 cm, height= 7cm]{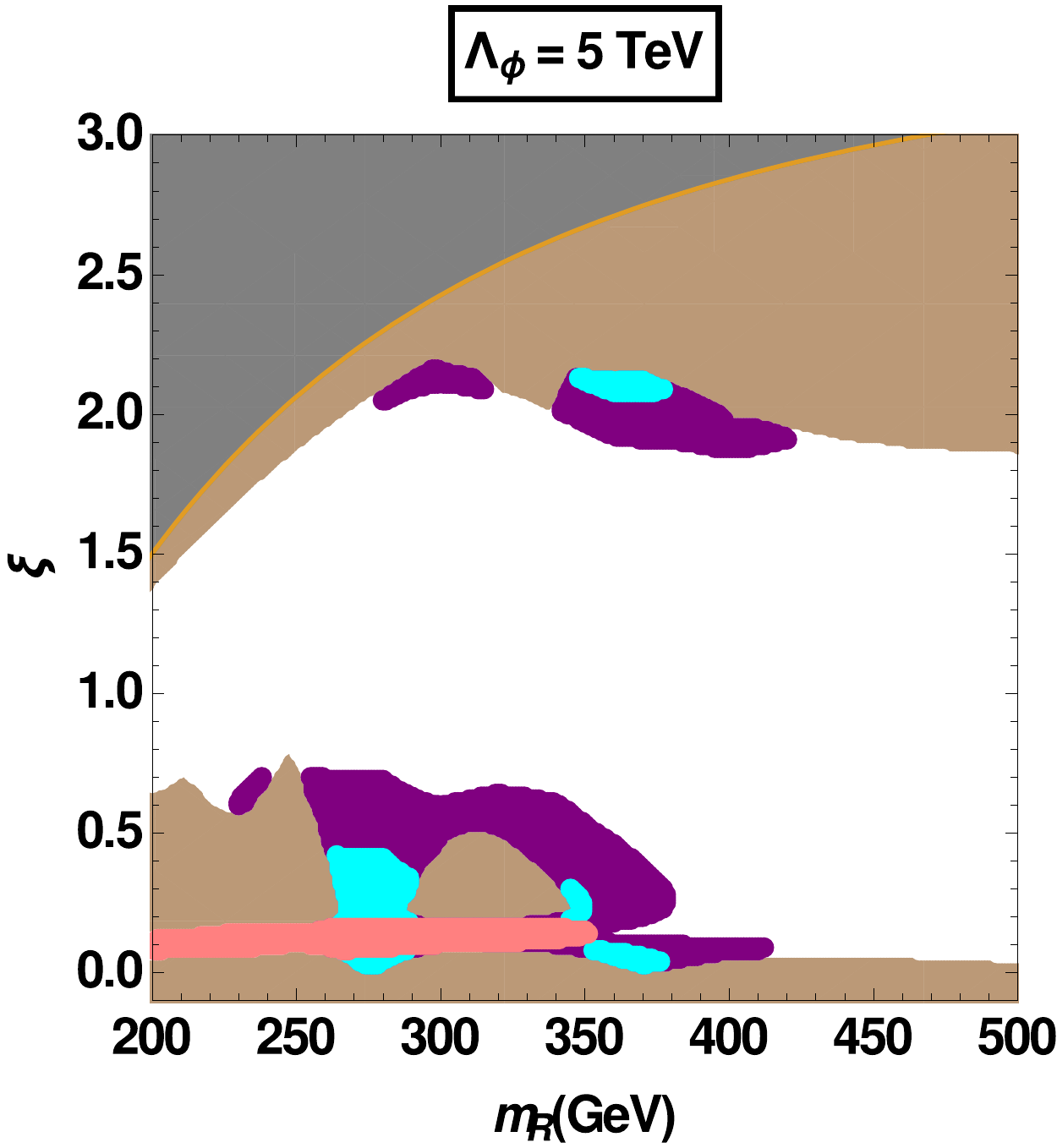}
    \caption{Projection of {\bf region-1 and region-2} parameter space at 14 TeV for $\Lambda_{\phi} = 4$ TeV (left panel) 
    and  for $\Lambda_{\phi} =5$ TeV (right panel). 
  The brown (grey) region is the theoretically allowed (disallowed) region; 
  the pink region represents the area probed by diphoton channel at 3000~fb$^{-1}.$
  Cyan and violet 
 colored regions represent regions which can be probed with  1000 and 3000 fb$^{-1}$ integrated 
 luminosity, respectively, by $ZZ\rightarrow 4\,{\rm leptons}$. The interior white region  represents the region which can not be probed at the LHC-14 TeV.}
  \label{fig:highmassZZ}
\end{figure}

For $m_{R}~<~300$~GeV, we demand that the two reconstructed $Z$'s have  invariant mass
lying within the window of 10 GeV centered about the radion mass. We plot the normalized $p_{T}$ distribution of the leading electron
and leading muon in Fig.~\ref{fig:pT_and_inv_mass_ZZ}.
As mass of the radion increases, the $Z$'s are boosted and hence leptons carry higher $p_{T}$. In order to increase the
signal significance further, we consider events with leading electron and muon having
$$p_{T}~>~50~{\rm GeV}~{\rm for}~300~{\rm GeV}~<~m_{R}~<~400~{\rm GeV}, ~p_{T}~>~80~ {\rm GeV}~{\rm for}~400~{\rm GeV}~<~m_{R}~<~500~{\rm GeV}$$
and $$~~p_{T}~>~100~ {\rm GeV}~{\rm for}~500~{\rm GeV}~<~m_{R}~<~800 ~{\rm GeV}.$$
In this mass range, we have loosened the invariant mass cut by 20 GeV. Instead of applying
a minimum cut on the transverse momentum of the leading lepton($p_{T}^{\rm l(e,\mu})$, one can apply a minimum cut on the transverse momentum of the reconstructed $Z$ boson, which
has the same distinct feature as shown in the Fig.~\ref{fig:pT_and_inv_mass_ZZ}. 
The reach of the 14 TeV LHC to probe the allowed region of the Higgs-radion mixing using $ZZ$ channel is shown in Fig.~\ref{fig:highmassZZ}. 
We consider integrated luminosity of 300~fb$^{-1}$,~1000~fb$^{-1}$~and~3000~fb$^{-1}$ at 14 TeV. The area denoted by cyan represents 
the region that can be probed at 1000~fb$^{-1}$. The area denoted by cyan + purple represents the region that can be probed at the 3000~fb$^{-1}$. We find that using the $ZZ$ channel, one can probe the radion mass up to 450 GeV.
The product of the coupling of the radion to gluon pair and radion to $Z$-pair is small in 
the parameter space allowed after LHC 8 TeV, and hence the production rate of the radion via gluon 
fusion and its decay to $ZZ$ is relatively suppressed in this region. Thus, we need high luminosity to probe this area. 
In spite of such high luminosity, 
only the boundary of the {\bf region-2} has been probed effectively. One can also consider the production
of the radion via vector boson fusion and its decay to $Z$ boson. However, the cross section is suppressed 
for $\Lambda_{\phi}=4$~TeV and demanding that the $Z$'s further decay to leptons decreases the production rate  further.
To probe this scenario more efficiently, we considered this region at the ILC in the next section. 
\section{Prospects for the searches at the ILC}
\label{sec:ilc_analysis}
\begin{figure}[htb]
\centering
\includegraphics[width=7.5 cm, height= 6cm]{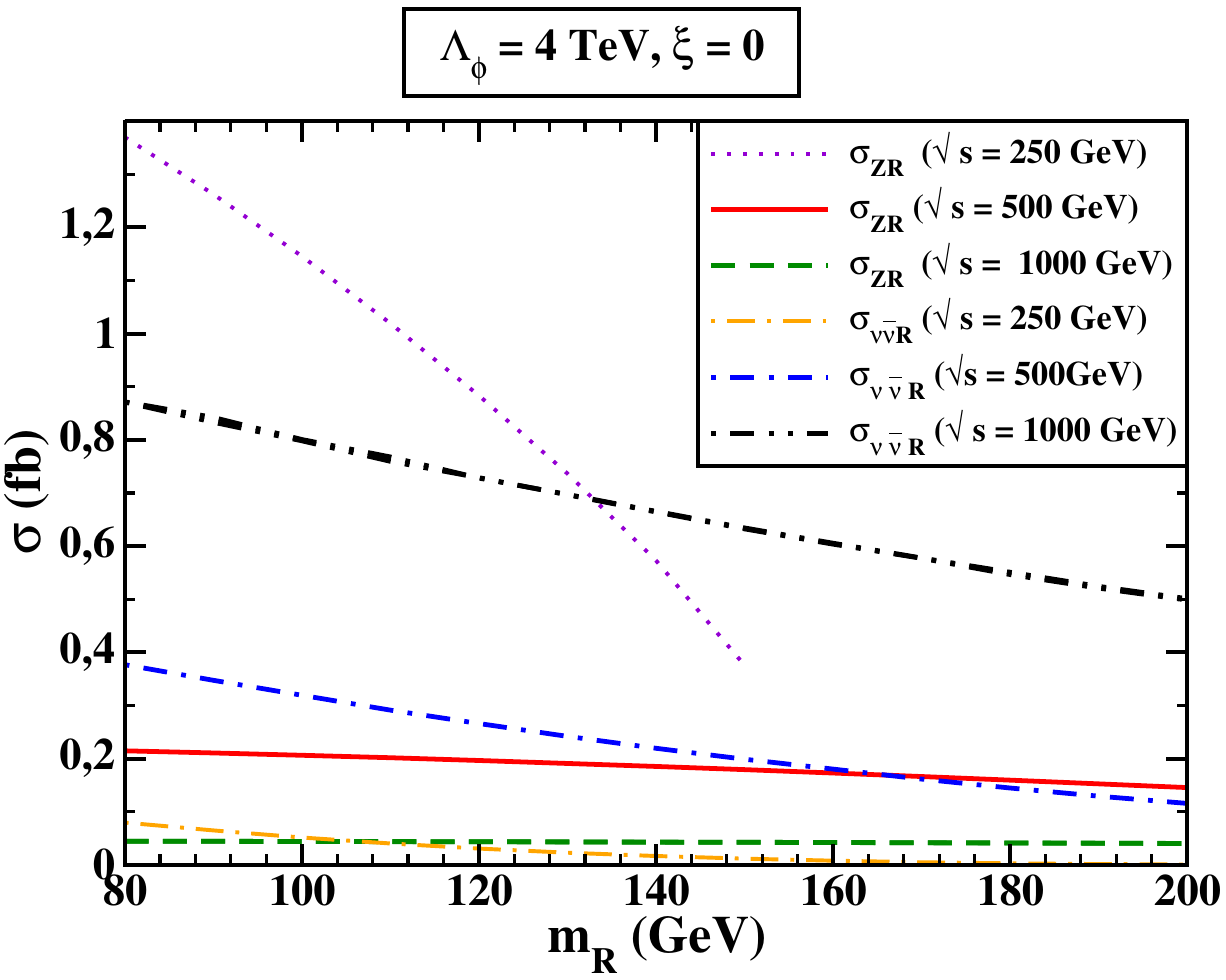}
\vspace{0.5cm}
\includegraphics[width=7.5 cm, height= 6cm]{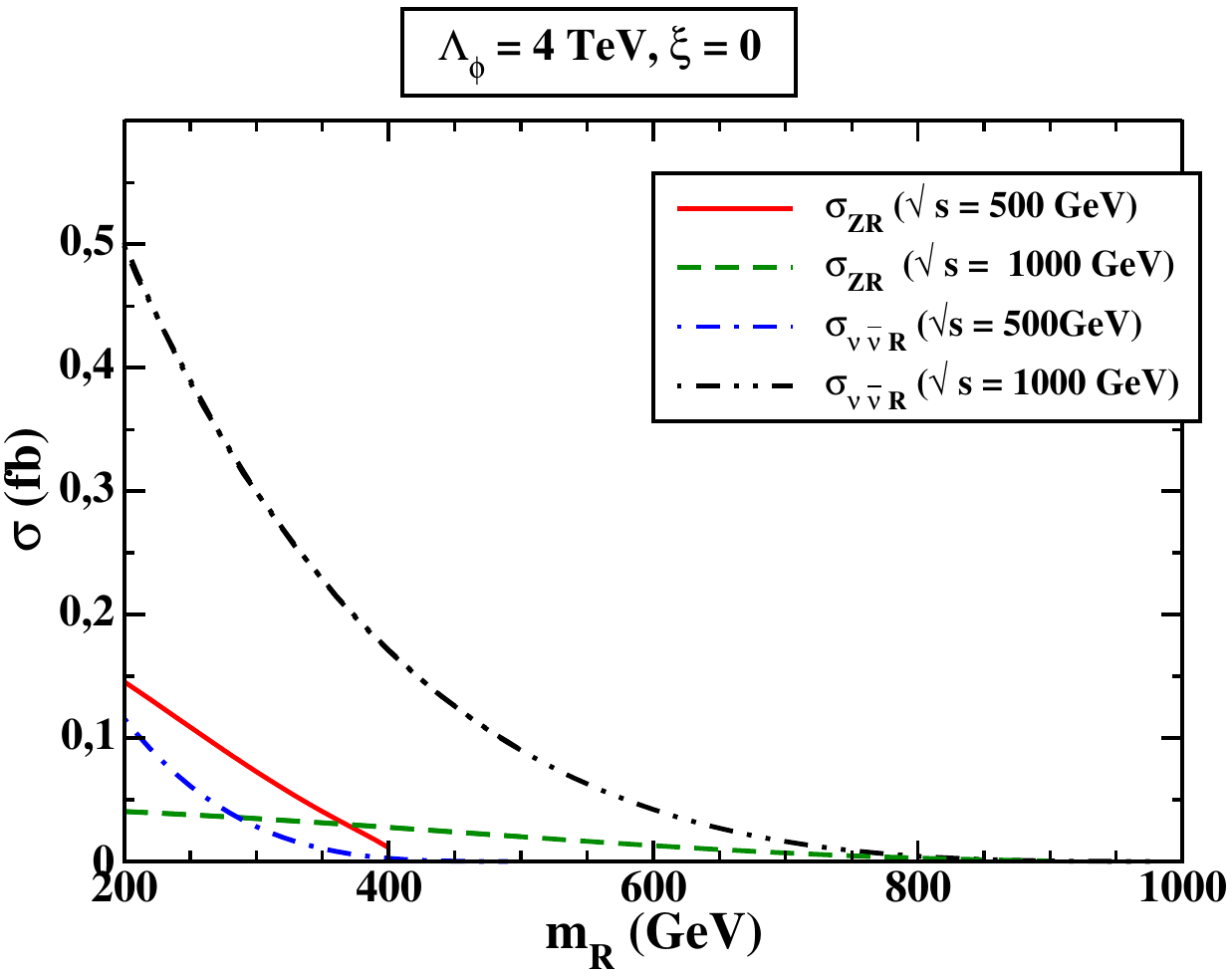}
\caption{The total production cross section of the unmixed radion at the ILC, as a function of the radion mass 
for $\Lambda_\phi$ = 4 TeV. Different contributions coming from $ZR$ and $WWR$ production at various ${\sqrt s} =250, 500$ and 
1000 GeV  are as indicated in legend on the right.}
\label{fig:ilc_cs}
\end{figure}

Next we perform a detailed study of the production cross section  of the mixed
radion at the ILC, and its subsequent decay into the different allowed final states in {\bf region-2}. 
As discussed  
before, the radion  at the LHC could be analyzed mainly through the $\gamma\gamma$
final state. The other prominent decay modes, $gg$ and $b\bar{b}$, for $m_R<$ 180 GeV, will have a 
large background in the LHC and therefore cannot be tested thoroughly. The direct search for
the radion in ILC can be performed as long as it is kinematically accessible, and through it
 decay either into a dijet final state or $WW/ZZ$ final state, depending on its mass. Since the  
radion couples much like the Higgs, the production process will be analogous with the Higgs. It will 
be produced through the channel 
$e^+e^-\rightarrow Z R$ (associated production with $Z$, denoted as $ZR$) and 
$e^+e^-\rightarrow \nu_e \bar{\nu}_e R$ ($WW$ fusion, denoted as $WWR$). We do not consider 
the $e^+e^-\rightarrow e^+ e^- R$ ($ZZ$ fusion) process
here, as it has a very small cross section and will therefore require a high luminosity for any meaningful results. We
show the total production cross section  for $\sqrt{s}$ of 250, 500 and 1000 GeV, with the respective
integrated luminosities of 250, 500 and 1000 fb$^{-1}$  in Fig.~\ref{fig:ilc_cs}, for $\xi$ = 0 and $\Lambda_\phi$ = 4 TeV,  with unpolarized beams.
It can be seen from Fig.~\ref{fig:ilc_cs} that $\sigma_{WWR}$ increases with $\sqrt{s}$, whereas 
$\sigma_{ZR}$, decreases. We consider the $ZR$ process, at $\sqrt{s}$ = 250 GeV,  as it is the dominant one. 
We first study the leptonic decay mode of $Z$, with $Z\rightarrow l\bar{l}$, where $l=e^-,~\mu^-$.
Since the leptonic branching ratio of $Z$ is small, we  also considered the hadronic decay mode. The analysis 
at $\sqrt{s}$ = 500 and 1000 GeV, is  through $WWR$, which is dominant at those energies.

\subsection{Analysis in the $e^+e^-\rightarrow Z R$, $Z\rightarrow l\bar{l},~R\rightarrow b\bar{b}$ decay channel}
\label{subsec:ZRll}

\begin{figure}[htb]
\centering
\includegraphics[width=8 cm, height= 6cm]{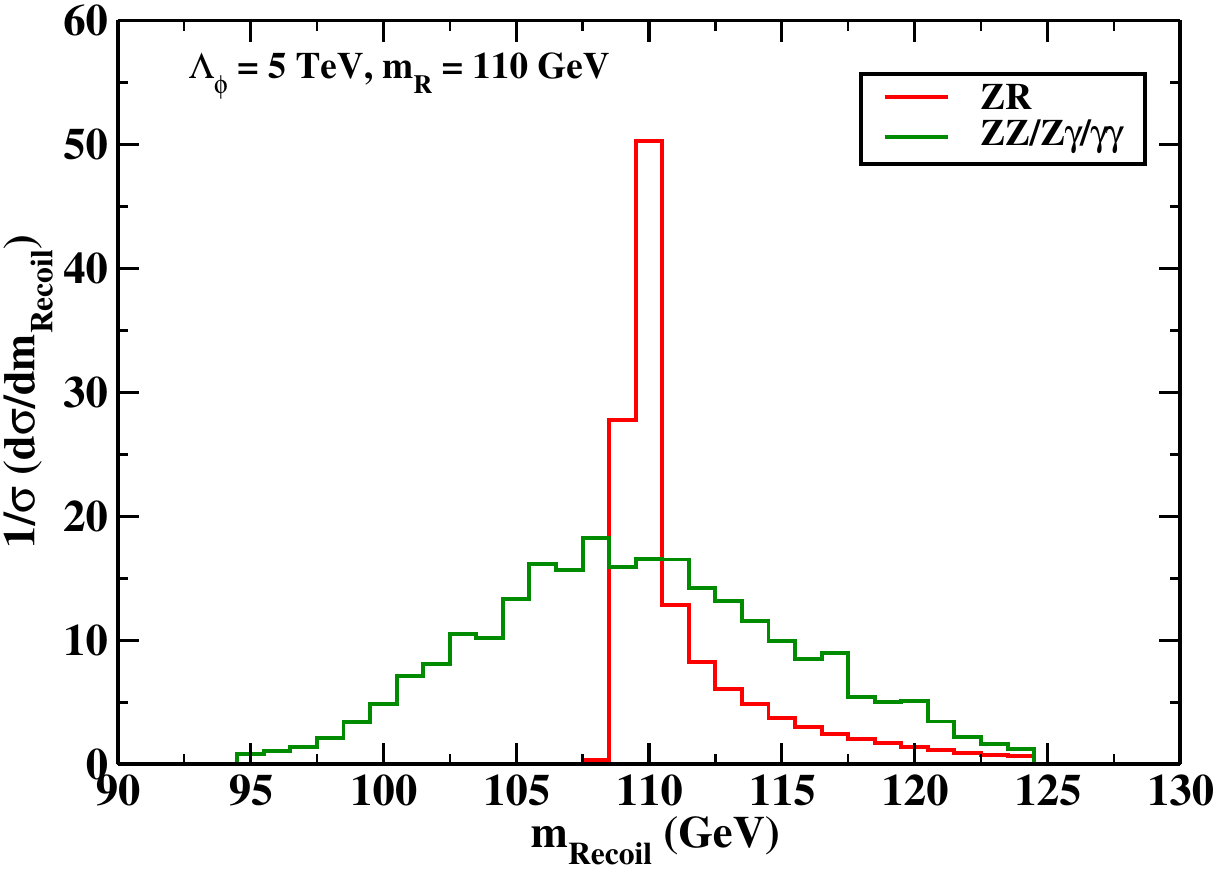}
\caption{The recoil mass distribution of the $e^+e^- \rightarrow ZR$ process at $\sqrt{s}$ = 250 GeV, with unpolarized beams.
The red line represents the signal, the green the background. }
\label{fig:recoil_mass}
\end{figure}

\begin{figure}[htb]
\centering
\includegraphics[width=7.5 cm, height= 6cm]{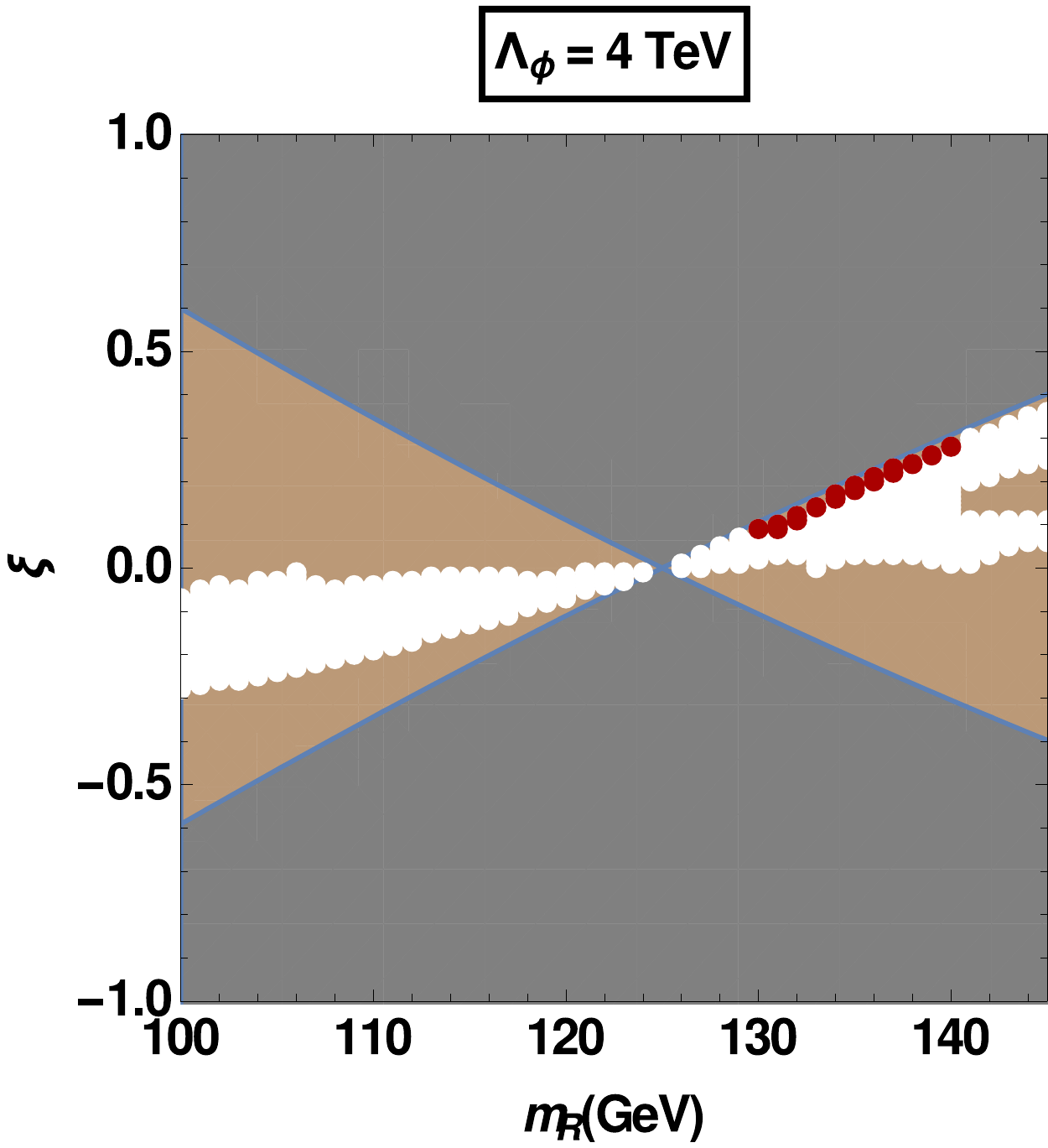}
\vspace{0.5cm}
\includegraphics[width=7.5 cm, height= 6cm]{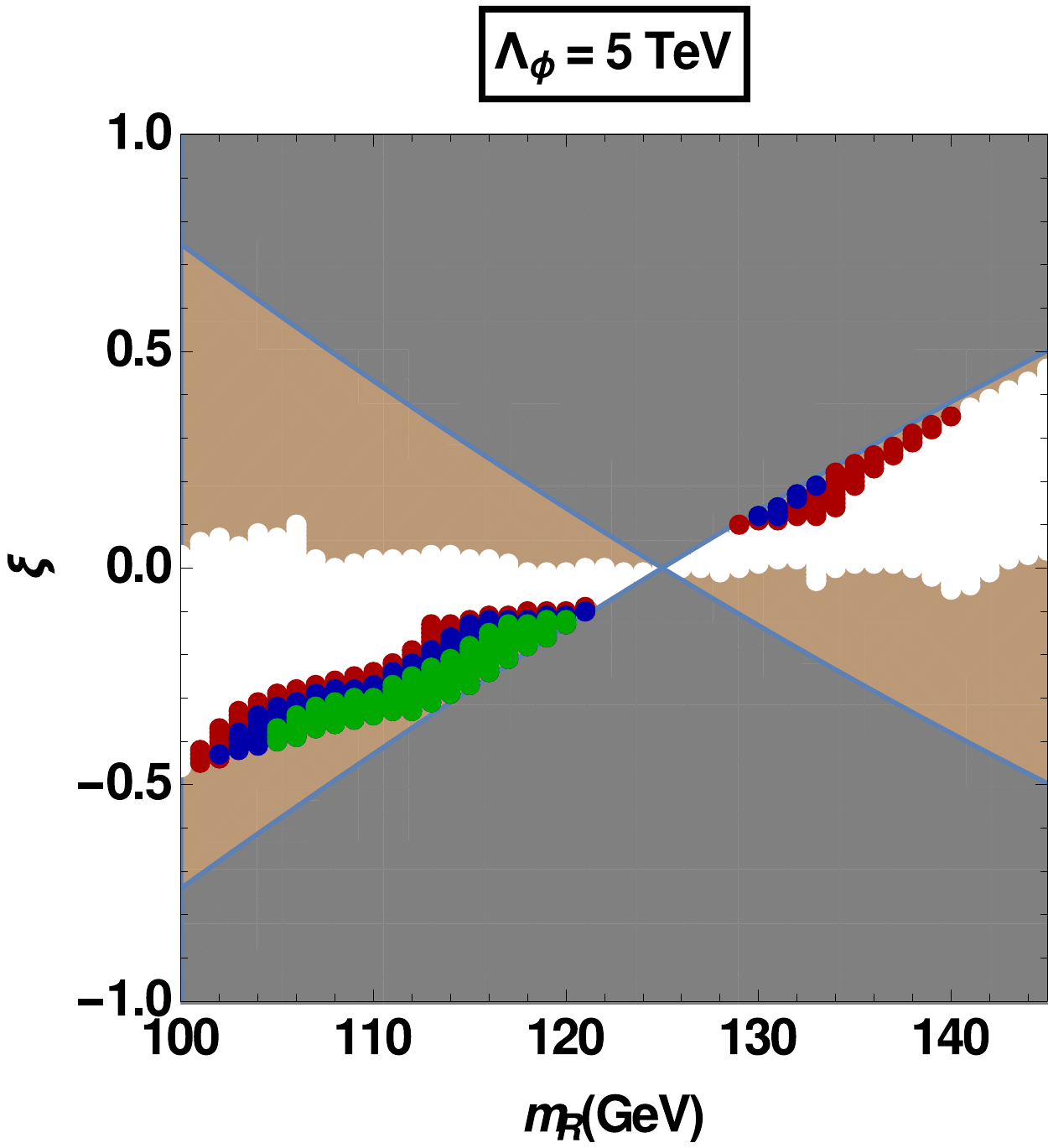}
\caption{Projected {\bf region-2} of the parameter space at $\sqrt{s}$ = 250 GeV for $\Lambda_{\phi} = 4$ TeV (left panel) and  for $\Lambda_{\phi} =5$ TeV (right panel), 
for a light mixed state through the $Z R$, $Z\rightarrow l\bar{l},~R\rightarrow b\bar{b}$ decay channels. 
The brown (dark grey) region is the theoretically allowed (disallowed) region. The green, green+blue, and green+blue+red 
colored regions represent areas  which can be probed at 250, 500 and 1000  fb$^{-1}$ integrated 
luminosity respectively. The interior white region is the one still allowed by the LHC 8 TeV results which
can not be probed by the ILC through this final state.}
\label{fig:exclusion_ZHll}
\end{figure}

The $ZR$ production process considered here will consist of two leptons and two $b$ jets in the final state. At the ILC, due to
a large cross section of the Bhabha Scattering $e^+e^-\rightarrow e^+e^-$ and the muon pair production 
$e^+e^- \rightarrow \mu^+\mu^-$, pre-selection cuts are applied on the leptons, in order to reduce 
the simulation time~\cite{Li:2012taa}. These cuts are as follows : 
\begin{eqnarray*}
(a)&&~~ |\cos\theta_{l}| < 0.95~~~~(b)~~ m_{l\bar{l}} \in (80, 100) ~{\rm GeV}~~~~(c)~~ P^T_{l\bar{l}} > 20 ~{\rm GeV} \\
(d)&&~~ m_{\rm recoil} \in m_R~ \pm 15 ~{\rm GeV}~~~~(e)~~ 0.2 <|\phi_l-\phi_{\bar{l}}|< 3. 
\end{eqnarray*}
Here $m_{l\bar{l}}$ is the invariant mass of the dilepton system,  $P^T_{l\bar{l}}$ is the transverse
momentum calculated from the vectorial sum of the two leptons, and the recoil mass $m_{\rm recoil}$ is expressed as
\begin{equation}
 m_{\rm recoil}^2 = s-2\sqrt{s} E_{l\bar{l}} + m_{l\bar{l}}^2,~~(E_{l\bar{l}} =E_l +E_{\bar{l}}),
\end{equation}
where $\sqrt{s}$ is the c.m. energy and $E_l,~E_{\bar{l}}$ are the energies of the two leptons. The last cut 
is on the difference between the azimuthal angle of the two leptons. The signal is 
selected by identifying two well measured leptons in the final state, which yield an invariant mass 
peak around the $Z$ boson mass. The recoil mass of the system should give the mass of the radion. Additionally
the signal event should consist of two $b$ tagged jets, with an invariant mass peak around $m_R~\pm$ 5 GeV.
The signal and the background events with the showering and the hadronization are generated in
PYTHIA 8 \cite{Sjostrand:2014zea}. 
Jet formation is done through Fastjet-3.2.0~\cite{Cacciari:2011ma} using the inbuilt $k_t$ algorithm 
for $e^+e^-$ collisions, which is similar to the Durham algorithm. A jet is tagged as a $b$ jet if it has a $b$
parton within a cone of $\Delta R <$ 0.4 with the jet axis, where 
$\Delta R =\sqrt{(\Delta \eta)^2+(\Delta \phi)^2}$. A tagging efficiency of 80\%~\cite{Asner:2013psa}
is also incorporated. The main backgrounds for the process under study are
\begin{eqnarray*}
(1)&&~~ e^+e^-\rightarrow \gamma \gamma,~~~~\gamma\rightarrow l\bar{l},~b\bar{b}
 ~~~~(2)~~ e^+e^-\rightarrow \gamma Z,~~~~Z,~\gamma\rightarrow l\bar{l},~b\bar{b} \\
 (3)&&~~ e^+e^-\rightarrow Z Z,~~~~Z\rightarrow l\bar{l},~b\bar{b}
 ~~~~(4)~~ e^+e^-\rightarrow Z H,~~~~Z\rightarrow l\bar{l},~H\rightarrow  b\bar{b}
\end{eqnarray*}
We added the contributions of the backgrounds for $\gamma \gamma,~\gamma Z,~ZZ$ and presented our result 
as a single background. The dominant background to this process
comes from the $ZH$ final state. In order to account for the detector effects, the momenta of the leptons and the jets
are smeared using the 
following parametrization. The jet energies are smeared~\cite{BrauJames:2007aa} with the different 
contributions being added in quadrature.
\begin{eqnarray}
 \frac{\sigma(E_{jet})}{E_{jet}}&=& \frac{0.4}{\sqrt{E_{jet}}} \oplus 2.5\%
\end{eqnarray}
The momentum of the lepton is smeared as a function of the momentum and the angle $\cos\theta$ 
of the emitted leptons~\cite{Li:2010ww}
\begin{eqnarray}
 \frac{\sigma(P_l)}{P_l^2} =\left(\begin{array}{cc}a_1\oplus\frac{b_1}{P_l},&|\cos\theta|< 0.78 \\
                \left(a_2 \oplus \frac{b_2}{P_l}\right)  \left(\frac{1}{\sin(1-|\cos\theta|)}\right) & |\cos\theta|> 0.78 
               \end{array}\right)
\end{eqnarray}
with
\begin{eqnarray}
(a_1,~b_1) &=& 2.08\times10^{-5}~({\rm{1/GeV}}),~~8.86\times10^{-4}, \nonumber \\
(a_2,~b_2) &=& 3.16\times10^{-6}~({\rm{1/GeV}}),~~2.45\times10^{-4}. 
\end{eqnarray}
The region below 100 GeV can be probed with a great precision at the LHC,  therefore we do not consider that region here.
Moreover, at the  ILC, the region below 100 GeV will have a large background from the $Z$ resonance peak.  Therefore we 
performed our analysis, for $m_R \geq$  100 GeV. The recoil mass distribution is shown in Fig.~\ref{fig:recoil_mass},
for a radion mass of 110 GeV and $\Lambda_\phi$ = 5 TeV. The recoil mass should be peaked near $m_R$, but
due to the initial state radiation and the bremsstrahlung effect, the distribution is spread out. 
In Fig.~\ref{fig:exclusion_ZHll} we show the allowed parameter space 
that can be probed by the ILC at $\sqrt{s}$ = 250 GeV through the $ZH$ final state, with the $Z$ decaying to two 
leptons. The theoretically allowed (disallowed) region, Eq.~\ref{eq:cond1} is shown in brown (grey), whereas
the region currently allowed by the LHC 8 TeV results is shown in white. The regions probed with different luminosities
are superimposed over each other and on the white region.  
The region in green denotes the values of the mixing parameter $\xi$
that can be observed with more than 3$\sigma$ significance level at 250 fb$^{-1}$, the regions in green and blue denote the 
probed regions at 500 fb$^{-1}$ and the red region along with green and blue can be probed with 1000 fb$^{-1}$.
As the branching ratio of $Z$ to the leptonic final state is small, only a narrow region can be probed for 
$\Lambda_\phi$ of 4 TeV, with a high luminosity of 1000 fb$^{-1}$. But this channel can be used as a good probe 
for $\Lambda_\phi$ of 5 TeV, where it can probe the region which can not be explored in LHC, shown in white in Fig.~\ref{fig:lowmass}. 
We next consider the hadronic decay mode of $Z$, with the aim to probe a larger region of the parameter space.

\subsection{Analysis in the $e^+e^-\rightarrow Z R$, $Z\rightarrow q\bar{q},~R\rightarrow b\bar{b}$ decay channel}
\label{subsec:ZRqq}
\begin{figure}[htb]
\centering
\includegraphics[width=7.5 cm, height= 6cm]{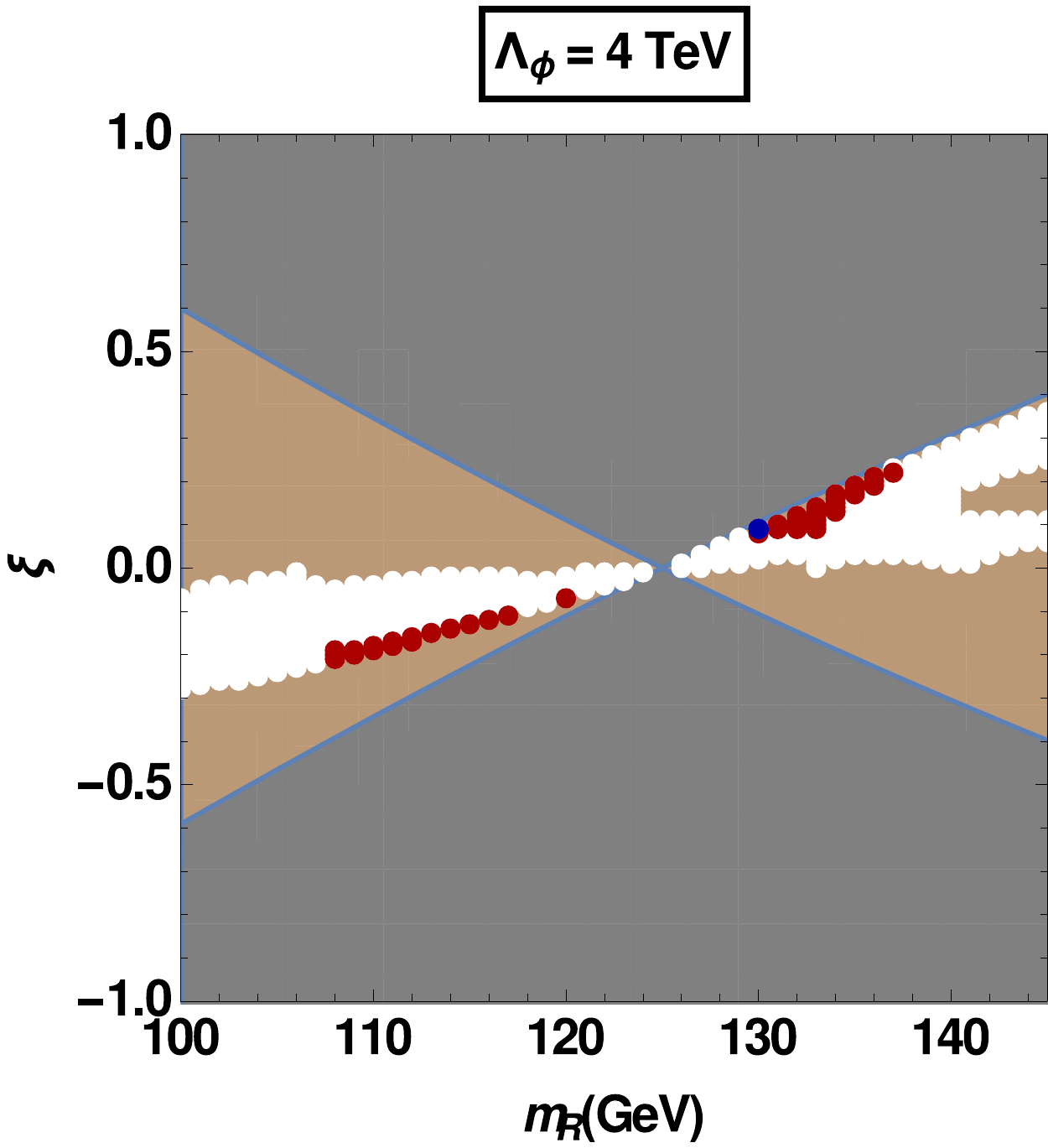}
\vspace{0.5cm}
\includegraphics[width=7.5 cm, height= 6cm]{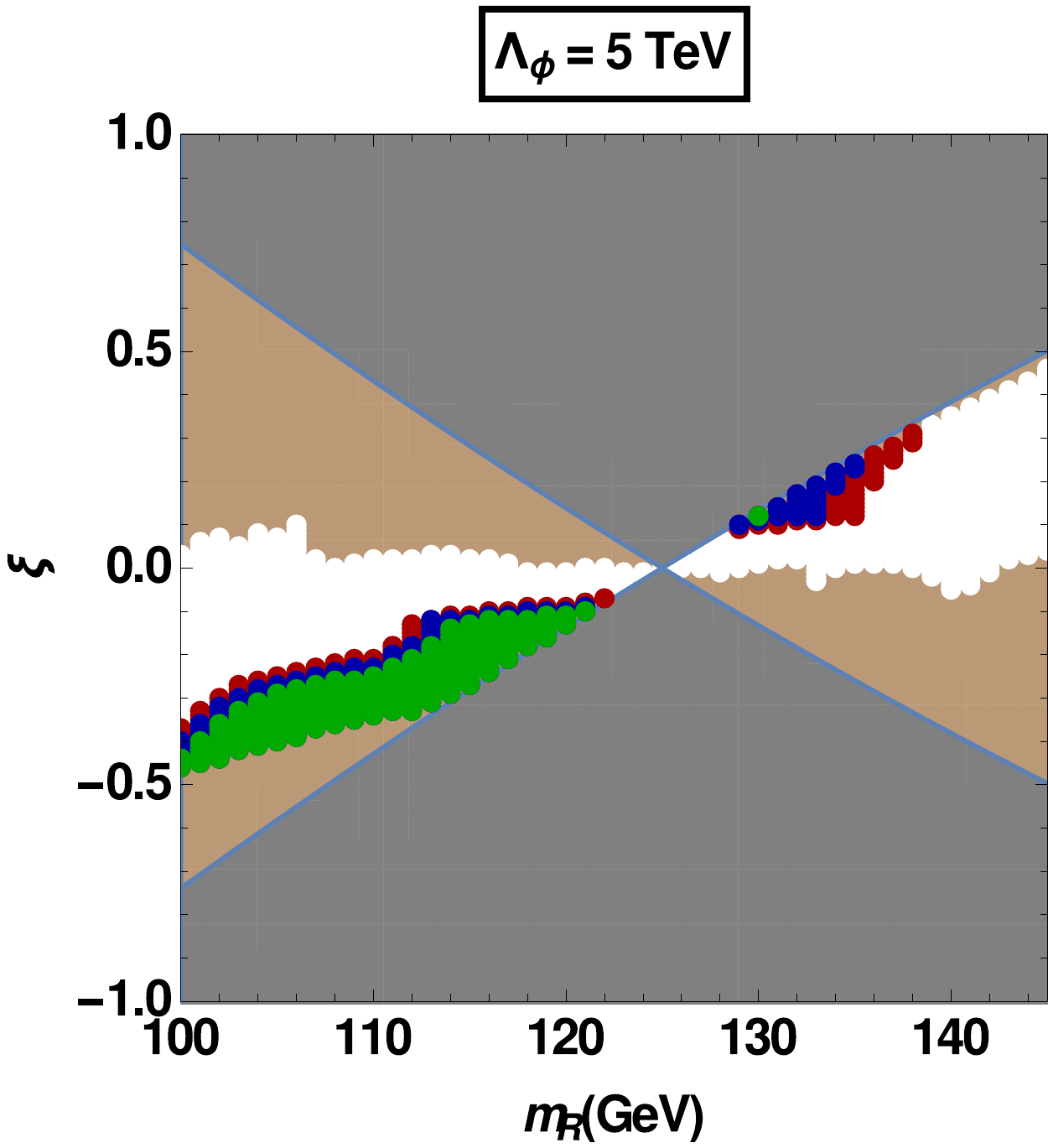}
\caption{Projected {\bf region-2} of  parameter space at $\sqrt{s}$ = 250 GeV for $\Lambda_{\phi} = 4$ TeV (left panel) and  $\Lambda_{\phi} =5$ TeV (right panel), 
 for a light mixed state through the $Z R$, $Z\rightarrow q\bar{q},~R\rightarrow b\bar{b}$ decay channel. The color 
 coding is the same as in Fig.~\ref{fig:exclusion_ZHll}.}
\label{fig:exclusion_ZHqq}
\end{figure}

The analysis for this decay channel is similar to the leptonic decay mode of $Z$. The signal consists of four jets,
therefore  events with isolated leptons are rejected. The jets are then constructed with the $k_t$ jet algorithm
implemented in Fastjet, with the jet radius of 1.2 and the $p^{min}_t$ = 1.0 GeV, without restricting the number of 
the reconstructed jets~\cite{Miyamoto:2013zva},~\cite{Thomson:2015jda}. Then, from all the jet pairs, a jet pair 
of mass consistent with $Z$ is selected. The recoil mass is calculated and the cut on recoil mass is applied. The 
remaining jets are then checked for $b$ tagging as discussed before. The events with two $b$ tagged jets with an invariant
mass peak around $m_R~\pm$ 5 GeV are then selected. The main backgrounds of the process are similar to the previous
process with the $Z, \gamma$ decaying to quarks. There is an additional background from the $WW$ final state, with the 
$W$'s decaying hadronically. We show in Fig.~\ref{fig:exclusion_ZHqq}, the region in the $\xi-m_R$ plane that can 
be probed at $\sqrt{s}$ = 250 GeV for different values of integrated luminosities. The color coding is similar to the
previous Fig.~\ref{fig:exclusion_ZHll}. The hadronic final state of $Z$ can probe a larger region of the parameter space compared to 
the leptonic decay mode. We do not consider this process for $\sqrt{s}$ = 500 GeV as the cross section falls 
with $\sqrt{s}$ and there the process $WWR$ yields better results. 

\subsection{Analysis in the $e^+e^-\rightarrow \nu_l \bar{\nu}_l R$, $R\rightarrow b\bar{b}$ decay channel}
\label{subsec:WW_fusion}

We next consider the measurement of the radion production cross section through $WW$ fusion 
at $\sqrt{s}$ = 500 and 1000 GeV. We do not consider this process at $\sqrt{s}$ = 250 GeV, as in the 
$\nu_l \bar{\nu}_l R$ final state, $ZR$, with $Z$ decaying to $\nu_l \bar{\nu}_l$ and $WWR$ cannot be 
treated separately. The $ZR$ cross section scales as $s^{-1}$, whereas the $WWR$ cross section increases
logarithmically to large $\sqrt{s}$. Therefore at low c.m. energies $ZR$ will act as one of the most 
challenging background for $WWR$. The signal event consists of two energetic very forward
$b$ jets along with missing 4 momentum. The main backgrounds in this search mode, apart from $ZR$, will be
($1$) two-jet production ($q\bar{q}$), ($2$) semi-leptonically and hadronically decaying $Z/W$ pairs 
($q\bar{q}q\bar{q}$, $q\bar{q}l\bar{l}$, $q\bar{q}\nu_l\nu_{\bar{l}}$, $q\bar{q}l\nu_{\bar{l}}$), 
($3$) single $Z/W$- boson production process ($l\nu_{\bar{l}} W$, $e^+e^-Z$) ($4$) $t\bar{t}$ production 
($b\bar{b}WW$) and ($5$) the SM Higgs production ($e^+e^-\rightarrow \nu_l \bar{\nu}_l H$, $H\rightarrow b\bar{b}$)
through  $WW$ fusion. Since the signal event consists of two $b$ jets along with missing energy, first the 
events with isolated leptons are rejected, and all the visible final particles are reconstructed into two jets. 
Jet formation is done with the same algorithm as discussed before for the $ZR$ process, with $Z$ decaying into 
two leptons. The lepton isolation cut subdues the background from semi-leptonically or leptonically decaying $W/Z$. 
The reconstructed jets are then tagged as $b$ jets, which significantly suppresses the hadronic background. 
The other selection cuts are: 
\begin{enumerate}
\item A cut on visible mass($m_{\rm vis}$) ($m_R \pm 15$ GeV), 
\item A cut on visible energy ($5 <$ $E_{\rm vis}$ $< m_R+150$ GeV), 
\item A cut on the missing mass ($m_{\rm missing}>$ 200 GeV for 
$\sqrt{s}$ = 500 GeV and $m_{\rm missing}>$ 400 GeV for $\sqrt{s}$ = 1000 GeV), and 
\item A cut on the reconstructed invariant 
radion mass $m_R \pm 10$ GeV. 
\end{enumerate}
These cuts are sensitive to $\sqrt{s}$ as well as the mass of the radion. The background
events from the hadronically decaying $W/Z$ have a lower missing mass, therefore a higher cut on missing mass 
is applied.
\begin{figure}[htb]
\centering
\vspace{0.6cm}
\includegraphics[width=.3\textwidth, height= 6cm]{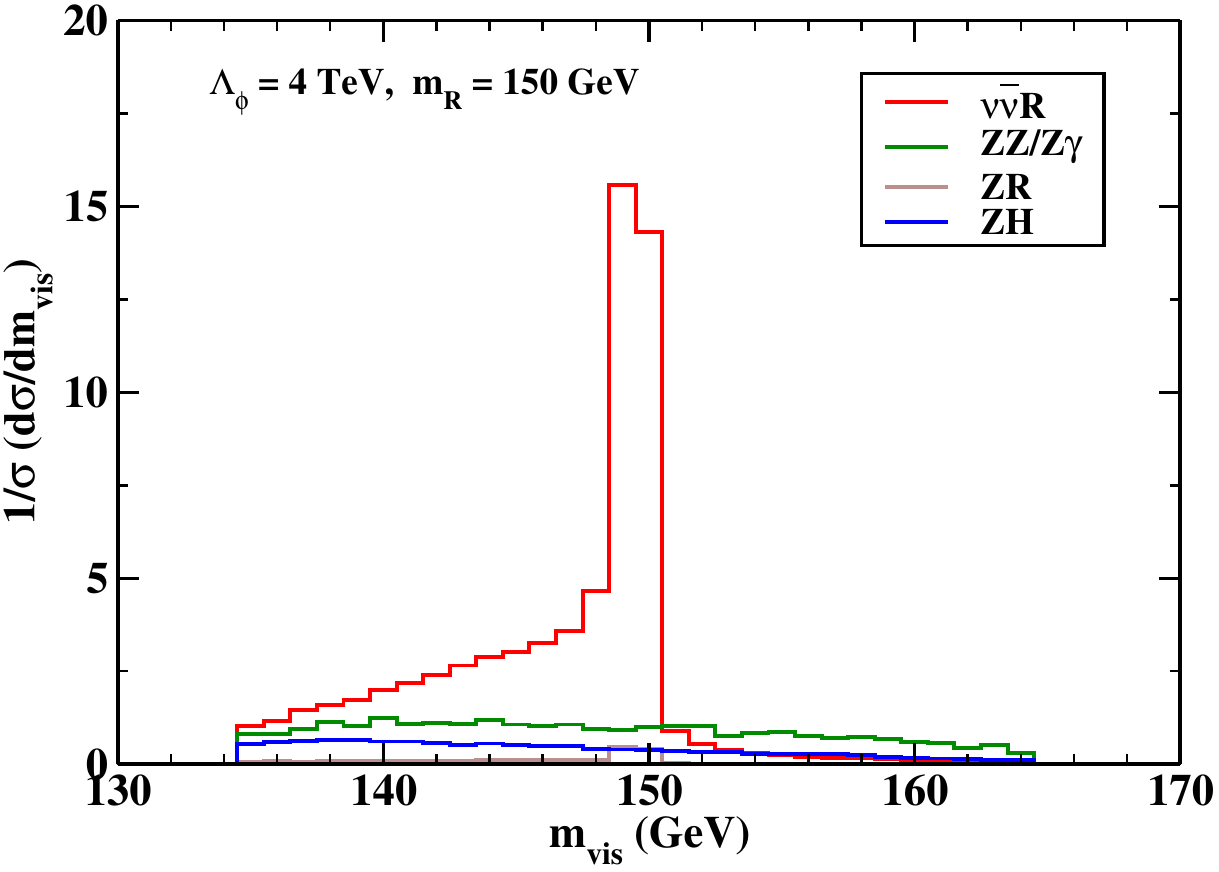}\hfill
\includegraphics[width=.3\textwidth, height= 6cm]{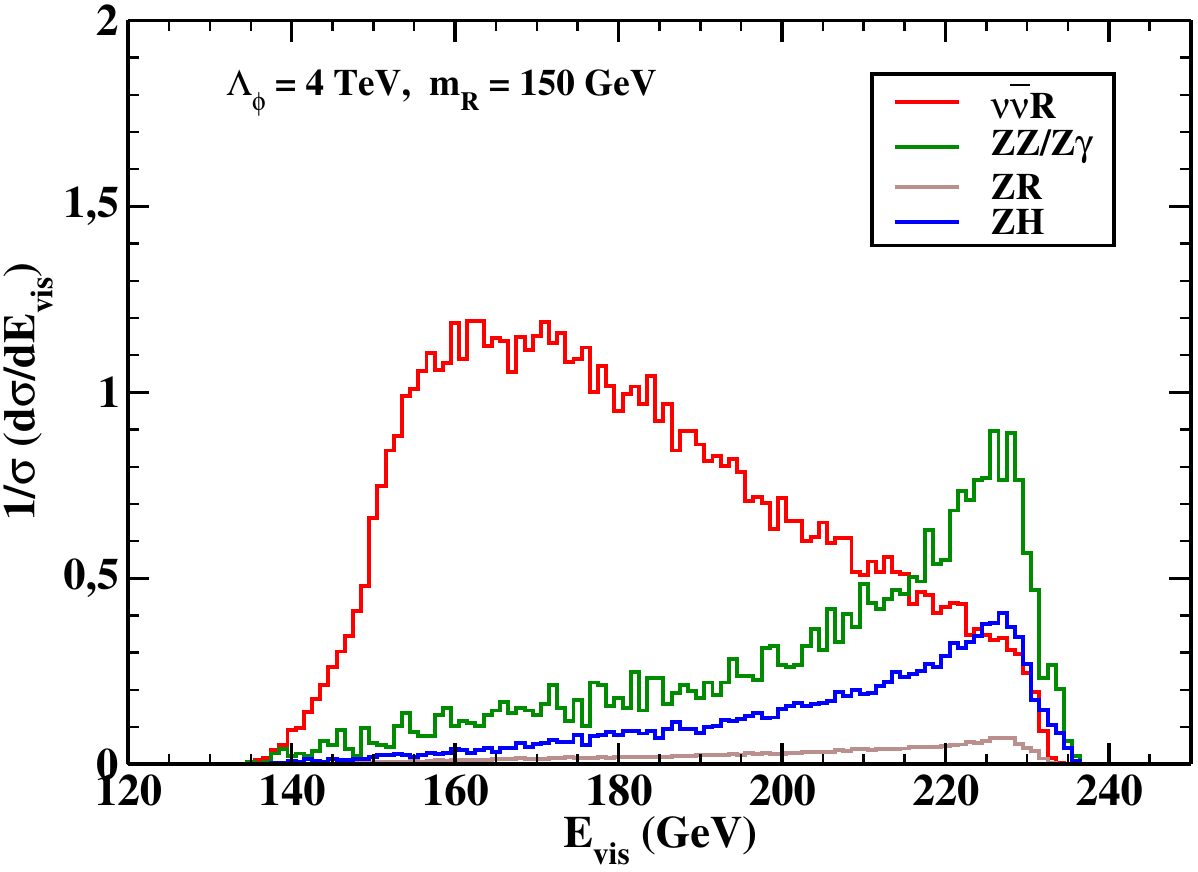}\hfill
\includegraphics[width=.3\textwidth, height= 6cm]{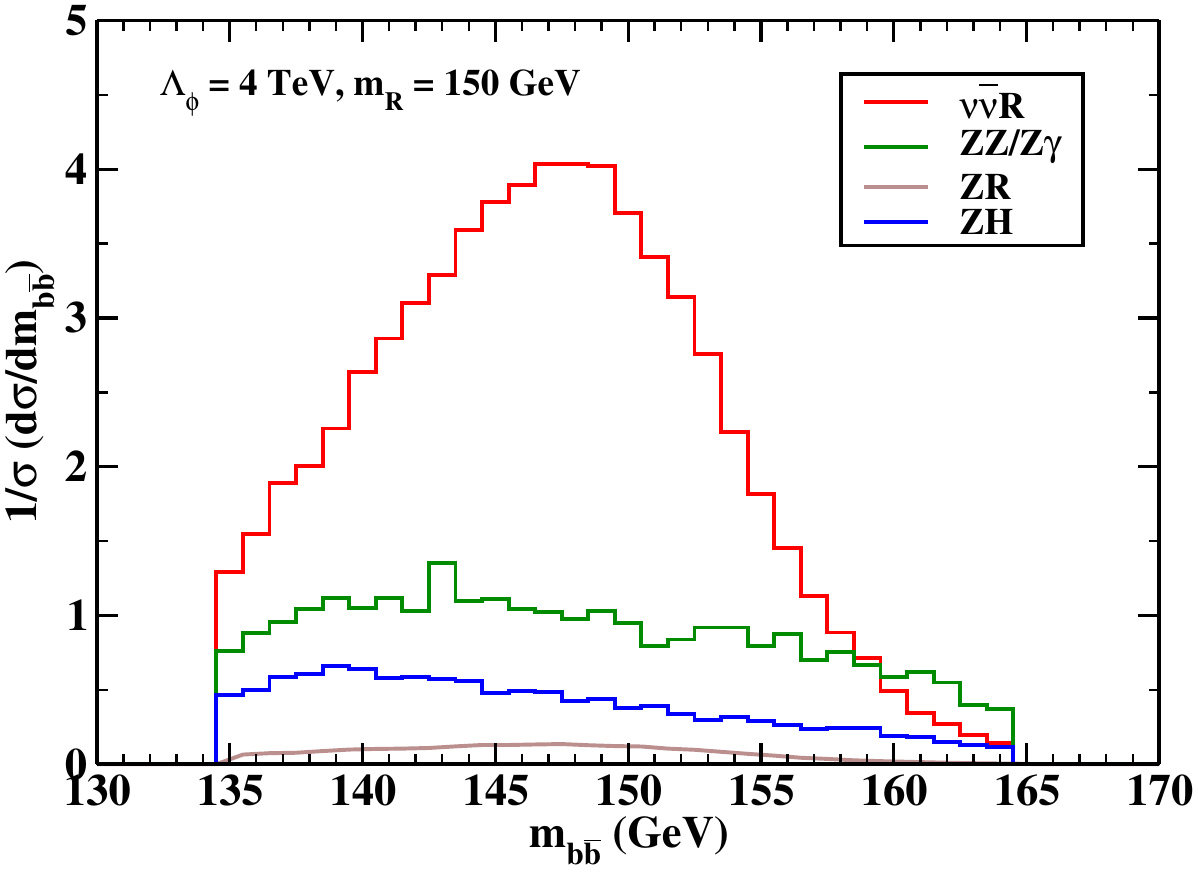}
 \caption{The  visible mass (left panel), visible energy (middle panel), and the invariant mass distribution (right panel) for a radion of mass 150 GeV at
 $\sqrt{s}$ = 500 GeV and a luminosity of 500 fb$^{-1}$. The red line 
 represents the signal, whereas the green, blue and the brown lines represent the background.}
 \label{fig:distribution_150GeV}
\end{figure}
 In Fig.~\ref{fig:distribution_150GeV} we show the visible mass, normalized visible energy,  and the invariant 
mass distribution for $m_R$ = 150 GeV. From the left-hand figure it can be seen  that the cut on the visible 
mass results in reducing most of the background events, as the background is more spread out than the signal. The signal
is not peaked at $m_R$, but is a bit spread in the lower end, due to the loss of the beam's energy from the 
bremsstrahlung and the initial state radiation effects. The same effect can be seen in the visible energy distribution (middle panel) 
of the signal. An upper cut on $E_{\rm vis}$, reduces most of the background as the background peaks towards higher $E_{\rm vis}$.
Finally, on the right hand panel of Fig.~\ref{fig:distribution_150GeV} we show the invariant mass reconstruction from the two $b$-tagged jets for both the signal and the background.
\begin{figure}[htb]
\centering
\includegraphics[width=.3\textwidth, height= 6cm]{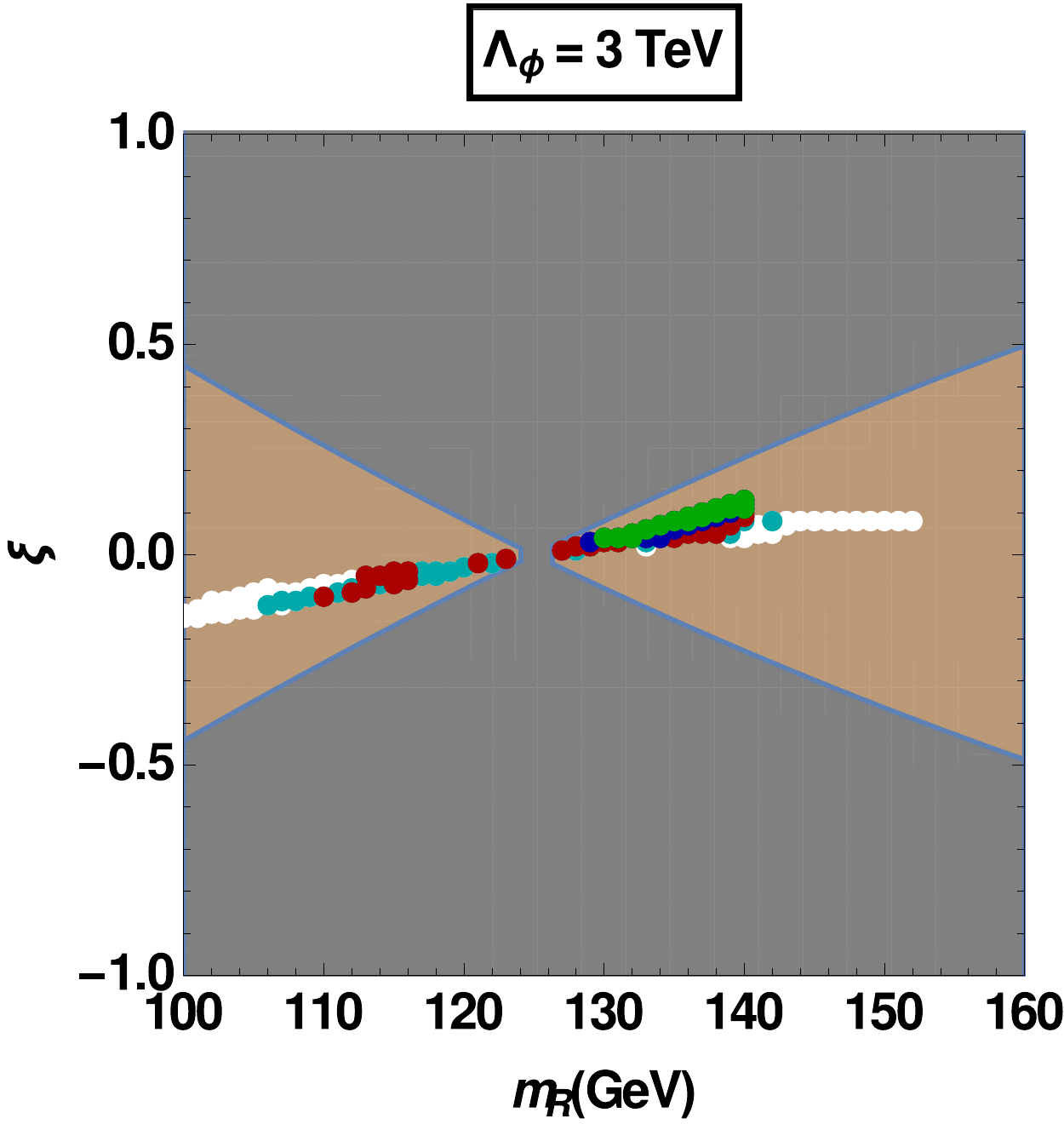}
\includegraphics[width=.3\textwidth, height= 6cm]{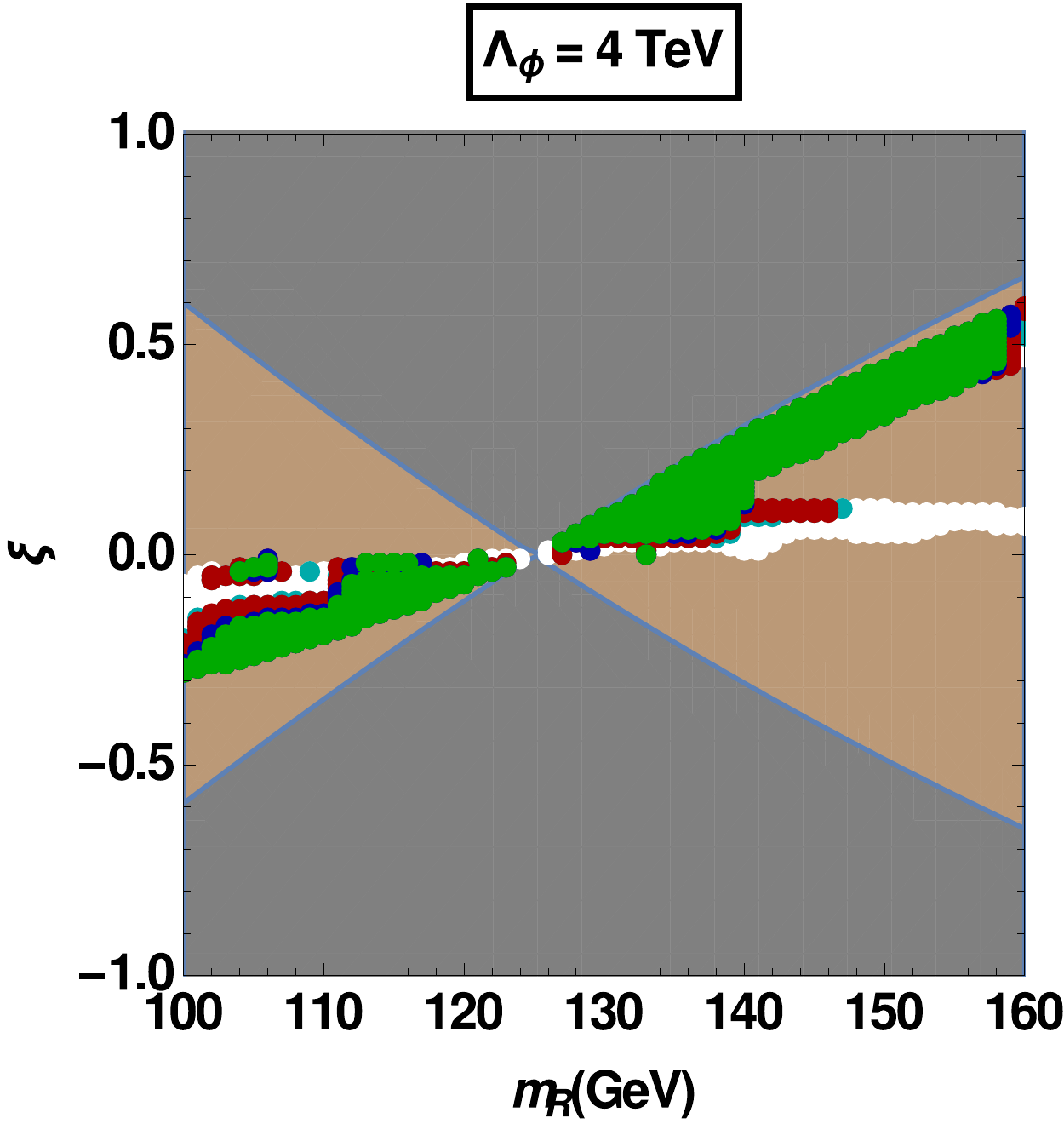}
\includegraphics[width=.3\textwidth, height= 6cm]{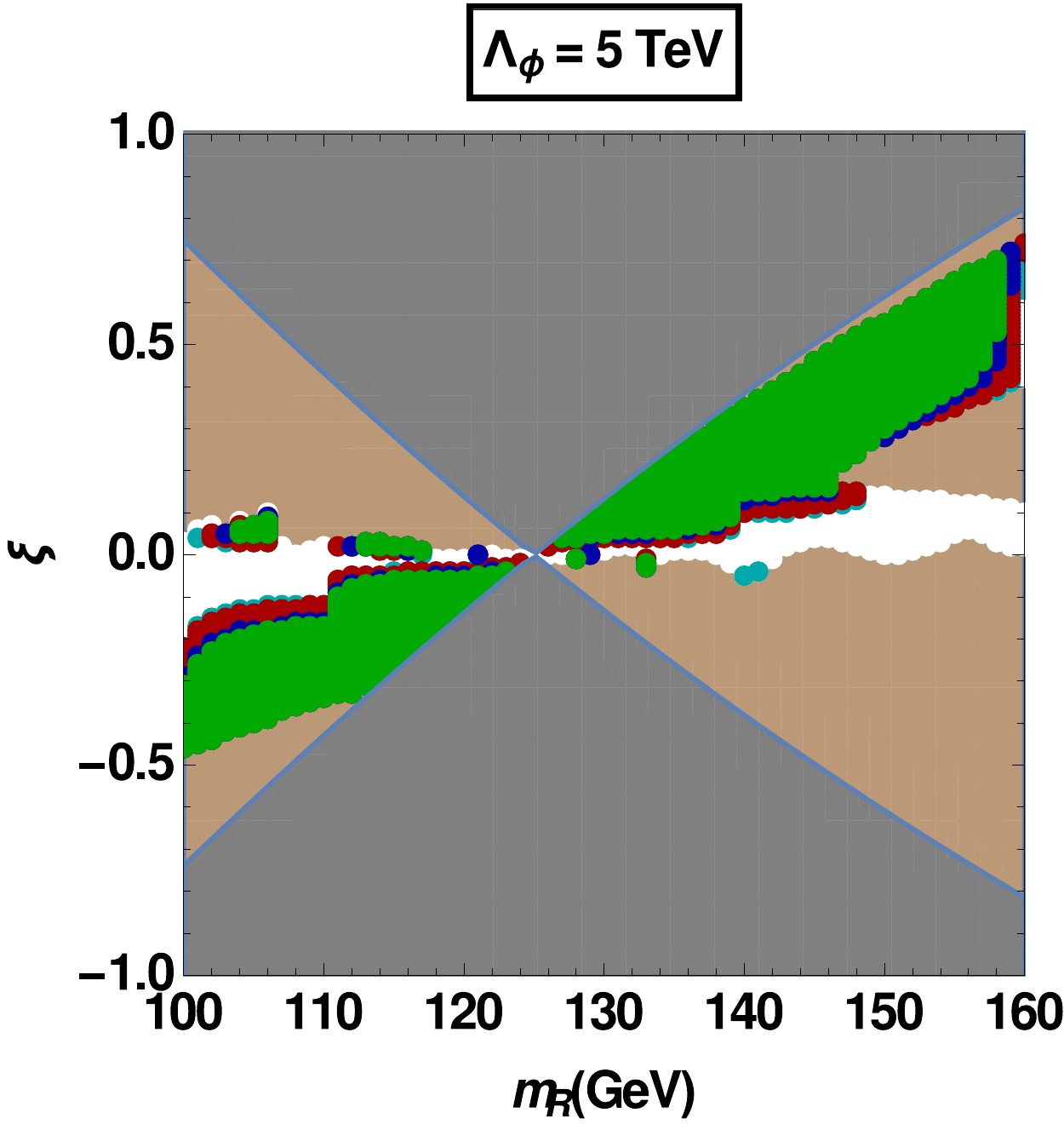}
\caption{Projected {\bf region-2} of the parameter space at $\sqrt{s}$ = 500 and 1000 GeV for 
$\Lambda_{\phi} = 3$ TeV (left panel), $\Lambda_{\phi}=4$ TeV (middle panel) and  $\Lambda_{\phi} =5$ TeV (right panel), for a light mixed state. The green, green+blue, green+blue+red
and green+blue+red+cyan colored regions represent parameter space regions which can be probed at $\sqrt{s}$ = 500, 
$\mathcal{L}$ = 500 fb$^{-1}$, $\sqrt{s}$ = 500 GeV, $\mathcal{L}$ = 1000 fb$^{-1}$, $\sqrt{s}$ = 1000 GeV, $\mathcal{L}$ = 500 fb$^{-1}$ 
and $\sqrt{s}$ = 1000, $\mathcal{L}$ = 1000 fb$^{-1}$, respectively. The interior white region is the one still allowed 
by the LHC 8 TeV results, but which can not be probed by the ILC through this final state.}
\label{fig:exclusion_nunubb}
\end{figure}
We show in Fig.~\ref{fig:exclusion_nunubb}, the projected parameter space in the $\xi-m_R$ plane, through $WW$ fusion
at $\sqrt{s}$ of 500 and 1000 GeV. The region in green is the region which can be probed at $\sqrt{s}$ = 500 GeV 
with an integrated luminosity of 500 fb$^{-1}$, blue and green regions is the one probed with $\sqrt{s}$ = 500 GeV 
and 1000 fb$^{-1}$. The red, blue and green region is the one which can be probed with $\sqrt{s}$ = 1000 GeV and an integrated 
luminosity of 500 fb$^{-1}$. Finally the colored region including cyan, red, blue and green is 
for $\sqrt{s}$ = 1000 GeV and 1000 fb$^{-1}$. The parameter 
space in the $\xi-m_R$ plane for $\Lambda_\phi$ = 3 TeV, which can not be probed by the $ZR$ process, can be probed by $WW$ fusion. The $WW$ fusion process can test most of the $\xi-m_R$ parameter space for 
$\Lambda_\phi$ = 3, 4 and 5 TeV. This is mainly because $WWR$ has a larger cross section compared
to $ZR$, and second, because $ZR$ is further suppressed through the selection of the $Z$ decay channel.
The decay channel of the final state radion  
$R \rightarrow b \bar{b}$  can only probe the $m_R<$ 160 GeV region. The $WW$, $ZZ$ decay channels open up for higher 
values of $m_R$ and have a larger branching ratios compared to $b\bar{b}$. We therefore next consider 
the $WW$ decay channel, so as to probe the sensitivity of ILC for higher radion mass.

\subsection{Analysis in the $e^+e^-\rightarrow \nu_l \bar{\nu}_l R$, $R\rightarrow W^+ W^-$, $W\rightarrow q\bar{q}$ decay channel}
\label{subsec:WW_fusion_WW}
\begin{figure}[htb]
\centering
\includegraphics[width=.3\textwidth, height= 6cm]{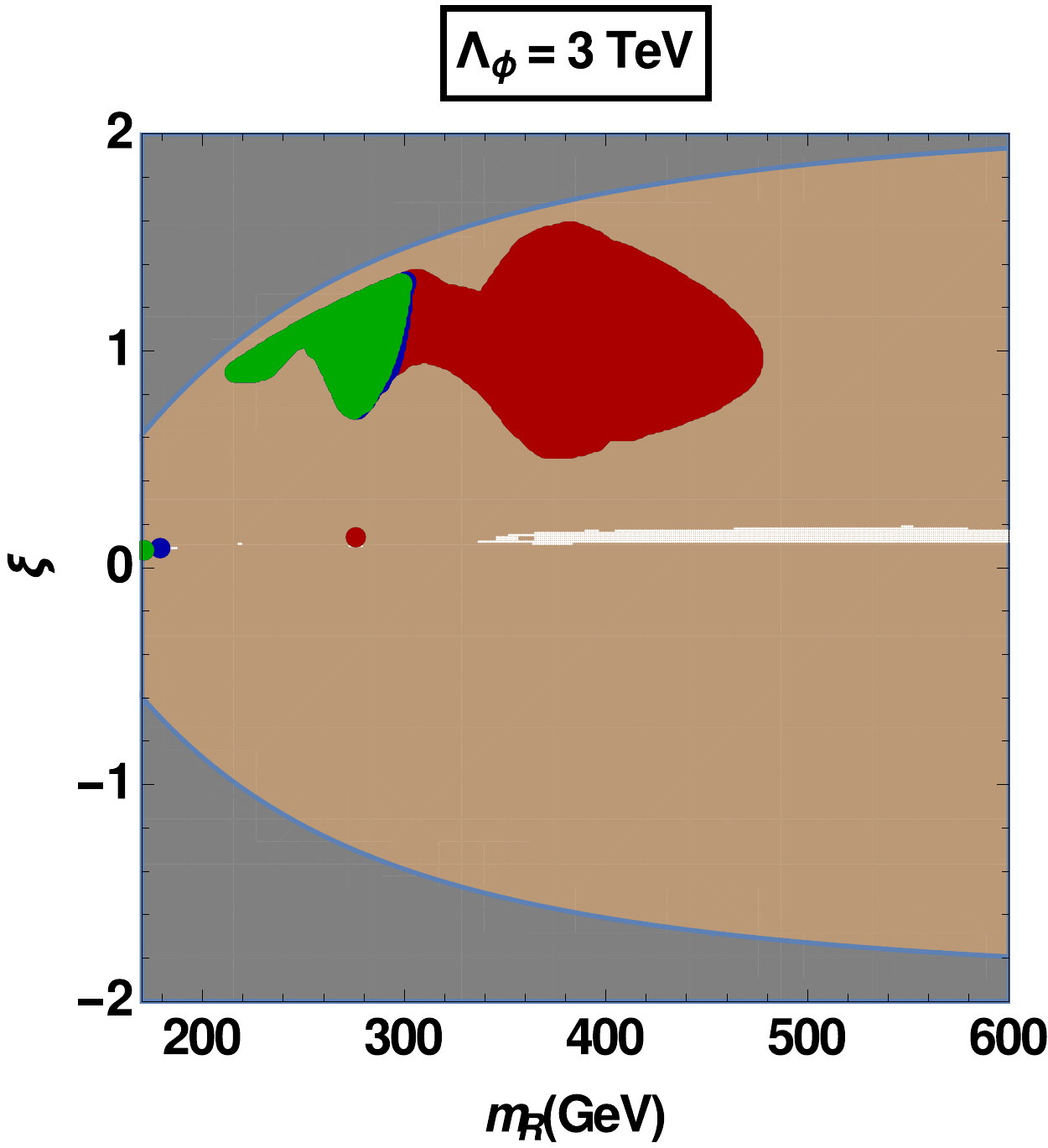}
\includegraphics[width=.3\textwidth, height= 6cm]{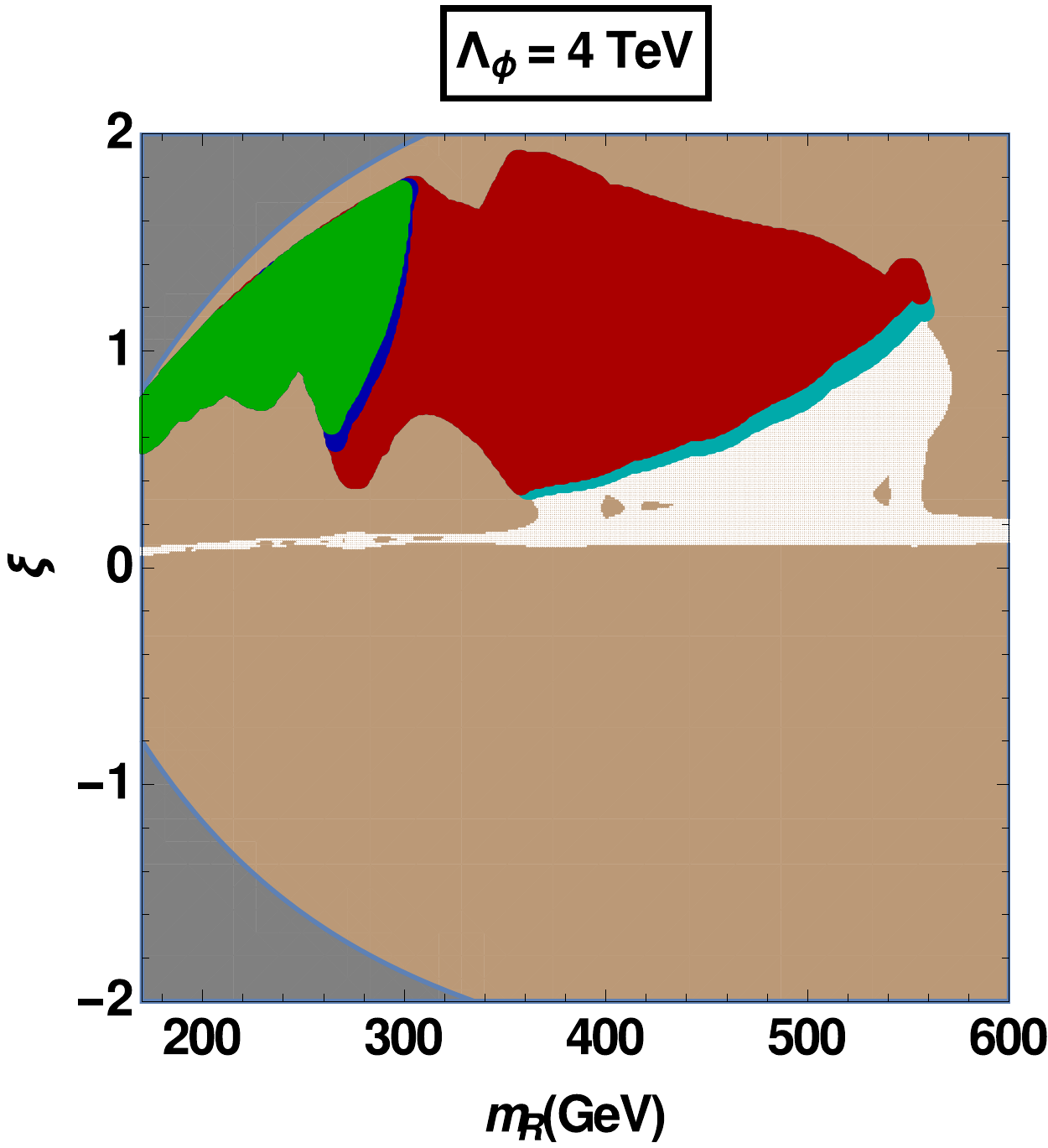}
\includegraphics[width=.3\textwidth, height= 6cm]{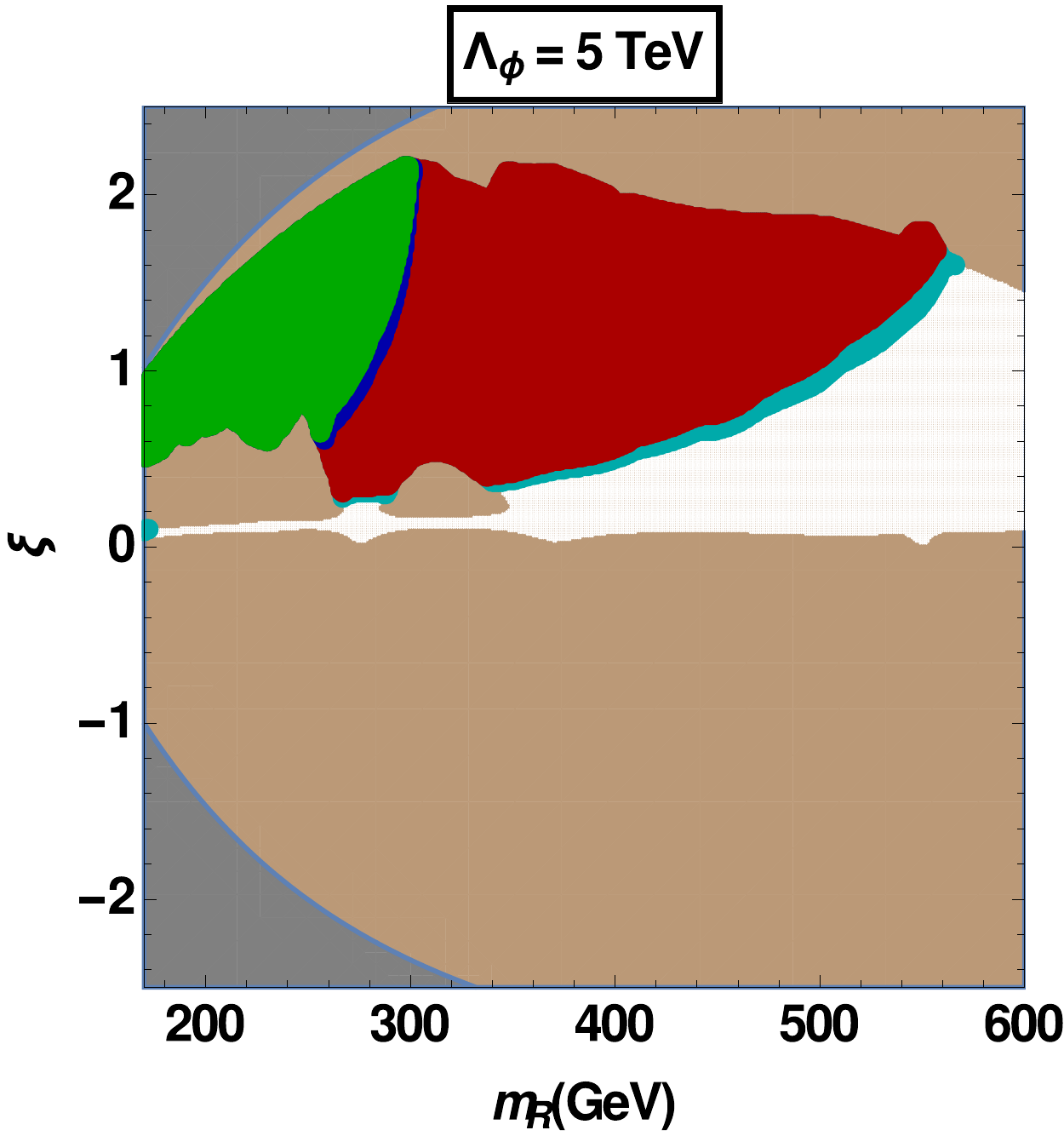}
\caption{Projected {\bf region-2} of the parameter space at $\sqrt{s}$= 500 and 1000 GeV for 
$\Lambda_{\phi} = 3$ TeV (left panel), $\Lambda_{\phi}=4$ TeV (middle panel) and  $\Lambda_{\phi} =5$ TeV (right panel), for a light mixed state. Green, Blue, Red and Cyan 
colored regions represent regions ruled out at ($\sqrt{s}$ = 500, $\mathcal{L}$ = 500 fb$^{-1}$), 
($\sqrt{s}$ = 500, $\mathcal{L}$ = 1000 fb$^{-1}$), ($\sqrt{s}$ = 1000, $\mathcal{L}$ = 500 fb$^{-1}$) 
and ($\sqrt{s}$ = 1000, $\mathcal{L}$ = 1000 fb$^{-1}$) respectively. The interior white region is the one still allowed 
by the LHC 8 TeV results, but can not be probed by the ILC, through this final state.}
\label{fig:exclusion_nunuWW}
\end{figure}

We finally analyze the $WW$ fusion process, with $R$ decaying to 2$W$'s and the $W$'s decaying hadronically.
The final state in this decay mode consists of two missing neutrinos and four jets, none of which is a $b$ jet. The 
events with isolated leptons are first rejected. Then the remaining particles are clustered into 4 jets
with the jets clustering algorithms. The main background processes are similar to the $WW$ fusion 
($R\rightarrow b\bar{b}$) analysis, dominated by $W^+W^-$ and $\nu\bar{\nu}Z$. The jets for the signal are
originating from $R$ which decays to 2$W$'s, therefore we demand that the total number of particles passing for the  
jet clustering algorithm be greater than 30~\cite{Durig:2014lfa}. This reduces the background processes with the jets constructed from
a single $W/Z$. The events with total $p_T<$ 20 GeV are rejected and the cuts are applied on the visible mass,
visible energy and missing mass, similar to what we discussed in~\ref{subsec:WW_fusion}. We do not show the $m_{\rm vis}, E_{\rm vis}$
distributions, as they will have the same pattern as Fig.~\ref{fig:distribution_150GeV}.  The 
four reconstructed jets are then paired to give two on-shell W bosons, with $m_W \pm 11$ GeV.  The reconstructed 
radion invariant mass from the on-shell $W$'s is required to be in the mass range $m_R \pm 10$ GeV. 
The $b$ jet veto is used to reduce the background from $H/R \rightarrow b\bar{b}, t\bar{t}$ final states 
or the $t\bar{t}$ pair production. It is difficult to study this mode in the LHC as it yields a hadronic final state.
The leptonically decaying $W$'s can be studied in LHC, but as discussed before this will prove difficult with  
radion mass reconstruction due to missing neutrinos. We therefore study the all hadronic modes at the ILC, and show,  in 
Fig.~\ref{fig:exclusion_nunuWW}, the parameter space that can be probed at ILC, at $\sqrt{s}$  500 and 1000 GeV. 
The color coding is the same as Fig.~\ref{fig:exclusion_nunubb}. There is a narrow white region, near $\xi$ = 0,
which cannot be probed at the ILC, this corresponds to {\bf region 1}, where the coupling of the radion with the
SM fermions and the massive gauge bosons vanishes, and only the coupling with massless gauge bosons exists. 
A through study of the hadronically decaying $WW$ final state in ILC will be able to constrain most of {\bf region-2}, for 
170 GeV $<m_R<$ 540 GeV. In the next section we discuss how the LHC and the ILC analysis will complement each other in constraining the entire $\xi-m_R$ parameter space.

\section{Discussions and Conclusions}
\label{sec:conclusion}
In what follows we summarize the results and indicate the conclusions to be drawn from our analysis.
We have studied the phenomenology of the radion like mixed state in
the custodial RS model, where all the SM particles except the Higgs
boson live in the bulk. We discussed the $\xi-m_R$ parameter space in the context of recent
LHC results from the heavy Higgs, Higgs signal strength and the 750 GeV resonance.
There are still allowed regions in the $\xi-m_R$ parameter space, which can be 
thoroughly examined in the LHC and the planned ILC.

The existing parameter space depending on $\Lambda_\phi$ is divided in two regions
for our analysis.  {\bf Region-1}, where the radion couples maximally to the massless
gauge bosons, can be studied thoroughly in the 13 and 14 TeV LHC, through the $\gamma\gamma$
final state.  We discussed the prospect of direct searches for the radion at the LHC through
the $\gamma\gamma$ decay mode and have shown the regions in the $\xi-m_R$ plane which can 
be probed this way. We find that with this decay channel the 14 TeV LHC with an integrated 
luminosity of 1000 fb$^{-1}$ can probe {\bf region 1} of a radion in the mass range 
80 $<m_R<$ 350 GeV with more than 3$\sigma$. There are also regions in the $\xi-m_R$ parameter
space, for $m_R\leq$ 200 GeV, which the 13 TeV LHC can probe with an integrated luminosity 
of 50 fb$^{-1}$ only. {\bf Region-2}, with the radion coupling maximally to the massive 
gauge bosons and $b$ quarks, can be studied in both the LHC and the ILC.
The massive gauge bosons from the radion decay further decay into quarks or leptons. 
Though the all hadronic final state has a large LHC background, it is possible to discover 
a heavy radion through the leptonic decay modes of the gauge bosons in a high luminosity LHC.
We therefore considered the $ZZ$ decay mode of the radion, with the $Z$ decaying leptonically, for
the 14 TeV LHC. We found that this channel extends the discovery reach of the LHC
in the  $\xi-m_R$ parameter space beyond what is tested through the diphoton mode, 
till $m_R$ of 450 GeV. 

As seen in Fig.~\ref{fig:lowmass},~\ref{fig:highmassZZ}, for $\Lambda_\phi$ = 4, 5 TeV, 
there are regions in the parameter space where the LHC cannot 
discover the radion or can not constrain the $\xi-m_R$ parameter space. This white region is 
{\bf region-2} where the radion decays dominantly to $b$ ($m_R<$ 160 GeV) or $WW/ZZ$ ($m_R>$ 160 GeV). 
This region can be thoroughly investigated in the  ILC, through the $ZR$ and the 
$WWR$ production mode, shown in Figs.~\ref{fig:exclusion_ZHqq},~\ref{fig:exclusion_nunubb},~\ref{fig:exclusion_nunuWW}.
In order to highlight the complementarity of the ILC and the LHC, we investigate some 
benchmark points for $m_R$ and show the range of $\xi$ ($\xi_{probed}$) that can be explored 
by both the machines. We again note that for $m_R<$ 100 GeV, the entire allowed parameter 
space of $\xi$ can be explored in the LHC for $\Lambda_\phi$ = 3 and 4 TeV.
In Table~\ref{tab:allowed4TeV} we list the range of $\xi$ that can be probed at 3$\sigma$
and 5$\sigma$ for selected values of $m_R$ and for $\Lambda_\phi$ = 4 TeV.  We also list 
as $\xi_{exp}$ the domain of $\xi$, which survives both the theoretical and the experimental 
constraints from 8 TeV LHC. The production and the decay modes of the radion  
considered for the analysis are also listed in the Table. We presented the results for the LHC 
for a luminosity of 150 fb$^{-1}$ for $\sqrt{s}$ = 13 TeV (LHC13) and 
3000 fb$^{-1}$ for $\sqrt{s}$ = 14 TeV (LHC14). In case of ILC, the smallest possible
combination of $\sqrt{s}$ and $\mathcal{L}$, which can probe the 
maximum $\xi - m_R$ parameter space is listed (ILC$^{\sqrt{s}}_{\mathcal{L}}$).
\begin{table}[htb]
\begin{center}
\begin{tabular}{|c|c|c|c|c|} \hline
$(m_R$, $\xi_{exp})$&Production mode &Decay mode &$\xi^{3\sigma}_{probed}$ & $\xi^{5\sigma}_{probed}$\\ \hline
 &&&& \\
                        &$gg \rightarrow R$ (LHC13)& $R\rightarrow \gamma\gamma$ & -0.09 $-$ -0.04 & -0.07 $-$ -0.04\\
110 GeV,                &$gg \rightarrow R$ (LHC14) & $R\rightarrow \gamma\gamma$ & -0.14 $-$ -0.04 & -0.13 $-$ -0.04\\
 (-0.19 $-$ -0.04)      &$ZR, Z\rightarrow q \bar{q}$ (ILC$^{250}_{1000}$)& $R\rightarrow b \bar{b}$ & -0.19 $-$ -0.17 &$\times \times$\\ 
                        &$WWR$ (ILC$^{1000}_{500}$)& $R\rightarrow b \bar{b}$  & -0.19 $-$ -0.11 &-0.19 $-$ -0.12\\  \hline
                        &$gg \rightarrow R$ (LHC13) & $R\rightarrow \gamma\gamma$ & 0.01 $-$ 0.06 & 0.01 $-$ 0.04\\                        
140 GeV,                &$gg \rightarrow R$ (LHC14)& $R\rightarrow \gamma\gamma$ & 0.01 $-$ 0.09 & 0.01 $-$ 0.08\\
 (0.01 $-$ 0.28)        &$ZR, Z\rightarrow q \bar{q}$ (ILC$^{250}_{1000}$) & $R\rightarrow b \bar{b}$ & 0.28 &$\times \times$   \\ 
                        &$WWR$ (ILC$^{500}_{500}$)& $R\rightarrow b \bar{b}$  &  0.13 $-$ 0.28  & 0.15 $-$ 0.28\\  
                        &$WWR$ (ILC$^{1000}_{1000}$)& $R\rightarrow b \bar{b}$  &  0.09 $-$ 0.28 & 0.1 $-$ 0.28\\\hline
                         &$gg \rightarrow R$ (LHC13)& $R\rightarrow \gamma\gamma$ & 0.09 $-$ 0.11 & 0.09 $-$ 0.11\\                        
200 GeV,                 &$gg \rightarrow R$ (LHC14)& $R\rightarrow \gamma\gamma$ &  0.09 $-$ 0.11 & 0.09 $-$ 0.11\\
(0.09 $-$ 0.11,  0.78 $-$ 1.05) &$gg \rightarrow R$ (LHC14)& $R\rightarrow ZZ$ &$\times \times$ &$\times \times$\\    
                        &$WWR$ (ILC$^{500}_{500}$)& $R\rightarrow W^+ W^-$ & 0.78 $-$ 1.05 & 0.78 $-$ 1.05\\ 
                        &$WWR$ (ILC$^{1000}_{500}$)& $R\rightarrow W^+ W^-$  & 0.78 $-$ 1.05 & 0.78 $-$ 1.05\\  \hline

                        &$gg \rightarrow R$ (LHC13)& $R\rightarrow \gamma\gamma$ & $\times \times$ & $\times \times$\\
280 GeV,                &$gg \rightarrow R$ (LHC14)& $R\rightarrow \gamma\gamma$ &  0.09 $-$ 0.16 &0.09$-$ 0.16,\\
(0.09 $-$ 0.16, 0.44 $-$ 1.64) &$gg \rightarrow R$ (LHC14)& $R\rightarrow ZZ$ &0.09$-$ 0.16,  0.44 $-$ 0.88 & 0.09$-$ 0.16,  0.44 $-$ 0.67 \\  
                        &$WWR$ (ILC$^{500}_{1000}$)& $R\rightarrow W^+ W^-$ & 1.67 $-$ 1.74 &$\times \times$\\ 
                        &$WWR$ (ILC$^{1000}_{500}$)& $R\rightarrow W^+ W^-$  & 0.73 $-$ 1.74  & 0.73 $-$ 1.74\\  \hline
                       &$gg \rightarrow R$ (LHC13)& $R\rightarrow \gamma\gamma$ &$\times \times$ &$\times \times$ \\
 400 GeV,              &$gg \rightarrow R$ (LHC14)& $R\rightarrow \gamma\gamma$ & $\times \times$ &$\times \times$\\
 (0.11 $-$ 1.72)       &$gg \rightarrow R$ (LHC14)& $R\rightarrow ZZ$ &0.2 $-$ 0.47, 1.67 $-$ 1.72 & $\times \times$\\   
                       &$WWR$ (ILC$^{500}_{1000}$)& $R\rightarrow W^+ W^-$ & $\times \times$ &$\times \times$\\ 
                       &$WWR$ (ILC$^{1000}_{1000}$)& $R\rightarrow W^+ W^-$  & 0.43 $-$ 1.72  & 0.52 $-$ 1.72\\  \hline
                       &$gg \rightarrow R$ (LHC13)& $R\rightarrow \gamma\gamma$ & $\times \times$ &$\times \times$\\
500 GeV,               &$gg \rightarrow R$ (LHC14)& $R\rightarrow \gamma\gamma$ & $\times \times$ &$\times \times$\\
(0.11 $-$ 1.49)        &$gg \rightarrow R$ (LHC14)& $R\rightarrow ZZ$ &$\times \times$ &$\times \times$\\   
                       &$WWR$ (ILC$^{1000}_{500}$)& $R\rightarrow W^+ W^-$ & 0.89 $-$ 1.49 & 1.07 $-$ 1.49\\ 
                       &$WWR$ (ILC$^{1000}_{1000}$)& $R\rightarrow W^+ W^-$  & 0.79 $-$ 1.49 & 0.95 $-$ 1.49\\  \hline                        
\hline
\end{tabular}
\caption{The range of $\xi$ ($\xi_{probed}$) that can be tested at the ILC and the LHC, 
in conjunction with considered values of $m_R$, for $\Lambda_\phi$ = 4 TeV. The allowed region
of $\xi$ from the theory and the 8 TeV LHC results is denoted by $\xi_{exp}$. The mass
and the $\xi$ region which cannot be probed through $\gamma\gamma$, $ZZ$ final state in the LHC 
and the $b\bar{b},~W^+ W^-$ final state in the ILC is denoted by ($\times \times$).}
\label{tab:allowed4TeV}
\end{center}
\end{table}

We briefly discuss the results for $\Lambda_\phi$ = 3 TeV before discussing the results for 4 TeV. The allowed range 
of $\xi - m_R$ for $\Lambda_\phi$ = 3 TeV is tightly constrained by the LHC results.
 From Fig.~\ref{fig:lowmass} it can be seen that a radion of mass up to 120 GeV can be fully explored by the LHC.
The ILC will be at a disadvantage for this low mass range because of a large background
from the $Z$ resonance and the SM Higgs. The complementarity of the ILC and the LHC is therefore 
visible for $m_R>$ 140 GeV. The experimentally allowed range of $\xi$ for $m_R$ = 140 GeV is (0.04 $-$ 0.13), 
and we find that the 14 TeV LHC with 3000 fb$^{-1}$ can probe at 3$\sigma$ the lower $\xi$ 
range ((0.04$-$0.09) {\bf region-1}), through the diphoton decay channel.
The 500(1000) GeV ILC with integrated luminosity of 500(1000) fb$^{-1}$ probes the $\xi$ range (0.11(0.08)$-$0.13).
This clearly shows how the two colliders will complement each other in probing the $\xi-m_R$ parameter space.
The radion of mass greater than 250 GeV can not be probed by the diphoton channel of LHC, but can be probed through the 
$ZZ$ final state, with the $Z$ decaying leptonically. The ILC running at a higher c.m. energy and higher 
luminosity can also probe this high mass region. 

We next present the results for $\Lambda_\phi$ = 4 TeV, in Table~\ref{tab:allowed4TeV}, where the 
complementarity of the ILC and the LHC is more pronounced for $m_R<$ 300 GeV. Let us consider $m_R$ = 140 GeV, for which the allowed 
region is $\xi_{exp}$ = (0.01 $-$ 0.28). It can be seen from the Table that the 14 TeV (3000 fb$^{-1}$) LHC, through 
the diphoton channel, and the ILC operating at $\sqrt{s} =1000$ GeV can fully explore the parameter space with 1000 fb$^{-1}$.
The same argument holds for the radion mass in the range of 200 to 400 GeV. The available parameter space 
for the radion with mass less than 400 GeV consists of two distinct regions, one near $\xi$ = 0 and another for $\xi >$ 0.5. 
The narrow band near $\xi$ = 0 is fully explored by the LHC through the diphoton channel. The $ZZ$ decay channel can help the 
LHC to probe the region with $\xi>$ 0.5, but it cannot fully explore the region. However, the 1000 GeV ILC, through the $WWR$
production mode, can fully probe the $\xi>$ 0.5 parameter space. The two distinct regions of $\xi$ merge for $m_R\geq$ 400 GeV
and this mass range cannot be tested at the LHC through the diphoton channel. The $WWR$ channel in the ILC can only probe the upper 
region of $\xi$. The LHC analysis of the radion in the $ZZ$ decay channel helps in probing the region for this mass, which
cannot be explored in the ILC. However, the $ZZ$ decay channel can only probe for a radion of mass less than 450 GeV, after 
which the  production cross section is too small to be observed in the LHC. The 1000 GeV ILC can also probe,
through the $WWR$ channel, regions of $\xi$  for heavier radions as long as
the radion mass is kinematically accessible. 

In conclusion, we showed that, using the coherent analysis at
the ILC and the LHC, the mixed Higgs-radion state can be revealed, its mass can be verified, and the mixing sector fully explored.

\section{Acknowledgments}

MF thanks NSERC  for partial financial support under grant 
number SAP105354. 
KH acknowledges the H2020-MSCA-RICE-2014 grant no. 645722 (NonMinimalHiggs).
UM and MP would like to thank Abhishek M. Iyer and Tousik Samui for discussions.
UM would like to thank the Department of Theoretical Physics, TIFR, for the use of its computational resources.
UM would also like to thank the hospitality of Concordia University, Montreal and Helsinki Institute of Physics,
Helsinki where this work was initiated.
MP would like to thank the Dept. of Theoretical Particle Physics, LMU, Munich for hospitality 
during the initial stage of this work. The work of MP is supported by the Croatian Science Foundation (HrZZ) 
project "Physics of Standard Model and Beyond", HrZZ 5169 as well as by the H2020 Twinning project,
"RBI-T-WINNING".

\end{document}